\documentclass[a4paper,12pt]{article}
\pdfoutput=1 
\usepackage[margin=1in,includefoot]{geometry}
\usepackage{amsmath,amssymb,amsthm,eufrak} 
\usepackage[pdfstartview=FitH,pdfpagemode=None]{hyperref}
\usepackage[noadjust]{cite} 
\usepackage{authblk} 

\usepackage{xfrac} 
\usepackage{cancel} 
\usepackage{comment} 
\usepackage{empheq} 

\usepackage[sans]{dsfont} 

\DeclareFontFamily{U}{dsss}{}
\DeclareFontShape{U}{dsss}{m}{n}{
  <-10>   s*[1] dsss8
  <10-12> s*[1] dsss10
  <12->   s*[1] dsss12
}{}
\DeclareMathAlphabet{\mathds}{U}{dsss}{m}{n}

\usepackage[scaled=1.2]{dsserif} 

\usepackage{color} 
\usepackage{soul}

\usepackage{graphicx}
\usepackage{subcaption}
\usepackage{booktabs}  

\usepackage{float} 

\usepackage{scalerel} 

\usepackage{multicol}

\usepackage{diagbox} 

\numberwithin{equation}{section} 

\def\be{\begin{equation}}
\def\ee{\end{equation}}
\def\bea{\begin{eqnarray}}
\def\eea{\end{eqnarray}}

\newcommand{\Tr}{{\rm Tr}}

\newcommand{\grlw}{\, \substack{>\vspace{-2.5pt}\\<} \,}
\newcommand{\menodet}[2]{{\scriptscriptstyle{-}} \hspace{1pt} #1_{{\hspace{-0.5pt}}_{\scaleto{(#2)}{5pt}}}}
\newcommand{\piudet}[2]{#1_{{\hspace{-0.5pt}}_{\scaleto{(#2)}{5pt}}}}

\renewcommand{\to}{\rightarrow}

\def\nb{\nonumber}

\newcommand{\rmd}{\,\mathrm{d}}

\newcommand{\ie}{i.e.,\ }
\newcommand{\eg}{e.g.,\ }

\newcommand{\pt}{\hspace{1pt}}

\def\({\left (}
\def\){\right )}

\let\benn\[
\let\eenn\]

\def\[{\left [}
\def\]{\right ]}

\let\oldlgraf\{ 
\renewcommand{\{}{\left \oldlgraf}

\let\oldrgraf\}
\renewcommand{\}}{\right \oldrgraf}

\title{\bfseries\LARGE Flux tube profile from Holography:\protect\\ 
finite size and strong coupling corrections}

\author[a,b]{Tommaso Canneti}

\affil[a]{Dipartimento di Fisica, Universit\`a di Torino %
\protect\\ Via Pietro Giuria 1; 10125 Torino, Italy}
\affil[b]{Istituto Nazionale di Fisica Nucleare, Sezione di Torino %
\protect\\ Via Pietro Giuria 1; 10125 Torino, Italy}

\date{\small tommaso.canneti@unito.it}
\begin{document}
\maketitle

\begin{abstract}
We use the holographic correspondence as a tool to study the classical flux tube profile connecting a static quark-antiquark pair in a $2+1$-dimensional strongly-coupled large $N$ QCD-like theory. The final result extends already known findings in the literature in several ways. First, it is an analytical function of both the space-like boundary coordinates; in other words, we keep track of what happens both along and transversely to the inter-quark axis. Then, we take into account the finiteness of the inter-quark distance and the first correction in the strong coupling expansion.  
To the same order, we also confirm the relation between the mass of the lightest glueball in the spectrum and the intrinsic width of the flux tube profile. We conclude by trying to gain some insights about the quantum fluctuations. Intriguingly, our proposal is in agreement with widespread expectations in the literature. En passant, we also derive a semi-analytical formula that gives the first correction to the scalar glueball masses in the strong coupling expansion.

\end{abstract}

\newpage
\tableofcontents


\section{Introduction}

A trademark of confining gauge theories is the emergence of a narrow, fluctuating flux tube connecting color charges. This picture is supported by both lattice simulations \cite{DiGiacomo:1989yp, DiGiacomo:1990hc, Singh:1993jj, Bali:1994de, Bali:2000gf, Takahashi:2000te, Bissey:2006bz} and indirect signatures, such as the properties of jets coming from high-energy collisions \cite{Andersson:1980vk, JADE:1981ofk, JADE:1983ihf, JADE:1983mvo, CELLO:1983pbu, TASSO:1985cia, TPCTwoGamma:1984tkq, Azimov:1985zta} and the observations of Regge trajectories \cite{Anisovich:2000ut, Anisovich:2001pn, Anisovich:2001pp, Anisovich:2002su, Anisovich:2002xoo, Bugg:2004xu}. 
Accordingly, the study of the flux tube physics plays a key role in advancing our understanding of the confining mechanism.

A successful description of the flux tube at large inter-quark distance $L$ is provided by an effective string theory (EST) approach to confinement \cite{Luscher:1980ac, Luscher:1980fr}. The flux tube is thus modeled as a one-dimensional vibrating string stretching between a static quark-antiquark pair, whose finite thickness arises purely from its quantum fluctuations. It follows that the squared width of the flux tube in a $d+1$-dimensional confining model at zero temperature is given by \cite{Luscher:1980iy}
\be\label{quantumbroad}
w^2 = \frac{d-2}{2\pi\sigma} \log \(\frac{L}{L_0}\) \, ,
\ee
where $\sigma$ is the string tension. This formula has been corroborated by several lattice studies \cite{Caselle:1995fh, Zach:1997yz, Koma:2003gi, Panero:2005iu, Giudice:2006hw, Rajantie:2012zn, Amado:2013rja, Pennanen:1997qm, Chernodub:2007wi, Bakry:2010zt, Cardoso:2013lla, Gliozzi:2010zv, Gliozzi:2010zt}. Here, $L_0$ is a parameter with units of length that cannot be predicted by EST itself. Therefore, it sets a lower bound on the validity of the effective description.

The above result derives from modeling the effective theory through the Nambu-Goto action \cite{nambu1970duality, Goto:1971ce, Hara:1971ur}. Nevertheless, because of inconsistencies at quantum level, it is well-known that the latter cannot capture the whole story in three or four dimensions. Notice that the flux tube originates from the squeezing of the color fields within a finite subregion of the space, as a result of confinement. Therefore, a more realistic description should deviate from the ideal one-dimensional representation of EST. In other words, the flux tube should feature a finite-size core regardless of the quantum fluctuations. This defines an ``intrinsic width'' --- completely missed by EST ---  
to be considered as the residual thickness of the flux tube whenever the inter-quark separation is pushed down to $L_0$. The existence of such a fundamental scale 
affects the behavior of the flux tube profile along the transverse direction to the inter-quark axis. Indeed, moving away from the sources, EST predicts a Gaussian shape, as recently confirmed in \cite{Caselle:2024ent}. Nevertheless, Monte Carlo simulations are compatible with an exponential drop-off (see, \eg \cite{Cea:2012qw, Verzichelli:2025cqc, Caselle:2026coc, Amorosso:2026mdo}).

An alternative description of the flux-tube dates back to the 1970s, when Nambu \cite{Nambu:1974zg}, Mandelstam \cite{Mandelstam:1974pi} and ’t Hooft \cite{tHooft:1979rtg} rephrased the confinement of quarks in terms of the Cooper-pair condensation in superconductors (see \cite{Baker:1989qp, Baker:1991bc} for recent reviews of the topic). In this sense, the flux tube is associated to the formation of an Abrikosov vortex in the dual model. Numerically, this proposal has been investigated in \cite{Cea:1992sd, Cea:1992vx, Cea:1993pi, Cea:1994ed, Cea:1994aj, Cea:1995zt, Cea:1995ga, Cardaci:2010tb, Cea:2012qw, Cea:2014uja, Cea:2015wjd, Cea:2017ocq}. Within this framework, the flux tube profile is expected to drop off exponentially as the distance from the inter-quark axis increases. Let us stress that the length scale of decay, related to the London length of the dual Abrikosov vortex, turns out to be independent of $L$. So, it is natural to identify it with the intrinsic width mentioned above. 

It would be certainly interesting to provide a full expression for the profile that bridges the Abrikosov-like behavior of the flux tube with the EST description of its quantum fluctuations. For instance, in \cite{Cardoso:2013lla}, the authors proposed as an ansatz the convolution of a typical classical profile with a Gaussian distribution that encodes the quantum oscillations of the flux tube. Intriguingly, the same structure emerges for long strings in massive QED in 2+1-dimensions \cite{Aharony:2024ctf},\footnote{Remarkably, some of the results in \cite{Aharony:2024ctf} have been confirmed from a numerical point of view in a very recent paper \cite{Caselle:2025vhx}.} where the classical component can be identified with the on-shell electric field configuration. This provides a precise measure for the intrinsic width of the profile and for its coherence length, \ie a measure of its curvature at the center of the flux tube. Anyway, for general strongly-coupled QCD-like models, such a classical solution is not known and the problem is highly non-trivial. 

Over the years, the holographic correspondence has been a truly powerful tool for the study of strongly coupled gauge theories in the large $N$ limit. 
The first realization of the holographic principle \cite{tHooft:1993dmi, Susskind:1994vu} in string theory resulted in the so-called $AdS$/CFT correspondence \cite{Maldacena:1997re, Gubser:1998bc, Witten:1998qj}. Later, this whole construction was also extended to certain non-supersymmetric and non-conformal gauge theories \cite{Witten:1998zw}, making some observables accessible in more realistic scenarios from the gravity side. In \cite{Polchinski:1991ax}, Polchinski and Strominger suggested that a fundamental string propagating in some curved background is expected to encode the description of the confining string at low energies. Nevertheless, they did not draw any conclusions about the underlying short-distance theory. Holography may fill this gap by providing some good candidates (see, \eg \cite{Aharony:2009gg, Aharony:2010cx}). 
Anyway, let us stress that the first indications about the duality between gauge and string theories dates back to well before the holographic principle. Indeed, in 1974, ‘t Hooft already noticed how the large $N$ expansion of a $SU(N)$ Yang-Mills theory resembles the weak coupling expansion of a string theory having string coupling
\be\label{gsN}
g_s \sim 1/N \, ,
\ee
keeping the `t Hooft coupling of the gauge theory fixed as $N$ goes to infinity \cite{tHooft:1973alw}.

In \cite{Witten:1998zw}, Witten considered a three-dimensional confining gauge theory describing the infrared dynamics of $N$ D3-branes wrapped along a compact space direction on a circle of radius $R_0$. This fixes the critical temperature of the confinement/deconfinement phase transition as
\be\label{TcR0}
T_c = \frac{1}{2\pi R_0} \, .
\ee
To get there, the strategy is to compactify the four-dimensional $(\mathcal N = 4)$ Super Yang Mills theory on the same circle with supersymmetry-breaking boundary conditions for the fermions. At low energies, the model reduces to a three-dimensional non-supersymmetric $SU(N)$ Yang-Mills theory coupled to a tower of massive Kaluza-Klein (KK) modes. In the strongly coupled large $N$ limit --- where the theory features a reliable Type IIB supergravity description --- the latter are at the same mass scale as the glueballs, namely
\be\label{MKKR0}
\Lambda_{\text{YM}} \sim M_{\text{KK}}=R_0^{-1} \, ,
\ee
where $\Lambda_{\text{YM}}$ is the Yang-Mills dynamical scale. This can be formalized by expressing the relation among these scales as~\cite{Gross:1998gk}
\be\label{running}
\lambda_3 \sim \frac{M_{\text{KK}}}{\log\(\frac{M_{\text{KK}}}{\Lambda_{\text{YM}}}\)} \, ,
\ee
$\lambda_3$ being the `t Hooft coupling of the three dimensional gauge theory at $M_{\text{KK}}$.

In this picture, the potential between two massive quarks in the gauge theory is encoded in the minimal world-sheet action of a fundamental string stretching between the color charges and diving into the bulk \cite{Rey:1998ik, Maldacena:1998im}. As the inter-quark distance $L$ increases, the fundamental string takes on a ``bathtub'' profile lying at the bottom of the geometry. 
Correspondingly, its energy is dominated by a term proportional to $L$ itself giving confinement \cite{Kinar:1999xu, Greensite:1999jw, Kinar:1998vq, Greensite:1998bp}. 
This is possible since a non-vanishing warp factor in front of the Minkowskian sector of the metric --- shared with the gauge theory --- prevents the string to probe arbitrarily infrared regions in the bulk \cite{Witten:1998zw}. 

The presence of such a string in the gravity theory, being a source for the dilaton field, translates into a non-vanishing vacuum expectation value for the dual boundary operator, \ie the Yang-Mills Lagrangian density \cite{Klebanov:1997kc, Klebanov:1999xv, Gubser:1998bc,Witten:1998qj}. The latter shall be confined within the flux tube of the three-dimensional theory. Moreover, it can be explicitly expressed as a function of the position, giving a measure of the profile we are looking for. Its quantum broadening has been deeply studied in the holographic literature \cite{Armoni:2008sy, Giataganas:2015yaa, Giataganas:2015xna, Greensite:2000cs, Loewy:2001pq, Ridgway:2009tca}, extending the prescription by Lusher et al.~\cite{Luscher:1980iy} and reproducing the logarithmic behavior in \eqref{quantumbroad}. On the other hand, quantum fluctuations can be easily frozen out by using as a source a classical fundamental string. In this way, we directly have access to the 
classical description of the flux tube, as well as to a measure of its intrinsic width. This resembles the approach in \cite{Cardoso:2006mf, Cardoso:2010kw}. There, the authors computed the very same parameter in two models for the magnetic confinement in superconductors by approximating the electromagnetic field at classical level. As a consequence, no quantum broadening of the flux tube appears. To fix notations, in the following we will refer to the flux tube induced by a classical open string joining the quark-antiquark pair at the boundary as the \emph{classical flux tube}.

In \cite{Danielsson:1998wt}, relying on the gauge/gravity correspondence as above, the authors provided an analytical prediction for the classical flux tube profile generated by an infinite fundamental string that probes the three-brane solution from before. This proposal has been derived in the large (strictly infinite) `t Hooft coupling limit. Nevertheless, from \eqref{running}, we expect that pure YM is realized in the opposite regime. Therefore, for a more quantitative comparison with the lattice data, the analysis should be pushed toward large but finite values of the coupling. In this sense, the first subleading stringy correction to the background at hand turns out to be crucial. Indeed, to anticipate, the holographic dictionary translates the higher derivative expansion in the gravity theory into the strong ’t Hooft coupling expansion in the dual gauge theory \cite{Witten:1998qj, Gubser:1998bc}. This class of corrections arises from the finiteness of the fundamental string, which can probe the curvature of the target space whenever the latter is comparable to the string length
\be
\ell_s = \sqrt{\alpha'} \, .
\ee
The above, in units of the characteristic radius of the background, is indeed used as the expansion parameter in the gravity model. Let us stress that each higher derivative term in the gravity action can be equipped with further corrections coming from the loop expansion in the string coupling. Clearly, from  \eqref{gsN}, this corresponds to the large $N$ expansion in the dual gauge theory. However, it will turn out that the tree level results in \cite{Gubser:1998nz, Pawelczyk:1998pb} will be enough for our purposes.

In this work, we extend the holographic prediction for the flux tube both at finite inter-quark distance\footnote{The reader can find some numerical results in \cite{Vyas:2019kvy}.} and at finite strong coupling. On one hand, this allows us to perform a more refined version of the computation and to clarify the perturbative scheme adopted. On the other, we provide a more realistic measure of the intrinsic width 
of the flux tube, ready to be compared with some large $N$ lattice simulations. In particular, we formally confirm at next-to-leading order (NLO) in the coupling expansion some expectations in the literature about the asymptotic behavior of the profile. Namely, we identify its intrinsic width with the (inverse) mass of the lowest lying state in the scalar glueball spectrum, up to NLO in the stringy corrections. En passant, we also provide a semi-analytical formula to compute the NLO corrections to the masses of all the other excited states, reproducing the numerical results in \cite{Ooguri:1998hq, Csaki:1998qr, deMelloKoch:1998vqw}. 

The paper is structured as follows. In section \ref{sec:setup}, we untangle the holographic relation between the on-shell dilaton field configuration and the measure of the classical flux tube profile we are interested in, up to NLO in the strong coupling expansion. This provides the reader with all the necessary tools to derive our prediction
. Those who are not interested in the technical details can skip ahead to section \ref{sec:fluxtube}, where the complete expression for the classical flux tube profile is presented. There, we also compute the intrinsic width of the profile together with our semi-analytical formula for the corrected glueball masses; furthermore, we analyze some interesting limiting cases and how it is possible to include the quantum fluctuations in the game. The discussion about our general proposal is deferred to the conclusions. In this way, we hope that the phenomenological significance of our findings will not be overshadowed by the technicalities of this paper. A comparison with lattice data will be available in a forthcoming related work \cite{toappear}. Complementary details can be found in the appendices.

\section{The setup}
\label{sec:setup}

This section is devoted to lay the foundations of our computation at NLO in the strong coupling expansion or, equivalently, in the higher derivative corrections to the gravity theory. More in details, in section \ref{sec:sugrabck}, we introduce the reader to the supergravity background probed by our source, \ie a fundamental string connecting the quarks at the boundary. Its classical configuration is deeply analyzed in section \ref{sec:fundstring}. There, we review well-known results in the literature, along with some missing details and novel NLO features. Then, in section \ref{sec:dilatoneom}, we solve the equation of motion for the dilaton field coupled to such a string. On one hand, we propose a LO method at finite inter-quark distance which generalizes some results in the literature.  On the other, at NLO, our computations are completely new. We conclude by clarifying the relation among gravity and field theory observables in section \ref{sec:holodic}.

\subsection{The supergravity background}
\label{sec:sugrabck}

Let us focus on the three-dimensional confining gauge theory discussed in the introduction at zero temperature. The dual holographic background, at leading order in the $\alpha'$ expansion, includes a trivial dilaton $\phi$, a self-dual Ramond-Ramond (RR) five-form field strength $F_{(5)}$ 
and an Einstein frame metric given by\begin{subequations}\label{LOback}
\begin{align}
&\label{backmetric}\rmd s^2 = \frac{u^2}{R^2}  \(-\rmd t^2 + \rmd x^2 + \rmd y^2\) + \frac{R^2 \rmd u^2}{u^2 f(u)} + \frac{u^2}{R^2} f(u) R_0^2 \rmd \theta^2 + R^2 \rmd \Omega_5^2 \, ,\\
&\label{backmetricdic}f(u)=1-\frac{u_0^4}{u^4} \, , \quad R^4 / \alpha'^2 = 4\pi g_s N = \lambda_3/T_c \, , \quad \lambda_3 = g_3^2 \, N \, , \quad u \in \[ u_0, +\infty \) .
\end{align}
\end{subequations}
Here, $u$ represents the holographic coordinate which describes, along with the angular variable $\theta \simeq \theta + 2\pi$, a disk sector $D_2$ resembling a \emph{cigar}. Indeed, its radius is asymptotically fixed by $R_0$ and shrinks to zero at the \emph{tip} placed at $u=u_0$. The latter must be identified as
\be\label{LOconical}
u_0 = \frac{R^2}{2 R_0} \, ,
\ee
so as to avoid conical singularities. Let us stress that the $u \sim u_0$ region is dual to the IR regime of the gauge theory and that the boundary lies at $u=\infty$. For completeness, $\lambda_3$ is related to the Yang Mills coupling $g_3$ as above and $\rmd \Omega^2_5$ is the metric of a unit five-sphere.

Let us stress that the geometry features a three-dimensional Minkowskian sector 
shared with the gauge theory and parameterized by $t, x, y$, as well as a five-sphere having constant radius $R$. In jargon, we would say that this solution corresponds to an $AdS_5$ soliton times a sphere supported by the RR flux of $F_{(5)}$,\footnote{Let us stress that the latter is proportional to $N$.} which comes from the double Wick rotation of an $AdS_5$ black hole metric. Moreover, notice that the relations in \eqref{backmetricdic} formalize the correspondence among the perturbative expansions for the dual theories mentioned in the introduction.

Back to us, the above background solves the supergravity equation of motions at leading order in the $\alpha'$ expansion, \ie in the higher derivative expansion. Here, we are interested in the effects coming from subleading corrections. In Einstein frame, the bosonic part of the (four-point) Type IIB effective action up to NLO in the high derivative corrections can be written as \cite{Grisaru:1986vi, Freeman:1986zh, Park:1987jp, Gross:1986iv, Tseytlin:1986zz, Green:1997tv, Green:1997di, Kiritsis:1997em, Green:1998by}\footnote{In \cite{Grisaru:1986vi, Freeman:1986zh, Park:1987jp, Gross:1986iv, Tseytlin:1986zz}, the authors proved  the existence of a scheme in which all the metric and dilaton dependent contributions to the $\mathcal O (\alpha'^3)$ four point effective action can be encoded in the only $W$ term above. This would not be true for a generic RR sector. Indeed, from one hand, the authors of \cite{Liu:2025uqu} recently clarified how the couplings to the RR axion and the RR three-form field strength represent an obstruction to such a procedure; nevertheless, they are both assumed to be zero here. On the other, in \cite{Paulos:2008tn, Green:2003an, Rajaraman:2005up}, the full set of higher derivative corrections including the curvature and the five-form field strength at $\mathcal O(\alpha'^3)$ has been provided; it turns out that the five-form dependent contributions to the equations of motion vanish on the three-brane solution at hand, making the results in \cite{Gubser:1998nz, Pawelczyk:1998pb} robust. Finally, let us stress that the $\alpha'^3 \mathcal R^4$ term alone satisfies all the duality constraints required at the four field level, even in the absence of the Kalb-Ramond field as here (\eg see \cite{Garousi:2013tca}). All in all, we conclude that the above supergravity action is totally enough for our purposes.}
\begin{subequations}\label{sugraaction}
\be
S_{\text{sugra}}^{(\text{bos})} = \frac{1}{2 \kappa^2} \hspace{-2pt} \int \hspace{-4pt} \rmd^{10} x \sqrt{\menodet{g}{10}} \[ \mathcal R -\frac12 \, g^{\mu\nu} \, \partial_\mu \phi \, \partial_\nu \phi - \frac14 \, g_s^2 \, |F_{(5)}|^2 +\gamma \, R^6 e^{-3\phi/2} \ell(\phi) \, W\] ,
\ee
where
\begin{align}
&\label{param}2 \kappa^2 = (2\pi)^7 g_s^2 \pt \alpha'^4 \, , \quad \gamma = \frac18 \zeta(3) \frac{\alpha'^3}{R^6} \, , \\[1ex]
&|F_{(5)}|^2 = F_{(5)}^{abcde} F^{(5)}_{abcde}/5! \, , \\[1ex]
&W = C^{hmnk} C_{pmnq} C_h{}^{rsp} C^q{}_{rsk} + \frac12 \, C^{hkmn} C_{pqmn} C_h{}^{rsp} C^q{}_{rsk} \, , \\[1ex]
&\label{ellphi}
\ell (\phi) = 1 + \frac{\pi^2 g_s^2}{3 \, \zeta(3)} \, e^{2\phi} + \frac{4 \pi g_s}{\zeta(3)} \, e^\phi \sum_{m \neq 0} \sum_{n \ge 1}  \left | \frac{m}{n} \right | K_1 \(\frac{2\pi |mn|}{g_s \, e^\phi}\)\, .
\end{align}
\end{subequations}
Moreover, $\mathcal R$ denotes the scalar curvature and $\piudet{g}{10}$ is the determinant of the ten-dimensional Einstein frame metric. Let us stress that {\small{\be
C^h{}_{mnk}=\mathcal R^{h}{}_{mnk} - \frac18 \( \delta^h{}_n \mathcal R_{mk} + \mathcal R^h{}_n g_{mk} - \mathcal R^h{}_k g_{mn} - \delta^h{}_k \mathcal R_{mn}\) + \frac{1}{72} \, \mathcal R \(\delta^h{}_n g_{mk} - \delta^h{}_k g_{mn}\)
\ee}}represents the conformal invariant components of the Weyl tensor.\footnote{E.~g., see formulae (11.76) and (11.94) of \cite{Blau2025notes}; here, we denote the Ricci and the Riemann tensors respectively as $\mathcal R_{mn}$ and $\mathcal R_{hmnk}$.}
Also, the leading and the first subleading terms in $\ell$ can be identified respectively as the tree and the one-loop contributions in the small $g_s$ expansion of the above $\mathcal O (\alpha'^3)$ high derivative correction. The following series, $K_1$ being a modified Bessel function of the second type, collects non-perturbative contributions which will be neglected hereafter. Finally, notice that we tacitly redefined $\phi$ by subtracting its constant vacuum expectation value $\log g_s$.

The metric solving the NLO supergravity equations of motion can be parameterized as\begin{subequations}\label{NLOmetricpiuparam}
\be\label{NLOmetric}
\rmd s^2 = H^2(u) \Bigl (-\rmd t^2 + \rmd x^2 + \rmd y^2 + P^2(u) \rmd u^2 + K^2 (u) R_0^2 \rmd \theta^2 \Bigr ) + R^2 \bigl (1 + 2 \pt \delta_\nu \bigr )\rmd \Omega_5^2 \, ,
\ee
where
\be
H^2(u) = \frac{u^2}{R^2} \(1-\frac{10}{3} \delta_\nu\) \, , \quad P^2(u) = \frac{R^4}{u^4 f(u)} \bigl (1+\delta_u  \bigr ) \, , \quad K^2(u) = f(u) \bigl (1 + \delta_\theta \bigr ) \, .
\ee
\end{subequations}
and the relation in \eqref{LOconical} gets corrected as
\be\label{correctedR0}
R_0 = \frac{R^2}{2 u_0} \, \(1 + \frac12 \, \delta_T\) \, ,
\ee
where $\delta_\nu$, $\delta_u$, $\delta_\theta$, $\delta_T \in \mathcal O(\gamma)$. In general, the latter can be expanded as
\be\label{deltaexpansion}
\delta_a = \gamma \, \delta_{a,1}^{\text{(tree)}} + \gamma \, g_s^2 \, \delta_{a,1}^{\text{(loop)}} \, ,
\ee
given $\delta_{a,1}^{\text{(tree)}}$, $\delta_{a,1}^{\text{(loop)}} \in \mathcal O(1)$, for $a=\nu,u,\theta,T$.

The tree level solution for the metric perturbations has been computed in \cite{Gubser:1998nz, Pawelczyk:1998pb} and reads\begin{subequations}\label{treedeltas}
\begin{align}
&\delta_{\nu,1}^{\text{(tree)}} = \frac{15}{32} \, \frac{u_0^8}{u^8} \(1 + \frac{u_0^4}{u^4}\) \, , \\
&\delta_{u,1}^{\text{(tree)}} = 15 \( 5 \, \frac{u_0^4}{u^4} + 5 \, \frac{u_0^8}{u^8} - 19 \, \frac{u_0^{12}}{u^{12}}\) \, , \\
&\delta_{\theta,1}^{\text{(tree)}} = - 15 \( 5 \, \frac{u_0^4}{u^4} + 5 \, \frac{u_0^8}{u^8} - 3 \, \frac{u_0^{12}}{u^{12}}\) \, , \\
&\label{deltaT}\delta_{T,1}^{\text{(tree)}} = - 30 \, .
\end{align}
To this order, the dilaton decouples from the above fluctuations. As a consequence, it has been possible to study its dynamics in the fixed background metric  \eqref{NLOmetric}. At tree level, the on-shell dilaton field turns out to be the background configuration
\be \label{phi1treebck}
\phi = \gamma \, \phi_{1,\text{bck}}^{\text{(tree)}}(u) \, , \quad \phi_{1,\text{bck}}^{\text{(tree)}}(u) = - \frac{45}{8} \(\frac{u_0^4}{u^4} + \frac{u_0^8}{2 \pt u^8} + \frac{u_0^{12}}{3 \pt u^{12}}\) \, .
\ee
\end{subequations}
At one-loop order, as far as we know, the explicit NLO solution for the metric tensor is not known. Nevertheless, the tree level solution will be enough for what concerns our problem.

For future convenience, let us also expand $W$ as
\be
W = W_0 + \gamma \, W_1^{\text{(tree)}} + \gamma \hspace{0.5pt} g_s^2 \, W_1^{\text{(loop)}} \, ,
\ee
with $W_0$, $W_1^{\text{(tree)}}$, $W_1^{\text{(loop)}} \in \mathcal O(1)$. It is easy to verify that
\be\label{W0}
W_0 = \frac{180}{R^8} \frac{u_0^{16}}{u^{16}} \, .
\ee
Finally, let us introduce the very useful combinations
\be
\delta_\pm = \frac12 \Bigl ( \pm \delta_u + \delta_\theta +\delta_T \Bigr ) \, , \quad \delta_{10} = \delta_+ - \frac{10}{3} \, \delta_\nu \, .
\ee
Of course, they can be expanded as in \eqref{deltaexpansion}, for $a=\pm, 10$. The solution in \eqref{treedeltas} immediately gives
\be
\delta_{+,1}^{\text{(tree)}} = - 15 \(1 + 8 \, \frac{u_0^{12}}{u^{12}} \)  \, , \quad \delta_{-,1}^{\text{(tree)}} = - 15 \( 1 + 5 \, \frac{u_0^4}{u^4} + 5 \, \frac{u_0^8}{u^8} - 11 \, \frac{u_0^{12}}{u^{12}}\) \, .
\ee
We will make extensive use of these quantities in the next sections.

\subsection{The classical string configuration}
\label{sec:fundstring}

Let us consider a static quark-antiquark pair lying  at the boundary along the $x$-axis and separated by a finite distance $L$. We can think about a classical fundamental string connecting the quarks and exploring the on-shell bulk geometry in \eqref{NLOmetric} with some profile
\be
u=u(x) \, ,  \quad |x| \leq + L/2 \, , \quad u(\pm L/2)=\infty \, . 
\ee
Here, by ``classical'', we mean that the latter solves the equations of motion of the Einstein frame Nambu-Goto action \cite{nambu1970duality, Goto:1971ce, Hara:1971ur}\footnote{Remember that, for a ten-dimensional target space, we can switch to the string frame metric through $g_{\mu\nu}^{\text{(str)}} = e^{\phi/2} g_{\mu\nu}$.}
\be \label{NGaction}
S_{NG} = - \frac{1}{2\pi\alpha'} \int \hspace{-4pt} \rmd \tau \rmd \sigma \, e^{\phi/2} \sqrt{\menodet{g}{2}} \, ,
\ee
where $\piudet{g}{2}$ stands for the determinant of the induced metric on the world-sheet, parameterized by the coordinates $\tau$, $\sigma$. This configuration has been deeply discussed in the literature whenever the string experiences the leading order bulk geometry in \eqref{LOback} \cite{Kinar:1999xu, Greensite:1999jw, Kinar:1998vq, Greensite:1998bp}. Nevertheless, as we will see, the latter is not enough for our purposes.

If our string propagates for a time $\mathrm T$, in the static gauge
\be \label{staticgauge}
\tau = t \, , \quad \sigma = x \, ,
\ee
we can write the above Nambu-Goto action as
\be
S_{\text{NG}} = - \frac{\mathrm T}{\pi\alpha'} \int_{u_m}^{\infty} \hspace{-6pt} \rmd u \, e^{\phi/2} H^2(u) \displaystyle{\sqrt{(\partial_u x)^2+P^2(u)} }\, ,
\ee
where $u_m$ refers to the minimum value of the holographic coordinate reached by the string at $x=0$. Indeed, by symmetry, we can assume that the first derivative of the profile with respect to $x$ vanishes there. 

The equation of motion for $x$ defines a conserved charge as
\be
\partial_u \(\frac{e^{\phi/2} H^2(u)}{\sqrt{1 + P^2(u) (\partial_x u)^2}}\) = 0 \, .
\ee
Therefore, it follows that the classical string profile
\be
u_c = u_c(x)
\ee
must satisfy
\be \label{fundeom}
\partial_x u_c = \pm \frac{1}{P(u_c)} \sqrt{\frac{e^{\phi(u_c)}H^4(u_c)}{e^{\phi(u_m)}H^4(u_m)} - 1} \, , 
\ee
where the positive (negative) sign holds for $x>0$ ($x<0$). Moreover, notice that
\be\label{genLum}
L= \int_{-L/2}^{+L/2} \hspace{-4pt} \rmd x = 2 \int_{u_m}^\infty \hspace{-6pt} \rmd u \, P(u) \[\frac{e^{\phi(u)}H^4(u)}{e^{\phi(u_m)}H^4(u_m)} - 1\]^{-1/2}
\ee
provides an implicit equation that links $u_m$ and $L$. Once solved for $u_m$, at least numerically, we can express the latter as a function of the inter-quark distance $L$. 

On the solution to equation \eqref{fundeom}, we thus get
\be\label{onshelldet}
{\textstyle{{\textstyle{\sqrt{\menodet{g}{2}^{\scaleto{\text{(on-shell)}}{5.5pt}}}}}}} = \frac{u_{c,0}^4(x)}{R^2 \pt u_{m,0}^2} \bigl (1 + \delta_g \bigr )\, ,
\ee
where we expanded the on-shell string profile and its minimum value as
\be \label{ucumexp}
u_c = u_{c,0} \bigl (1 + \delta_c\bigr ) \, , \quad u_m = u_{m,0} \bigl (1 + \delta_m\bigr ) \, ,
\ee
given $u_{c,0}$, $u_{m,0} \in \mathcal O(1)$, $\delta_c$, $\delta_m \in \mathcal O (\gamma)$, and we collected in $\delta_g \in \mathcal O (\gamma)$ all the subleading corrections as
\be
\delta_g = \textstyle{\frac12} \, \phi(u_{c,0}) - \textstyle{\frac12} \, \phi(u_{m,0}) - \textstyle{\frac{20}{3}} \, \delta_\nu(u_{c,0}) + \textstyle{\frac{10}{3}} \, \delta_\nu(u_{m,0}) + 4 \, \delta_c - 2 \, \delta_m \, .
\ee
Notice that, from the last section, the first non-trivial contribution to the dilaton is expected to appear at order $\gamma$. Moreover, the definition of $u_{c,0}$, $u_{m,0}, \delta_c, \delta_m$ holds at some fixed $L$. For the sake of simplicity, in this section we will omit such a dependence.

\begin{figure}[t]
    \centering
    \begin{subfigure}{0.425\textwidth} 
       \centering
        \includegraphics[width=\linewidth]{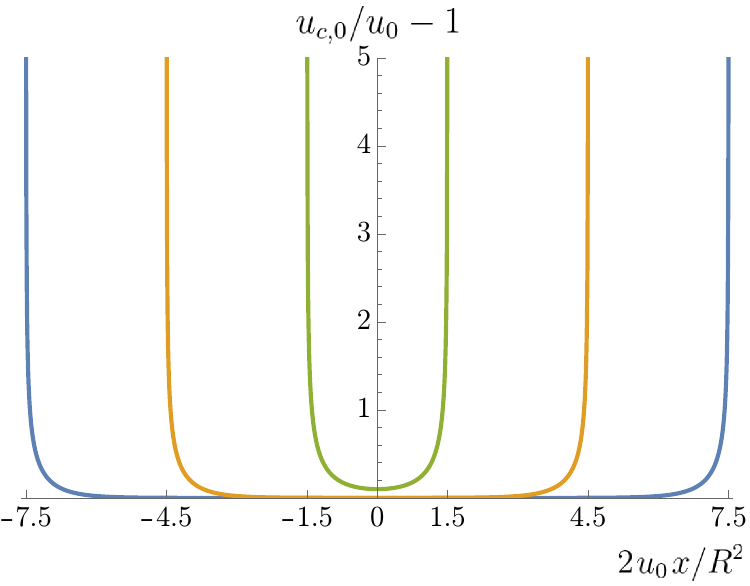} 
        \caption{}
        \label{fundshape}
    \end{subfigure}
 \hspace{25pt}
    \begin{subfigure}{0.43\textwidth} 
       \centering
        \includegraphics[width=\linewidth]{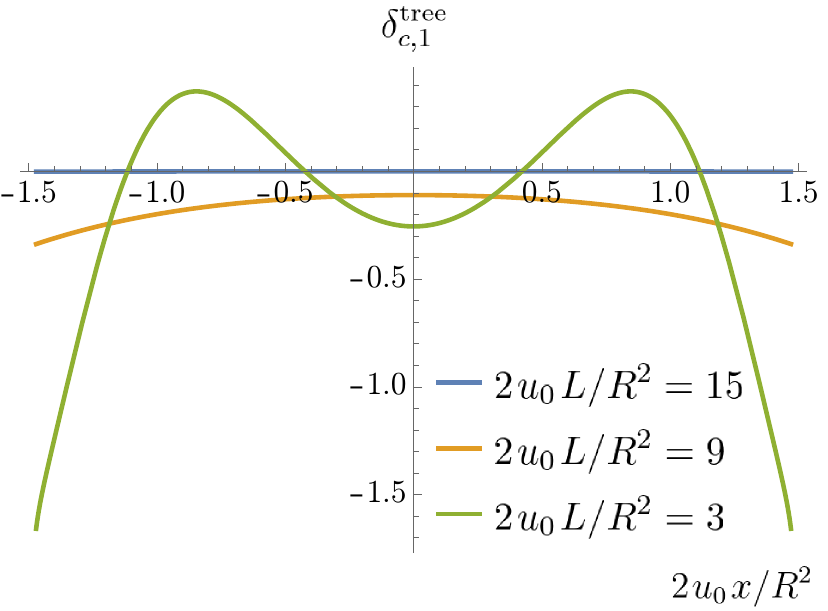} 
        \vskip 5pt
        \caption{}
        \label{deltac}
    \end{subfigure}
    \caption{(a) Shape of the fundamental string diving into the leading order bulk geometry \eqref{LOback} for different values of $2 u_0 L / R^2 = 3 \text{ (green line)}, 9 \text{ (yellow line)}, 15 \text{ (blue line)}$. We reported the solution $u_{c,0}$ to equation \eqref{equc0} in units of $u_0$, removing a constant offset, as $2 u_0 x/ R^2$ varies. (b) Plot of the tree level deviations $\delta_{c,1}^{\text{tree}}$ from the leading order profiles on the left as functions of $2 u_0 x /R^2$. We got there by solving equation \eqref{eqdeltac} numerically on the tree level background provided in \eqref{treedeltas}.}
    \label{plotuc0deltac}
\end{figure}
Perturbatively, we can fix $u_{c,0}$ and $\delta_c$ as the solutions of
\begin{subequations}\label{equc0deltac}
\begin{empheq}[left=\empheqlbrace]{align}
&\label{equc0}\scaleto{\displaystyle \hspace{2pt} \partial_x u_{c,0} = \pm \frac{u_0^2}{R^2} \, \sqrt{\(\frac{u_{c,0}^4}{u_0^4}-1\)\(\frac{u_{c,0}^4}{u_{m,0}^4}-1\)} \, , \quad x \grlw 0 \, ,}{33pt}\\[2ex]
&\label{eqdeltac}\scaleto{\displaystyle \hspace{2pt} \partial_x \delta_c = \frac{\partial_x u_{c,0}}{2 \pt u_{c,0}} \hspace{-2pt}\[\frac{\delta_h(u_{c,0}) - 4 \pt \delta_m}{1-u_{m,0}^4/u_{c,0}^4} - \delta_u(u_{c,0}) -\hspace{-2pt} \(1-2 \frac{u_{0,c}^4}{u_{m,0}^4} \frac{2 u_{c,0}^4 - u_{m,0}^4-u_0^4}{R^4 \(\partial_x u_{c,0}\)^2}\) \hspace{-2pt} 2 \, \delta_c \] \hspace{-2pt} ,}{33pt}
\end{empheq}
\end{subequations}
where we defined
\be\label{deltah}
\delta_h(u) = \phi(u) - \phi(u_{m,0}) - \textstyle{\frac{20}{3}} \, \delta_\nu(u) + \textstyle{\frac{20}{3}} \, \delta_\nu(u_{m,0}) \, .
\ee
Notice that $\delta_h$ above vanishes at $u=u_{m,0}$ and so $\delta_h \in \mathcal(u-u_{m,0})$ as $u \to u_{m,0}$. It follows that, on the solution of \eqref{equc0},  equation \eqref{eqdeltac} is well-defined as $u_{c,0}$ and $\delta_c$ respectively go to $u_{m,0}$ and $\delta_m$. Specifically, it vanishes in this limit in agreement with the definition of $u_m$. Furthermore, let us stress that $\delta_c$ is a functional of $u_{c,0}$ at any fixed $x$. Indeed, equation \eqref{eqdeltac} can be rephrased as
\be
\partial_x \delta_c (x) + \mathcal F [u_{c,0}(x)] \delta_c(x) = \mathcal G [u_{c,0}(x)] \, ,
\ee
for some functionals $\mathcal F$ and $\mathcal G$. Its formal solution reads
\be
\delta_c(x) = e^{-\int_0^x \hspace{-4pt} \rmd \eta \, \mathcal F [u_{c,0}(\eta)]} \(\delta_m + \int_0^x \hspace{-4pt} \rmd \zeta \, e^{\int_0^\zeta \hspace{-4pt} \rmd \eta \, \mathcal F [u_{c,0}(\eta)]} \, \mathcal G [u_{c,0}(\zeta)]\) \, ,
\ee
which, at any fixed $x$, depends just on the classical profile of the string at leading order.

\begin{figure}[t]
        \hspace{18pt}
    \begin{subfigure}{0.42\textwidth} 
       \centering
         \includegraphics[width=\linewidth]{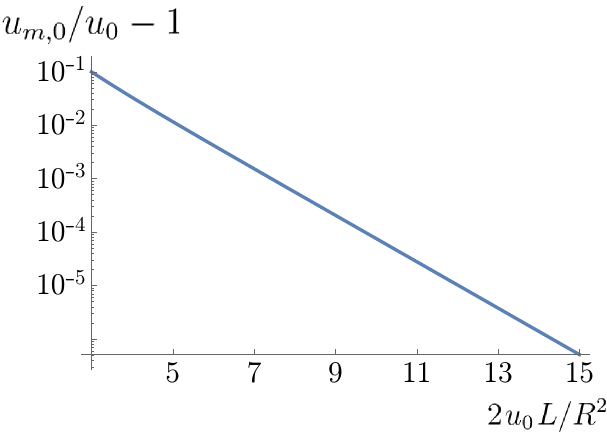} 
        \caption{}
        \label{umzero}
    \end{subfigure}
  \hspace{32.5pt}
   \begin{subfigure}{0.42\textwidth} 
       \centering
        \includegraphics[width=\linewidth]{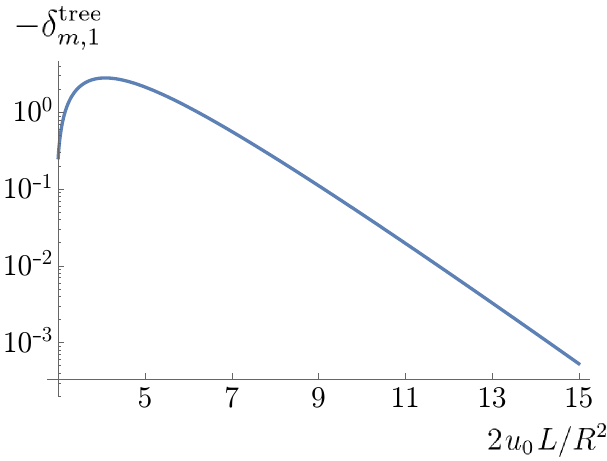} 
        \caption{}
        \label{deltam}
    \end{subfigure}
     \\
     \centering
         \begin{subfigure}{0.47\textwidth} 
       \centering
        \includegraphics[width=\linewidth]{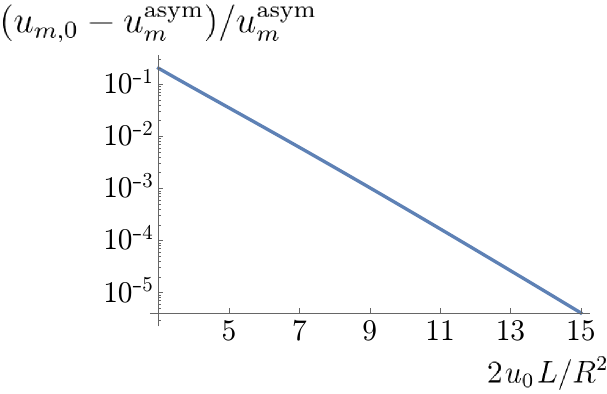} 
        \caption{}
        \label{relerrum}
    \end{subfigure}
        \hfill
    \begin{subfigure}{0.47\textwidth} 
       \centering
        \includegraphics[width=\linewidth]{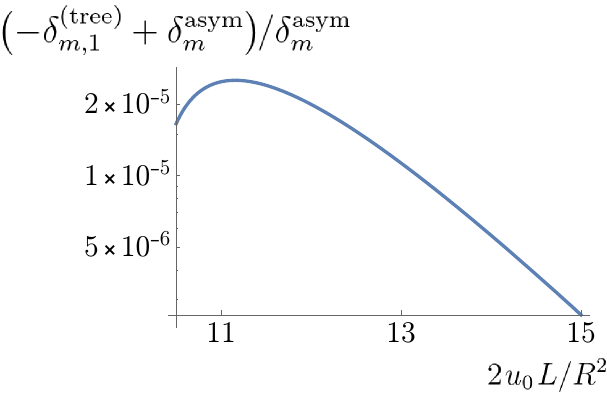} 
        \caption{}
        \label{relerrdeltam}
    \end{subfigure}
    \caption{(a) Plot of $u_{m,0}/u_0 - 1$ as $2u_0 L / R^2$ varies. To get this result, we inverted equation \eqref{LOLum} numerically. (b) Plot of the tree level perturbation $\delta_{m,1}^{\text{tree}}$ computed evaluating equation \eqref{NLOLum} on the solution \eqref{treedeltas}. Notice that it is a function of $L$ thorough $\varepsilon = f (u_{m,0}(L))$. (c) Relative error between $u_{m,0}$ coming from the numerical inversion of the relation in \eqref{LOLum} and its analytical asymptotic behavior given by $u_{m}^{\text{asym}}$ in \eqref{umasym}. (d) Relative error between the tree level value of $\delta_m$ computed from the numerical analysis of equation \eqref{NLOLum} and its analytical asymptotic behavior given by $\delta_{m}^{\text{asym}}$ in \eqref{deltamasym}.}
    \label{plotumL}
\end{figure}
On the other hand, changing variable in \eqref{genLum} as $\eta=u/u_m$ and then expanding it according to \eqref{ucumexp}, we have
\begin{subequations}\label{equmdeltam}
\begin{empheq}[left=\empheqlbrace]{align}
&\label{LOLum} \scaleto{\displaystyle \hspace{2pt} \frac{2 \hspace{2pt} u_0}{R^2} \, L =  \frac{4 \pt u_0}{u_{m,0}} \int_1^\infty \hspace{-8pt} \rmd \eta \, \frac{1}{\sqrt{\bigl (\eta^4-1+\varepsilon \bigr )\bigl (\eta^4-1 \bigr )}} \, , \quad \varepsilon = f(u_{m,0}) = 1 - \frac{u_0^4}{u_{m,0}^4} \, ,}{37pt}\\[1ex]
&\label{NLOLum} \scaleto{\displaystyle \hspace{2pt} \delta_m = \frac{1}{\displaystyle 2 \hspace{-2pt}\int_1^\infty \hspace{-8pt} \rmd \eta \, \frac{\eta^4 + 1 - \varepsilon}{\bigl (\eta^4 - 1 + \varepsilon \bigr )^{3/2} \sqrt{\eta^4-1}}} \int_1^\infty \hspace{-8pt} \rmd \eta \, \frac{\bigl ( \eta^4-1 \bigr ) \delta_u(\eta) - \eta^4 \delta_h(\eta)}{\bigl (\eta^4-1 \bigr )^{3/2} \sqrt{\eta^4-1+\varepsilon} } \, .}{49pt}
\end{empheq}
\end{subequations}
The leading order equation \eqref{LOLum} already appears in \cite{Kinar:1999xu, Greensite:1999jw, Kinar:1998vq, Greensite:1998bp}, where $\varepsilon$ is defined above. To the contrary, \eqref{NLOLum} is new and quantifies the NLO deviations from $u_{m,0}$. The latter depends on $L$ through $u_{m,0}$ itself. Let us stress that $L$ is a zero-order parameter which defines the setup along with $u_0$.

Of course, $\delta_g$ and $\delta_h$ can be expanded as in \eqref{deltaexpansion}. Thus, at tree level we can make use of the solution in \eqref{treedeltas} to produce the plots in figure \ref{plotuc0deltac} and \ref{plotumL}. We conclude that the larger $L$ is, the more the string shape resembles a rectangular configuration lying at $u=u_0$. Notably, to this order, the same technology developed in \cite{Kinar:1999xu, Greensite:1999jw, Kinar:1998vq, Greensite:1998bp} can be applied here to evaluate what happens in the large $L$ regime. In appendix \ref{app:largeL}, we review such a strategy to study \eqref{LOLum} in this regime, including also some subleading contributions which turn out to affect the asymptotic result significantly; then we extend it to the analysis of \eqref{NLOLum} at tree level. Our results are
\begin{subequations}\label{asympLum}
\be\label{asympLum1}
L = - \frac{R^2}{2 \pt u_0} \(\log \frac{\varepsilon}{4}  - \log c \) + \mathcal O \(\varepsilon \log \varepsilon\) \, , \quad \varepsilon \to 0 \, ,
\ee
and
\be\label{asympLum2}
\delta_{m,1}^{\text{tree}} = \frac\varepsilon4 \(\frac{335}{4} \log \frac\varepsilon4 + \frac58 d\) + \mathcal O(\varepsilon^2 \log^2 \varepsilon) \, , \quad \varepsilon \to 0 \, ,
\ee
\end{subequations}
where we defined
\be\label{defcd}
c = e^{-\frac12(\pi - 6 \log 2)} \, , \quad d = 26349/77 + 67 \pi - 402 \log 2 \, .
\ee
In other words,
\begin{subequations}
\begin{align}
&\label{umasym}u_{m,0} \approx u_{m}^{\text{asym}} = u_0 \(1 +  c \, e^{-2 u_0 L / R^2 }\) \, ,   \quad L\to\infty \, ,\\
&\label{deltamasym} \delta_m \approx \delta_{m}^{\text{asym}} = \frac54 c \, e^{-2 u_0 L / R^2 } \[67\(- \frac{2 u_0 L}{R^2}+\log c\) + \frac{d}{2} \] \, , \quad L\to\infty \, .
\end{align}
\end{subequations}
Notice that the factor $c$ is missing in \cite{Kinar:1999xu, Greensite:1999jw, Kinar:1998vq, Greensite:1998bp} (\eg cf.~(99) of \cite{Kinar:1999xu}). Given its large deviation from one, namely
\be
c \approx 1.66 \, ,
\ee
it is crucial to get the numerical agreement in the asymptotic regime showed in figure \ref{relerrum} and figure \ref{relerrdeltam}. As a final remark, let us take a closer look to the plots in figure \ref{umzero} and figure \ref{deltam}. For small enough values of $\gamma$, it is clear that we can neglect $\gamma\delta_{m,1}^{\text{(tree)}}$ with respect to $u_{m,0}/u_0 - 1$. Anyway, if $\gamma$ is small, but not that small, then the contribution coming from $\delta_{m,1}^{\text{tree}}$ can affect the deviation of $u_m$ from one significantly in the large $L$ limit. In any case, let us stress that $\gamma$ should be small enough so that the correction can be treated as a perturbation.

\subsection{The dilaton equation of motion}
\label{sec:dilatoneom}

At this point, we have everything we need to derive the dilaton equation of motion. Let us stress that the Einstein frame supergravity action in \eqref{sugraaction} displays a canonically normalized kinetic term for the dilaton field. On the other hand, the Einstein frame Nambu-Goto action in \eqref{NGaction} makes its coupling to the fundamental string of the previous section explicit. Moreover, it has the features of a one-loop correction to the $\mathcal O (\gamma)$ higher-derivative term in the low energy effective action \eqref{sugraaction}, as follows from the relations in \eqref{param}. This new source term encodes the root of new phenomena that are usually neglected.

All in all, a fundamental string probing the supergravity background at hand acts like a source for the dilaton field. Moreover, it is expected to produce a field configuration at order $\gamma \hspace{0.5pt} g_s^2$. So, in principle, the first higher derivative correction in the supergravity action is crucial to compute even the leading new contribution to the on-shell dilaton field. Actually, it is enough also for the next-to-leading one at order $\gamma^2 g_s^2$ (cf.~equation \eqref{onshelldet}). Indeed, here we are not interested in the dilaton background configuration decoupled from the stringy source. Rather, we aim to compute the field configuration related to a position-dependent profile at the boundary. Perturbatively, any source term for the dilaton field coming from a $\mathcal O(\gamma^2 g_s^2)$ higher derivative correction to the effective action would produce just a $u$-dependent contribution to the full solution. Nevertheless, the latter can be shifted away. Hence the claim. At this level, the argument may sound cumbersome. For this reason, we will clarify this point later on with equations at hand.

Up to $O(\gamma^2 g_s^2)$, the dynamics of the dilaton field configuration sourced by a classical fundamental string in the static gauge \eqref{staticgauge} is ruled by
\begin{subequations}\label{Stot}
\be
S[\phi] = S_{\text{dil}}[\phi] + S_{\text{source}} [\phi]\, ,
\ee 
where
\begin{align}
&\label{scalarsectorsugra}\hspace{-8pt}S_{\text{dil}}[\phi] = - \frac{1}{2 \kappa^2} \hspace{-2pt} \int \hspace{-4pt} \rmd^{10} x \sqrt{\menodet{g}{10}} \[ \frac12g^{\mu\nu} \partial_\mu \phi \partial_\nu \phi - \gamma R^6 W \sum_{n=1}^\infty \[(-3)^n + \frac{\pi^2 g_s^2}{3 \zeta(3)}\] \frac{\phi^n}{2^n n!} \] \hspace{-2pt} , \\
&\hspace{-8pt}S_{\text{source}} [\phi] = - \frac{1}{2 \kappa^2} \hspace{-2pt} \int \hspace{-4pt} \rmd t \rmd x \, \gamma \hspace{0.5pt} g_s^2 \, \frac{(2\pi)^6 8 R^6}{\zeta(3)} {\textstyle{\sqrt{\menodet{g}{2}^{\scaleto{\text{(on-shell)}}{5.5pt}}}}} \,  \sum_{n=1}^\infty \frac{1}{2^n n!} \(\left . \phi \right |_{u=u_c(x)}\)^n \, .
\end{align}
\end{subequations} 
Notice that $S_{\text{source}}$ corresponds to the on-shell Nambu-Goto action in \eqref{NGaction}. Moreover, remember that the on-shell induced metric on the world-sheet is provided in \eqref{onshelldet}. Reducing over the compact spaces, the final form of the action is
\begin{subequations}\label{Stotclass}
\be
S[\phi] = - \frac{\pi^4 R^5}{2\kappa^2} \hspace{-6pt} \int \hspace{-4pt} \rmd t \rmd x \rmd y \rmd u \, \Bigl (\mathcal L_{\text{dil}}[\phi] + \mathcal L_{\text{source}} [\phi]\Bigr )  \, ,
\ee
where
\begin{align}
&\mathcal L_{\text{dil}}[\phi] = (1+ \delta_{10}) \, \sqrt{\menodet{g}{5}{}_{{\hspace{-1.5pt}}_{\scaleto{, 0}{5pt}}}} \{ g^{\mu\nu} \partial_\mu \phi \pt \partial_\nu \phi - 2 \pt \gamma R^6 W\sum_{n=1}^\infty \[(-3)^n + \frac{\pi^2 g_s^2}{3 \zeta(3)}\] \frac{\phi^n}{2^n n!}\} , \\
&\label{sourceterm}\mathcal L_{\text{source}} [\phi] = \gamma \hspace{0.5pt} g_s^2 \, \frac{512 \pi^2 R}{\zeta(3)} \, {\textstyle{{\textstyle{\sqrt{\menodet{g}{2}}}}}} \,  \sum_{n=1}^\infty \frac{\phi^n}{2^n n!} \, \delta(y) \, \delta(u-u_c(x))  ,
\end{align}
\end{subequations}
with $\mu, \nu = t,x,y,u$. To get there, we used that the volume of a unit five sphere is given by $\text{Vol}(S^5) = \pi^3$ and $\piudet{g}{5}{}_{{\hspace{-1.5pt}}_{\scaleto{, 0}{5pt}}}$ stands for the determinant of the five-dimensional $AdS$-sector in the leading order metric \eqref{backmetric}. Furthermore, 
let us stress that $\phi$ now refers to the projection of the original ten-dimensional field onto the constant five-spherical harmonic. Indeed, higher harmonic modes 
are responsible for the non-zero expectation values of boundary operators which are different from the Yang-Mills Lagrangian density \cite{Klebanov:1997kc, Witten:1998qj, Gubser:1998bc, Klebanov:1999xv}. As a consequence, we can neglect them (for a similar discussion see \cite{Callan:1999ki}). Moreover, we are interested in what survives the KK mode decoupling limit (see the discussion in the introduction). Therefore, we can also ask for a $\theta$-independent dilaton field.

The related equation of motion is, up to $\mathcal O(\gamma^2g_s^2)$,
\be\label{fulleom}
\scaleto{\begin{split}
\partial_\mu \Bigl ( (1+\delta_{10} ) &\sqrt{\menodet{g}{5}{}_{{\hspace{-1.5pt}}_{\scaleto{, 0}{5pt}}}} \hspace{0.5pt}g^{\mu\nu} \partial_\nu \phi\Bigr ) = \frac32 \gamma \hspace{0.5pt} R^6 \hspace{0.5pt} W (1+\delta_{10} ) \sqrt{\menodet{g}{5}{}_{{\hspace{-1.5pt}}_{\scaleto{, 0}{5pt}}}} \sum_{m=0}^\infty \[(-3)^m - \frac{\pi^2 g_s^2}{9 \zeta(3)}\] \frac{\phi^m}{2^m m!} +\\
&+ \gamma \hspace{0.5pt} g_s^2 \, \frac{128 \pi^2 R}{\zeta(3)} {\textstyle{\sqrt{\menodet{g}{2}^{\scaleto{\text{(on-shell)}}{5.5pt}}}}} \(1+\textstyle\frac12 \, u_{c,0} \, \partial_u \phi \, \delta_c\) \,  \sum_{m=0}^\infty \frac{\phi^m}{2^m m!} \, \delta(y) \, \delta(u-u_{c, 0}(x)) \, ,
\end{split}}{69pt}
\ee
with $\mu, \nu = t,x,y,u$. Here, we expanded the second Dirac delta for small deviations $\delta_c$ from the classical string profile $u_{c,0}$ (cf. \eqref{ucumexp}), using that $\delta'[\phi] = - \delta[\phi']$. In general, the above dilaton equation leads to PDEs. This represents an important departure from the traditional ODE approach, which usually focuses on background field configurations.

Let us stress that the source we are dealing with is static and stretches just along $x$. Therefore, we can assume that the on-shell dilaton configuration explicitly depends at most on $u$ and $y$. Moreover, we can expand it as 
\be
\phi = \phi_0 + \gamma \, \phi_1^{\text{(tree)}} + \gamma \hspace{0.5pt} g_s^2 \, \phi_1^{\text{(loop)}} + \gamma^2 \, \phi_2^{\text{(tree)}} + \gamma^2 g_s^2 \, \phi_2^{\text{(loop)}} \, ,
\ee
where $\phi_0$, $\phi_1^{\text{(tree)}}$, $\phi_1^{\text{(loop)}}$, $\phi_2^{\text{(tree)}}$, $\phi_2^{\text{(loop)}} \in \mathcal O(1)$. 

Then, let us solve the equation of motion \eqref{fulleom} order by order. In terms of the dimensionless variables\footnote{Remember that the minimum value of $u$ is $u_m > u_0$ at any finite $L$. Therefore, the above inequality holds strictly.}
\be\label{dimensionlessvar}
v=(u/u_0)^2 > 1 \, , \quad z = 2 \pt u_0 \pt  y / R^2 \, ,
\ee
the first two equations are
\begin{subequations}\label{eomphi0phi1}
\begin{empheq}[left=\empheqlbrace]{align}
&\label{eomphi0}\displaystyle \hspace{2pt} D_v \phi_0 = 0 \, , \\
&\label{eomphi1}\displaystyle \hspace{2pt} D_v \phi_1^{\text{(tree)}} = \frac38 \pt \frac{R^8 \pt W_0}{v^2-1} \sum_{m=0}^{\infty} \frac{(-3)^m \phi_0^m}{2^m m!} - \frac{\delta_{+,1}^{\text{(tree)}} \partial_z^2 \phi_0 + \partial_v \hspace{-2pt} \[ \delta_{-,1}^{\text{(tree)}} v \pt (v^2-1) \pt \partial_v \phi_0 \]}{v \pt (v^2-1)} ,
  \end{empheq}
\end{subequations}
where
\be
D_v = \partial_v^2 + \(\frac1v + \frac{1}{v-1} + \frac{1}{v+1}\) \partial_v + \frac{1}{v(v^2-1)} \partial_z^2
\ee
is a linear differential operator. Trivially, they are solved by
\be\label{backphi0phi1}
\phi_0 = 0 \, , \quad \phi_1^{\text{(tree)}} = \phi_{1,\text{bck}}^{\text{(tree)}}(v) \, .
\ee
These are the same expressions found in \cite{Gubser:1998nz, Pawelczyk:1998pb} and introduced in the section \ref{sec:sugrabck} (cf.~\eqref{phi1treebck}). Here, the decoupling of the tree level dilaton field from the metric perturbations is manifest.

On the background solution \eqref{backphi0phi1}, the other relevant equations at any fixed $x$ read
\be
\begin{cases}
\displaystyle \hspace{2pt} v \pt (v^2-1) \pt D_v \phi_1^{\text{(loop)}} = F(v) + \xi \, v^2 \, \delta(z) \, \delta(v-v_{c,0}(x))  \, , \\[2ex]
\displaystyle \hspace{2pt} v \pt (v^2-1) \pt D_v \phi_2^{\text{(loop)}} = G(v) + J\bigl [\phi_1^{\text{(loop)}}\bigr ] + \xi \, v^2 \Bigl ( \delta_{g,1}^{\text{(tree)}} + \textstyle{\frac12} \phi_{1,\text{bck}}^{\text{(tree)}} \Bigr ) \delta(z) \, \delta(v-v_{c,0}(x)) \, ,
\end{cases}
\ee
where we defined the dimensionless parameter
\be\label{xi}
\xi = \frac{256 \pi^2}{\zeta(3)} \frac{u_0^2}{u_{m,0}^2} \, ,
\ee
the functions $F$ and $G$ as
\begin{subequations}
\begin{align}
&F(v) = -\frac{\pi^2}{24 \hspace{0.5pt} \zeta(3)} \pt v \pt R^8 \pt W_0 \, , \\
&G(v) = -\frac{\pi^2}{48 \hspace{0.5pt} \zeta(3)} \pt v \pt R^8 \pt W_0 \, \phi_{1,\text{bck}}^{\text{(tree)}} - \partial_v \hspace{-2pt} \[ \delta_{-,1}^{\text{(loop)}} v \pt (v^2-1) \pt \partial_v \phi_{1,\text{bck}}^{\text{(tree)}} \] + \\
& \hspace{40pt}+ \frac38 \pt v \pt R^8 \hspace{-2pt} \(W_0 \, \delta_{10,1}^{\text{(loop)}} + W_1^{\text{(loop)}}\) - \frac{\pi^2}{24 \hspace{0.5pt} \zeta(3)} \pt v R^8 \hspace{-2pt} \(W_0 \, \delta_{10,1}^{\text{(tree)}} + W_1^{\text{(tree)}}\) \, , \nonumber
\end{align}
\end{subequations}
and the current $J$ given by
\be
J \bigl [\phi_1^{\text{(loop)}} \bigr ] = - \frac{9}{16} \pt v \pt R^8 \pt W_0 \, \phi_1^{\text{(loop)}} - \delta_{+,1}^{\text{(tree)}} \partial_z^2 \phi_1^{\text{(loop)}} - \partial_v \hspace{-2pt} \[ \delta_{-,1}^{\text{(tree)}} v \pt (v^2-1) \pt \partial_v \phi_1^{\text{(loop)}} \] \, .
\ee
Remember that $u_{m,0}$ has been introduced in the previous section.

Let us stress that here we are just interested in the $z$-dependent dilaton configuration generated by the string-like source. In other words, we are not concerned with solving for the background components of $\phi_1^{\text{(loop)}}$ and $\phi_2^{\text{(loop)}}$, which depend just on $v$. So, we can shift them away. More in details, our ansatz for the full solution is
\begin{subequations}\label{shifts}
\begin{align}
&\label{FourierexpLO}\phi_1^{\text{(loop)}}(v,z)=\phi_{1,\text{bck}}^{\text{(loop)}}(v)+\frac{1}{2\pi} \int_{-\infty}^{+\infty} \hspace{-6pt} \rmd k \, e^{i \pt k \pt z} \, \Phi_{\text{LO}}(v,k)  \, ,\\
&\phi_2^{\text{(loop)}}(v,z)=\phi_{2,\text{bck}}^{\text{(loop)}}(v)+\frac{1}{2\pi} \int_{-\infty}^{+\infty} \hspace{-6pt} \rmd k \, e^{i \pt k \pt z} \, \Phi_{\text{NLO}}(v,k)  \, ,
\end{align} 
\end{subequations}
where $\phi_{1,\text{bck}}^{\text{(loop)}}$ and $\phi_{2,\text{bck}}^{\text{(loop)}}$ are such that
\be
v \pt (v^2-1) \pt D_v \phi_{1,\text{bck}}^{\text{(loop)}} = F(v) \, , \quad v \pt (v^2-1) \pt D_v \phi_{2,\text{bck}}^{\text{(loop)}} = G(v) + J\bigl [\phi_{1,\text{bck}}^{\text{(loop)}}\bigr ] \, ,
\ee
whatever they are. Let us stress that $k$ is the conjugate momentum to $z$. Finally, notice that, on the solution for $\Phi_{\text{LO}}$, we have
\begin{align}
\hspace{-8pt}\delta_{+,1}^{\text{(tree)}} k^2 \Phi_{\text{LO}} \hspace{-2pt} - \hspace{-2pt} \partial_v \hspace{-2pt} \[ \delta_{-,1}^{\text{(tree)}} v \pt (v^2-1) \pt \partial_v \Phi_{\text{LO}} \] \hspace{-2pt} = \delta_{u,1}^{\text{(tree)}} k^2 & \Phi_{\text{LO}}  \hspace{-2pt} - \hspace{-2pt} \partial_v \delta_{-,1}^{\text{(tree)}} v \pt (v^2-1) \pt \partial_v \Phi_{\text{LO}} + \\ 
&\hspace{30pt} - \xi \, v^2 \delta_{-,1}^{\text{(tree)}} \delta(v-v_{c,0}(x)) \, . \nonumber
\end{align}
Hence, the dependence on the constant $\delta_{T,1}^{\text{(tree)}}$ cancels out from the non-localized source contributions.

All in all, the equation of motion for the dilaton field configuration produced by a classical fundamental string at any fixed $x$ can be encoded in the system\footnote{To get there, we make use of the well-known integral representation of the Dirac delta, that is $2\pi\delta(z)= \int_{-\infty}^{+\infty} \text{Exp}[i \pt k \pt z]$. Moreover, we make $W_0$ explicit according to its definition in \eqref{W0}. Finally, remember that $\phi_{{1,\text{bck}}}^{\text{(tree)}}$ is the function of $u=u_0\sqrt{v}$ given in \eqref{phi1treebck}.}
\begin{subequations}
\begin{empheq}[left=\empheqlbrace]{align}
&\label{LOeom}\displaystyle \hspace{2pt} v \pt (v^2-1) \pt \mathcal D_v \Phi_{\text{LO}} = \xi \,  v^2 \pt \delta(v-v_{c,0}(x))  \, , \\[2ex]
&\label{NLOeom}\displaystyle \hspace{2pt} v \pt (v^2-1) \pt \mathcal D_v \Phi_{\text{NLO}} = \mathcal J\bigl [\Phi_{\text{LO}}\bigr ] + \xi \,  v^2 \, \delta_d \, \delta(v-v_{c,0}(x)) \, ,
  \end{empheq}
\end{subequations}
where
\be
\mathcal D_v = \partial_v^2 + \(\frac1v + \frac{1}{v-1} + \frac{1}{v+1}\) \partial_v - \frac{k^2}{v(v^2-1)} \, ,
\ee
and
\be\label{J}
\mathcal J \bigl [\Phi_{\text{LO}} \bigr ] = - \frac{405}{4 \, v^7}\, \Phi_{\text{LO}} + \delta_{u,1}^{\text{(tree)}} k^2 \pt \Phi_{\text{LO}} - \partial_v \delta_{-,1}^{\text{(tree)}} v \pt (v^2-1) \pt \partial_v \Phi_{\text{LO}} \, .
\ee
Moreover, we defined the combination
\be
\delta_d = \delta_{g,1}^{\text{(tree)}} - \delta_{-,1}^{\text{(tree)}} +\phi_{1,\text{bck}}^{\text{(tree)}}/2 \, .
\ee
Notice that equation \eqref{LOeom}, in the $L\to\infty$ limit, reduces to formula (4.92) of \cite{Danielsson:1998wt}. Here, we got it through a more refined perturbative expansion. On the other hand, equation \eqref{NLOeom} is new.

Now we can clarify some of the observations discussed at the beginning of this section. First of all, the leading contribution to the dilaton profile sourced by the string appears at order $\gamma g_s^2$, as expected. It is clear from equation \eqref{LOeom}. Then, perturbatively, any $k$-dependent source term in equation \eqref{NLOeom} is linear in $\Phi_{\text{LO}}$. This means that any contribution coming from an $\mathcal O(\gamma^n)$ higher derivative correction to the effective action would not affect the equation for $\Phi_{\text{NLO}}$ at order $\gamma^2 g_s^2$ if $n>1$. We conclude that the supergravity action in \eqref{sugraaction} is completely enough for what concerns this paper. Furthermore, let us stress that, on the leading vanishing background solution in \eqref{backphi0phi1}, the subleading configuration $\Phi_{\text{NLO}}$ does not mix with the metric perturbations of the same order. This is the same mechanism that decouples $\phi_1^{\text{(tree)}}$ from the tree level corrections to the background metric in \eqref{treedeltas}. Finally, the latter are the only geometrical ingredients that appear in the above equations. Indeed, the unknown loop corrections to the metric tensor are removed by the shifts in \eqref{shifts}. Hence, we can explicitly and safely solve both the above equations for $\Phi_{\text{LO}}$ and $\Phi_{\text{NLO}}$. 

\subsubsection{The leading order problem}
\label{sec:LOdil}

Let us start by solving the leading order equation of motion \eqref{LOeom}. Its associated homogeneous equation is an example of Heun equation (\eg see \cite[\href{https://dlmf.nist.gov/31}{chapter 31}]{NIST:DLMF}) and it has been deeply discussed in the literature about the computation of the glueball mass spectrum from supergravity \cite{Witten:1998zw, deMelloKoch:1998vqw, Csaki:1998qr, Zyskin:1998tg, Minahan:1998tm, Brower:2000rp}. Here, we focus on a particular solution satisfying certain boundary conditions, namely normalizability and regularity at $u=u_0$ (\ie $v=1$). Let us stress that the differential equation at hand features regular singularities at $v=0, \pm 1, +\infty$. Thus, its solution can be expanded as power series about each of these points. The respective radii of convergence are at least equal to the distance from the nearest singularity.

In appendix \ref{app:homoeom}, we provide the derivation of the analytic expansions for the solution to the homogeneous problem around the tip of the cigar and infinity. To sum up, we find\begin{subequations}\label{Heundilaton}
\begin{align}
&\label{h1}\scaleto{\Phi_{\text{LO}}(v,k) = c_1(k) \, h_1 (v,k) \, , \quad h_1 (v,k) = \frac{1}{v^2} + \sum_{n=1}^{\infty} \frac{a(n,k)}{v^{n+2}} \, , \quad 1<v<\infty \, ,}{33pt}\\
&\label{h2}\scaleto{\Phi_{\text{LO}}(v,k) = c_2(k) \, h_2 (v,k) \, , \quad h_2 (v,k) = 1 + \sum_{n=1}^{\infty} b(n,k) (v-1)^{n} \, , \quad 1\leq v<2 \, ,}{33pt}
\end{align}
\end{subequations}where\footnote{The analogue of formula \eqref{recb} in \cite{Zyskin:1998tg} displays a typo.}
\begin{subequations}\label{recrel}
\begin{align}
&\scaleto{a(-1,k)=0 \, , \quad a(0,k)=1 \, ,}{11.75pt} \nb\\
&\label{reca}\scaleto{a(n+1,k) = \frac{1}{(n+2)^2-1} \Bigr [ a(n-1,k) (n+1)^2+k^2 a(n,k) \Bigl ] \, ,}{26pt}\\[2ex]
&\scaleto{b(-1,k)=0 \, , \quad b(0,k)=1 \, ,}{11.75pt} \nb\\
&\label{recb}\scaleto{b(n+1,k) = -\frac{1}{2(n+1)^2} \Bigr [ b(n-1,k) (n^2-1)+\(3n(n+1)-k^2\) b(n,k) \Bigl ] \, ,}{26pt}
\end{align}
\end{subequations}
while $c_1$ and $c_2$ are, at this level, arbitrary $v$-independent functions of $k$. Let us stress that, in each of the neighborhood discussed here, the homogeneous problem would admit another independent solution. Nevertheless, the latter is non-normalizable or non-regular at the tip.

In general, the solution to equation \eqref{LOeom} shall be of the form presented in \eqref{Heundilaton} away from the source. However, the presence of the Dirac delta introduces some constraints that our global solution must satisfy. First, at any fixed $x$ such that
\be \label{barvc}
\bar v_c \equiv v_{c,0}(x) = (u_{c,0}(x)/u_0)^2\in(1,2) \, ,
\ee
we can ask for continuity as
\be\label{LOcont}
c_1 \hspace{-1pt}\cdot\hspace{-1pt} h_1(\bar v_c, k) = c_2 \hspace{-1pt}\cdot\hspace{-1pt} h_2(\bar v_c, k) \, .
\ee
The above domain corresponds to all the values of $x$ where both $h_1$ and $h_2$ converges (see \eqref{Heundilaton}). Notice that the larger $L$ is, the wider the set of accessible $x$ is. Moreover, the first derivative of a continuous function built from $h_1$ and $h_2$ can have at most a finite jump. The latter can be computed integrating equation \eqref{LOeom} along an infinitesimal neighborhood of $\bar v_c$. Then, recalling that $\xi$ has been defined in \eqref{xi}, we get
\be\label{NLOjump}
c_1 \hspace{-1pt}\cdot\hspace{-1pt} \partial_v \pt h_1(\bar v_c, k) - c_2 \hspace{-1pt}\cdot\hspace{-1pt} \partial_v h_2(\bar v_c, k) = \frac{\xi \, \bar v_c}{\bar v_c^2 - 1} \, .
\ee

Solving the above constraints, the global solution to equation \eqref{LOeom} reads
\be\label{soldilLO}
\Phi_{\text{LO}}(v,k;\bar v_c) =
\begin{cases}
 c_1(k;\bar v_c) \, h_1(v,k) \, , &  v \ge \bar v_c \, , \\[0.5ex]
 \displaystyle c_1(k;\bar v_c) \, \frac{h_1(\bar v_c,k)}{h_2(\bar v_c,k)} \, h_2(v,k) \, , &  v < \bar v_c \, ,
\end{cases}
\ee
where
\be\label{c1}
c_1(k;\bar v_c) = -\frac{1}{w(k)}  \, \xi \, \bar v_c^2 \, h_2(\bar v_c,k) \, .
\ee
Here, $w$ is a function of $k$ defined as
\begin{subequations}\label{W}
\be\label{reducedW}
w(k) = v \, (v^2-1) \, W(v,k) \, ,
\ee
where
\be\label{Wronskian}
W(v,k) = h_1(v,k) \, \partial_v h_2(v,k) - h_2(v,k) \, \partial_v h_1(v,k)
\ee
\end{subequations}
is the Wronskian of $h_1$ and $h_2$. The independence of $w$ on $v$ follows from the Abel identity (\eg see section 16.512 of \cite{GradshteynRyzhik}). Notice that a parametric dependence on $\bar v_c$ appears as a consequence of the gluing conditions at the position of the fundamental string in the bulk.

\subsubsection{The next-to-leading order problem}
\label{sec:NLOdil}

The following step is to solve the NLO equation of motion in \eqref{NLOeom}. This time, a non-trivial source term survives also far from the stringy source. 
Therefore, requiring the same boundary conditions as before, we can parameterize the full solution as
\be\label{soldilNLO}
\Phi_{\text{NLO}}(v,k;\bar v_c) =
\begin{cases}
 d_1(k;\bar v_c) \, h_1(v,k) + c_1(k;\bar v_c) P_1(v,k; \bar v_c) \, , &  v \ge \bar v_c \, , \\
  d_2(k; \bar v_c) \, h_2(v,k) + c_1(k; \bar v_c) P_2(v,k; \bar v_c) \, , &  v < \bar v_c \, ,
\end{cases}
\ee
where $\bar v_c$ is defined in \eqref{barvc}. In principle, $d_1$ and $d_2$ are two arbitrary functions of $k$; as we will see, the $\bar v_c$-dependent gluing conditions will fix their values unambiguously. Moreover, remember that $h_1$ and $h_2$ are the analytic solutions to the homogeneous problem defined in \eqref{Heundilaton}. On the other hand, $c_1 P_1$ ($c_1 P_2$) is a particular solution to equation \eqref{NLOeom} without the delta-like source, having support in $1<v<\infty$ ($1\leq v<2$). 
Nevertheless, from a perturbative point of view, both $P_1$ and $P_2$ are expected to inherit a parametric dependence on $\bar v_c$ from the leading order solution in \eqref{soldilLO}. Also, we require that $P_1$ is normalizable and that $P_2$ is regular at the tip. Finally, let us stress that $c_1$ has been fixed by the leading order problem as in \eqref{c1} and that $\bar v_c >1$ strictly.

The full solution
\be
\Phi_c (v, k; \bar v_c) = \gamma \pt g_s^2 \Bigl [ \Phi_{\text{LO}}(v,k; \bar v_c) + \gamma \pt \Phi_{\text{NLO}}(v,k; \bar v_c)\Bigr ]
\ee
should be continue at the NLO physical location of the stringy source, namely
\be\label{expandedvc}
v_c = \bar v_c \pt \bigl (1+ 2 \pt \gamma \pt \delta_{c,1}^{\text{(tree)}} \bigr ) \, ,
\ee
where $\delta_{c,1}^{\text{(tree)}}$ is the tree level value of the fluctuation $\delta_c$ introduced in \eqref{ucumexp} (see also \eqref{dimensionlessvar}). Hence, the condition 
\be
\lim_{v\to v_c^+}\Phi_c(v,k; \bar v_c) = \lim_{v\to v_c^-}\Phi_c(v,k; \bar v_c) \, ,
\ee
on the leading order gluing conditions \eqref{LOcont} and \eqref{NLOjump}, translates into
\be
d_1 \hspace{-1pt}\cdot\hspace{-1pt} h_1(\bar v_c,k) + c_1 \hspace{-1pt}\cdot\hspace{-1pt} P_1(\bar v_c,k;\bar v_c) = d_2 \hspace{-1pt}\cdot\hspace{-1pt} h_2(\bar v_c,k) + c_1 \hspace{-1pt}\cdot\hspace{-1pt} P_2(\bar v_c,k; \bar v_c) -  \delta_{c,1}^{\text{(tree)}} \pt \frac{2 \pt \xi \pt  \bar v_c^2}{\bar v_c^2 - 1} \, .
\ee
On the other hand, the integration of the equation of motion \eqref{NLOeom} along an infinitesimal neighborhood of $\bar v_c$ produces a finite jump in the first derivative of $\Phi_{\text{NLO}}$ given by
\be\label{NLOfinitejump}
\lim_{v\to\bar v_c^+}\partial_v\Phi_{\text{NLO}}(v,k; \bar v_c) - \lim_{v\to\bar v_c^-}\partial_v\Phi_{\text{NLO}}(v,k; \bar v_c) = \delta_d (\bar v_c) \pt \frac{\xi \, \bar v_c}{\bar v_c^2 - 1} \, .
\ee
Again, $\xi$ is the same parameter defined in \eqref{xi}.

All in all, the solution for $d_1$ is
\be\label{d1}
d_1(k; \bar v_c) = c_1(k; \bar v_c) \( \Delta_J (k;\bar v_c)/w(k) + \delta_d(\bar v_c) + 2 \, \bar v_c \, \delta_{c,1}^{\text{(tree)}} \, \partial_v \hspace{-1pt}\log h_2(\bar v_c, k)\) \, ,
\ee
where
\be\label{deltaJ}
\Delta_J(k; \bar v_c) = v(v^2-1) \(h_2(v,k)\)^2 \left . \partial_v \hspace{-2pt} \[\frac{P_1(v,k; \bar v_c) - P_2(v,k; \bar v_c)}{h_2(v,k)}\] \right |_{v=v^*} \, .
\ee
Here, $v^*$ can take any fixed value between $1$ and $2$ (not necessarily $\bar v_c$, as it would be imposed by the Dirac delta). Indeed, $\Delta_J$ is actually $v$-independent on the solution to the leading order problem. This can be proved regardless of what the explicit expressions of $P_1$ and $P_2$ are, by taking its derivative with respect to $v$. 
On the other hand, $\delta_d$ codifies the finite jump in the first derivative of $\Phi_{\text{NLO}}$ at $\bar v_c$ introduced in \eqref{NLOfinitejump}. Finally, the last term encodes the NLO deviation of the physical location of the stringy source from $\bar v_c$. A similar expression for $d_2$ follows easily.

Now, let us look for an explicit expression of $P_1$ and $P_2$. Remember that, modulo the prefactor $c_1$, they are particular solutions to equation \eqref{NLOeom} without the stringy source and in different neighborhoods. So, by definition, it holds
\be\label{eomP}
v \pt (v^2-1) \pt \mathcal D_v P_n (v,k;\bar v_c)= \mathcal K[\Phi_{\text{LO}} (v,k;\bar v_c)\bigr ] \, , \quad n=1,2 \, ,
\ee
where the leading order solution $\Phi_{\text{LO}}$ is given in \eqref{soldilLO} and we defined the reduced current
\be
\mathcal K[\Phi_{\text{LO}} (v,k; \bar v_c)\bigr ]=\mathcal J\bigl [\Phi_{\text{LO}} (v,k; \bar v_c)\bigr ]/c_1(k; \bar v_c)
\ee
to declutter equations.

Let us start by defining the ansatz
\be\label{ansatzP}
P_n (v,k; \bar v_c)= p_n(v,k; \bar v_c) \,h_n(v,k) \, , \quad \mathcal D_v  h_n = 0 \, , \quad n=1,2 \, .
\ee
Then, equation \eqref{eomP} reduces to
\be
\partial_v^2 \pt p_n (v,k; \bar v_c)+ \partial_v \log \hspace{-2pt} \[(h_n(v,k))^2 \pt v \pt (v^2-1)\] \partial_v \pt p_n (v,k; \bar v_c) = \frac{\mathcal K[\Phi_{\text{LO}} (v,k; \bar v_c)\bigr ]}{v \pt (v^2-1) \pt h_n(v,k)} \, ,
\ee
which is solved by
\be\label{pn}
p_n(v,k; \bar v_c) = \hspace{-4pt} \int_{b_n}^v \hspace{-8pt} \rmd v' \frac{1}{(h_n(v',k))^2 v' \pt (v'^2-1)} \[ q_n + \hspace{-4pt} \int_{b_n}^{v'} \hspace{-10pt} \rmd v'' \pt h_n(v'',k) \pt  \mathcal K[\Phi_{\text{LO}} (v'',k; \bar v_c)\bigr ]\] ,
\ee
where $b_n$, $n=1,2$, are some fixed constants. Looking at the solutions \eqref{soldilLO} and \eqref{soldilNLO}, it is clearly convenient to set
\be
b_1 = \bar v_c \, , \quad b_2 = 1 \, .
\ee
Let us stress that we tacitly omit an additive $v$-independent function of $k$. Indeed, the latter can be reabsorbed within $d_1$ and $d_2$ in the parameterization \eqref{soldilNLO}. All in all, the prefactor in front of the homogeneous part of the full solution is fixed as in \eqref{d1}. The only remaining free parameters introduced in \eqref{pn}, namely $q_1$ and $q_2$, can be fixed requiring the solution to be respectively normalizable and regular at the tip.

Notice that\begin{subequations}
\begin{align}
&\frac{1}{(h_1(v,k))^2 v \pt (v^2-1)} = v - 2 \pt a(1,k) + \mathcal O (1/v) \, , \quad v \to \infty \, ,\\[1ex]
&\int_{\bar v_c}^{v} \hspace{-4pt} \rmd v \, h_1(v,k) \pt \mathcal J\bigl [h_1(v,k)\bigr ] = \mathcal Q (k; \bar v_c) + \mathcal O (1/v^4) \, , \quad v \to \infty \, ,
\end{align}
where
\be
\mathcal Q (k; \bar v_c) = \int_{\bar v_c}^{\infty} \hspace{-10pt} \rmd v \, h_1(v,k) \pt \mathcal J\bigl [h_1(v,k)\bigr ] \, .
\ee
\end{subequations}
Therefore, since $h_1 \in \mathcal O (1/v^2)$ as $v\to\infty$ (see formula \eqref{h1}), it thus follows that $P_1$ is normalizable if
\be
q_1 = -\mathcal Q(k; \bar v_c) \, .
\ee
On the other hand, in the opposite regime we have
\be
\int_1^{v} \hspace{-4pt} \rmd v \, h_2(v,k) \pt \mathcal J\bigl [h_2(v,k)\bigr ] \in \mathcal O (v-1) \, , \quad v\to1^+ \, .
\ee
As a consequence, the wanna-be divergent behavior in the outer integral in $p_2$ is canceled out from the second term in \eqref{pn}. To the contrary, in the first contribution there is nothing that prevents such a divergence. So we are forced to set
\be
q_2 = 0 \, .
\ee
In this way, we get a regular solution at the tip of the cigar.

\begin{figure}[t]
    \centering
    \begin{subfigure}{0.45\textwidth} 
       \centering
        \includegraphics[width=\linewidth]{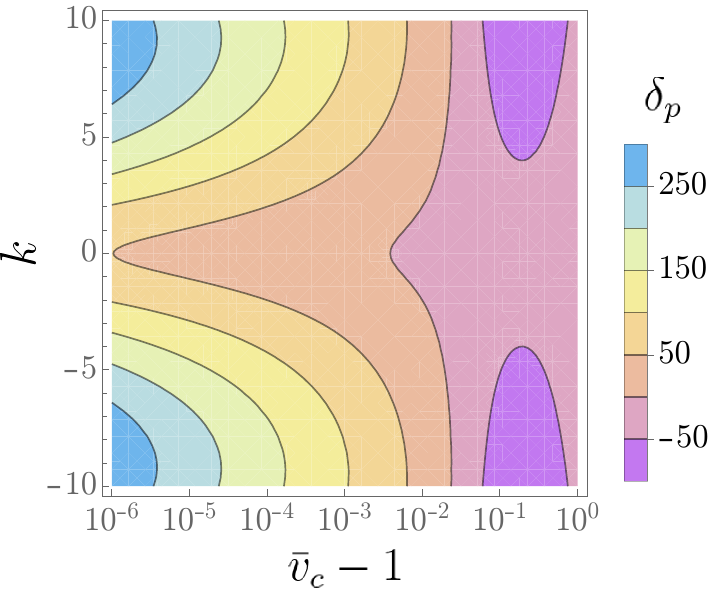} 
        \caption{}
        \label{plotdeltap}
    \end{subfigure}
   \hfill
   \begin{subfigure}{0.48\textwidth} 
       \centering
        \includegraphics[width=\linewidth]{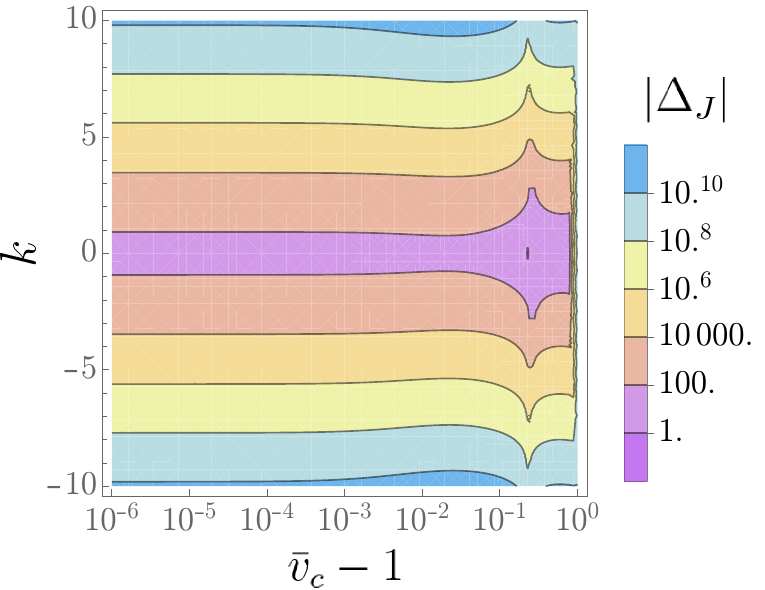} 
        \caption{}
        \label{plotDeltaJ}
    \end{subfigure}
    \caption{Plot of $\delta_p$ (left) and $|\Delta_J|$ (right) as $\bar v_c$ and $k$ vary. They are respectively defined in \eqref{deltap} and \eqref{finaldeltaJ}.}
    \label{}
\end{figure}
All in all, a normalizable solution to the NLO problem in $v\in[1,\infty)$, that is also regular at the tip of the cigar, is provided by plugging the expressions for $d_1$ and $d_2$ discussed above and
\be
\begin{cases}
\displaystyle p_1(v,k; \bar v_c) = - \hspace{-4pt} \int_{\bar v_c}^v  \hspace{-4pt} \frac{\rmd v'}{(h_1(v',k))^2 \pt v' \pt (v'^2-1)} \hspace{-3pt} \int_{v'}^{\infty} \hspace{-11pt} \rmd v'' \pt h_1(v'',k) \pt \mathcal J\bigl [h_1(v'',k)\bigr ] , \, & v \ge \bar v_c \, ,\\[2ex]
\displaystyle p_2(v,k; \bar v_c) = \frac{h_1(\bar v_c,k)}{h_2(\bar v_c,k)} \hspace{-1pt} \int_1^v \hspace{-6pt} \frac{\rmd v'}{(h_2(v',k))^2 \pt v' \pt (v'^2-1)} \hspace{-3pt} \int_1^{v'} \hspace{-11pt} \rmd v'' \pt h_2(v'',k) \pt \mathcal J\bigl [h_2(v'',k)\bigr ] , \, & v < \bar v_c \, ,
\end{cases}
\ee
into the ansätze in \eqref{soldilNLO} and \eqref{ansatzP}. In particular, the solution for $v \ge \bar v_c$ reads
\be\label{Philargev}
\scaleto{\frac{\Phi_{\text{NLO}}(v,k; \bar v_c)}{c_1(k; \bar v_c) h_1(v,k)} = \Delta_J (k; \bar v_c)/w(k) + \delta_d(\bar v_c) + 2 \, \bar v_c \, \delta_{c,1}^{\text{(tree)}} \, \partial_v \hspace{-1pt}\log h_2(\bar v_c, k) + p_1(v,k; \bar v_c) \, .}{28pt}
\ee
It thus follows that
\be
\scaleto{\lim_{v\to\infty} \frac{v^3 \, \partial_v \Phi_{\text{NLO}} (v,k; \bar v_c)}{-2 \, c_1(k;\bar v_c)} = \Delta_J (k; \bar v_c)/w(k) + \delta_d(\bar v_c) + 2 \, \bar v_c \, \delta_{c,1}^{\text{(tree)}} \, \partial_v \hspace{-1pt}\log h_2(\bar v_c, k) + \delta_p(k; \bar v_c) \, ,}{28pt}
\ee
where
\be\label{deltap}
\delta_p(k; \bar v_c) = - \hspace{-4pt} \int_{\bar v_c}^\infty \hspace{-8pt} \rmd v \frac{1}{(h_1(v,k))^2 v \pt (v^2-1)} \hspace{-4pt} \int_{v}^{\infty} \hspace{-10pt} \rmd v' \pt h_1(v',k) \pt \mathcal J\bigl [h_1(v',k)\bigr ] \, .
\ee
This quantity will be useful in the following. We have plot it in figure \ref{plotdeltap}.

Finally, with this solution at hand, in appendix \ref{app:deltaJ} we show how the dependence on $v$ disappears from $\Delta_J$. Its final expression is
\be\label{finaldeltaJ}
\Delta_J(k;\bar v_c) = - \frac{h_2(\bar v_c,k)}{h_1(\bar v_c,k)} \hspace{-1pt} \int_{\bar v_c}^{\infty} \hspace{-6pt} \rmd v \, h_1(v,k) \pt \mathcal J\bigl [h_1(v,k)\bigr ] \hspace{-1pt} - \frac{h_1(\bar v_c,k)}{h_2(\bar v_c,k)} \hspace{-1pt} \int_1^{\bar v_c} \hspace{-6pt}\rmd v \, h_2(v,k) \, {\mathcal J}\bigl [h_2(v,k)\bigr ]\, .
\ee
It is now clear how the parametric dependence on $\bar v_c$ is realized. For a graphical representation see figure \ref{plotDeltaJ}. Recalling the definition of $\mathcal J$ in \eqref{J}, it easily follows that
\be\label{derDeltaJ}
\partial_{\bar v_c} \Delta_J(k;\bar v_c) \propto w(k) \, .
\ee
As a consequence, $\Delta_J$ turns out to be completely independent of $\bar v_c$ --- and so of $x$ and $L$ --- on the zeros of $w$. The reader can find an explicit power series representation of the integrals appearing above in \eqref{finalintegrals}.

\subsection{The holographic dictionary}
\label{sec:holodic}

The observables of dual theories must be related to each other. The set of the prescriptions that codify this map is called \emph{holographic dictionary} and it was proposed in \cite{Witten:1998qj, Gubser:1998bc}. Roughly speaking, a local operator $\mathcal{O}$ in the quantum field theory (QFT) corresponds to a bulk field $\phi$ sharing the very same quantum numbers. Now, in the QFT side, it is natural to deform the theory by introducing a source for $\mathcal{O}$, let us say $\phi_0$. In this way, the action describing the boundary theory becomes
\be
S_{\text{QFT}} \, \mapsto \, S_{\text{QFT}} + \int_{\partial AdS_5} \hspace{-8pt} \rmd^4 \vec x \, \phi_0(\vec x) \, \mathcal{O}(\vec x) \, ,
\ee
where $\partial AdS_5$ denotes the boundary of the $AdS_5$ soliton in the bulk geometry \eqref{NLOmetric}. The conclusion is that $\phi$ at the boundary acts like a source $\phi_0$ for the dual operator $\mathcal{O}$.

The duality thus realizes in the equality of the partition functions from the two sides of the correspondence, given some boundary conditions which constraint the path integral of the bulk theory. Remember that supergravity is sometimes a good approximation of weakly coupled string theory in the low energy limit. The latter is dual to the strongly coupled large $N$ limit of the QFT. In other words, we can state that
\be
Z_{\text{QFT}}\[\phi_0\] =  \int_{\left . \phi \right |_{\partial AdS_5} = \phi_0} \mathcal D \phi \, \, e^{i \pt S[\phi]} \, ,
\ee
where $Z_{\text{QFT}}$ is the generating functional of the strongly coupled boundary field theory and $S$ is the supergravity action in the dual model. 

As a consequence, in the saddle point approximation, the $n$-point correlation function of a local boundary operator $\mathcal{O}$ in the strongly coupled large $N$ limit of a (non-deformed) QFT is given by
\be
\left \langle \mathcal{O}(\vec x_1) ... \mathcal{O}(\vec x_n) \right \rangle_{\text{QFT}} \approx \frac{1}{i^{n-1}} \left . \frac{\delta^{n} \,  S\[\phi_c\]}{\delta \phi_0(\vec x_1) ... \delta \phi_0(\vec x_n)} \right |_{\left . \phi \right |_{\partial AdS_5} = 0} \, ,
\ee
where $\phi_c$ is the solution to the bulk equations of motion for $\phi$ satisfying the required boundary conditions. This is an example about how holography can translate very difficult QFT tasks into classical computations in the gravity dual.

In our case, the dilaton field $\phi$ discussed in the previous sections turns out to be dual to the Yang-Mills Lagrangian density \cite{Klebanov:1997kc, Klebanov:1999xv}, that is
\be
\mathcal O(\vec x) = - \frac{1}{2g_3^2} \Tr F^2(\vec x) \, ,
\ee
$g_3$ being the Yang-Mills coupling constant of the three-dimensional gauge theory. Therefore, it holds that \cite{Gubser:1998bc}
\be\label{dictprofile}
\frac{1}{2g_3^2} \left\langle \Tr F^2 \right \rangle \approx - \frac{\delta S\[\phi_c\]}{\delta \phi} \biggr |_{\left . \phi \right |_{\partial AdS_5} = 0} \, ,
\ee
where $S$ is the action introduced in \eqref{Stot}. Let us focus on the classical dilaton field produced by the static fundamental open string of section \ref{sec:fundstring}. This configuration --- or rather, its Fourier transform --- has been computed in section \ref{sec:dilatoneom} up to NLO in the strong coupling expansion. 
Let us stress that the request of normalizability for such a solution realizes into the vanishing boundary condition $\left . \phi \right |_{\partial AdS_5} = 0$. The expectation value in \eqref{dictprofile} is exactly the measure of the classical flux tube profile we are looking for.

The problem is thus reduced to the computation of the functional derivative of a classical gravity action. To begin with, let us compute the variation of the functional defined in \ref{Stot}, that is
\be
\delta S[\phi] = -\frac{\pi^4 R^5}{2\kappa^2} \int \hspace{-4pt} \rmd^3 \vec x \rmd u \[\frac{\partial \mathcal L}{\partial \phi} \delta \phi + \frac{\partial \mathcal L}{\partial(\partial_\mu \phi)} \delta(\partial_\mu \phi)\] \, .
\ee
Remember that $u\in[u_0,\infty)$ denotes the holographic direction in the background \eqref{NLOmetric}. Integrating by parts, we can require vanishing boundary condition in the Minkowskian sector. Then, on-shell, we get
\be\label{varS}
\delta S[\phi_c] = -\frac{\pi^4 R^5}{2\kappa^2} \int \hspace{-4pt} \rmd^3 \vec x \[\left . \frac{\partial \mathcal L}{\partial(\partial_u \phi)} \delta\phi \right |_{\phi=\phi_c}\]^{u=\infty}_{u=u_0} \, .
\ee
We conclude that\footnote{The contribution coming from the lower extreme of integration at $u=u_0$ vanishes since $g^{uu}(u_0) \sim f(u_0)=0$, not because $\partial_u\phi_c(u_0)=0$ as supposed in \cite{Vyas:2019kvy}.}
\be \label{Sprime}
\frac{\delta S\[\phi_c\]}{\delta \phi} \biggr |_{\left . \phi \right |_{\partial AdS_5} = 0} = - \frac{\pi^4 R_0}{\kappa^2} \lim_{u\to\infty}  u^5 f(u) \, \partial_u \phi_c \, .
\ee
All the quantities here have been introduced in the previous sections. In particular, $R_0$ is the corrected parameter defined in \eqref{correctedR0} and related to the critical temperature in the model as in \eqref{TcR0}. Plugging the above formula back into  \eqref{dictprofile}, we get our holographic prediction for the classical profile of the flux tube connecting the quarks in the boundary theory. Notice that, in the large $u$ limit, the only contribution at NLO to $\phi_c$ comes from \eqref{Philargev}.

\section{Flux tube profile from Holography}
\label{sec:fluxtube}

This section is devoted to our holographic proposal for the classical flux tube profile in a strongly-coupled large $N$ three-dimensional $SU(N)$ gauge theory, taking into account both the finiteness of the inter-quark distance and the first subleading correction in the strong `t Hooft coupling expansion. All the technical aspects of its derivation have been covered in section \ref{sec:setup}. Here, we aim to put all the pieces together and to present our final result, along with some interesting observations. 

In particular, in section \ref{sec:classprof} we will report our general prediction together with a brief recap of the notations. In this way, we hope that the reader which is not interested in the details of the computation can understand what follows independently from section \ref{sec:setup}. Then, in section \ref{sec:physint}, we provide a relationship among the decay length scale of the profile along the transverse direction to the inter-quark axis and the mass of the lowest lying scalar glueball of the confining gauge theory. We conclude with the analysis of some interesting limiting case in section \ref{sec:largeLprofile} and by giving some intuitions about the role of the quantum fluctuations in section \ref{sec:quantfluc}.

Despite good intentions, we understand that the formulae in this section may be difficult to digest. For this reason, we defer all the plots about the general prediction, along with the related discussion, to the conclusions. In this way, we hope to convey the take-away messages without getting lost in technical details.

\subsection{The classical profile}
\label{sec:classprof}

Let us consider a static quark-antiquark pair placed at a distance $L$ at the boundary of the background in \eqref{NLOmetric}, in the Minkowskian sector. There, we can imagine to define a large $N$ confining gauge theory having a critical temperature $T_c$ and a `t Hooft coupling $\lambda_3=g_3^2 N$. Let us suppose that the color charges are connected by a fundamental string diving into the bulk up to a radial coordinate
\be
v_m=(u_m/u_0)^2>1 \, ,
\ee
where $u_0$ denotes the bottom of the geometry. 

The strongly-coupled regime of the boundary theory can be parameterized as\footnote{The parameter $\gamma$ has been defined in \eqref{param}. We can rephrase it as above thanks to the entry of the holographic dictionary in  \eqref{backmetricdic}.}
\be
\gamma = \pt \zeta(3) \bigl(T_c/4\pt\lambda_3\bigr)^{3/2} \ll 1 \, .
\ee 
In this limit, we can express the classical profile of the flux-tube established between the quarks in units of $T_c^3\sqrt{\lambda_3/T_c}$ as\footnote{To get the above result, the reader has just to plug the explicit expressions for $\Phi_{\text{LO}}$ and $\Phi_{\text{NLO}}$ (see section \ref{sec:LOdil} and section \ref{sec:NLOdil} respectively) into formulae \eqref{dictprofile} and \eqref{Sprime}.}
\be \label{completeprofile}
\scaleto{
\boxed{
\frac{\left\langle \Tr F^2(x,z) \right \rangle}{2 \pt g_3^2 \pt T_c^3 \sqrt{\lambda_3/T_c}}  \approx \frac\pi2 \frac{v_c^2(x)}{v_m} \hspace{-4pt}\int^{+\infty}_{-\infty} \hspace{-12pt} \rmd k \, e^{i \pt k \pt z} \frac{ h_2( v_c(x), k)}{w(k) - \gamma \pt \Delta_J(k;v_c(x))} \Bigl ( 1 +5 \pt \gamma \pt \Delta_{\text{vev}}(v_c(x),k) \Bigr ) + \mathcal O(\gamma^2)
} \, ,
}{37pt}
\ee
for any position $x$ along the inter-quark axis such that the LO value of the classical shape $v_c$ of the fundamental string takes values in $(1,2)$ (cf.~\eqref{barvc}). Hence, the prediction holds at \emph{any} finite $L$, as long as the latter condition is met; the reader can find a solution for $v_c=(u_c/u_0)^2$ and $v_m=v_c(0)$ in section \ref{sec:fundstring}.\footnote{Let us stress that, from $v_c=v_{c,0}(1+2\gamma\delta_{c,1}^{\text{(tree)}})+\mathcal O(\gamma^2)$ (cf.~\eqref{expandedvc} and \eqref{barvc}), it follows that
\benn
h_2(v_c,k)=h_2(v_{c,0},k) \bigl(1+2\pt\gamma \pt v_{c,0}\pt\delta_{c,1}^{\text{(tree)}} \partial_v \log h_2(v_{c,0},k)\bigr)+\mathcal O(\gamma^2) \, .
\eenn
Similarly, we have $v_m=v_{m,0}\bigl(1+2\gamma\delta_{m,1}^{\text{(tree)}}\bigr)+\mathcal O(\gamma^2)$ (cf. \eqref{ucumexp}).} To the contrary, the dimensionless transverse coordinate $z$ to the inter-quark axis, along with its conjugate momentum $k$, can take any value. The latter are expressed in units of $2\pi T_c \(1-15\gamma\)$.\footnote{See \eqref{TcR0}, \eqref{correctedR0}, \eqref{deltaT} and \eqref{dimensionlessvar}.} For completeness, $h_2$ and $w$ are defined in \eqref{h2} and \eqref{W} respectively, while $\Delta_J$ has been deeply discussed in appendix \ref{app:deltaJ}; its final expression is reported in formula \eqref{finaldeltaJ}. Finally, we also introduced
\be
\Delta_{\text{vev}}(v_c,k)=-9+ \frac{111}{8 \pt v_c^2} +\frac{211}{16 \pt v_c^4} - \frac{34}{v_c^6} + \frac{9}{16 v_m^2} + \frac{19}{32 \pt v_m^4} + \frac{1}{2 \pt v_m^6} + \frac15 \pt \delta_p(k; v_c) \, ,
\ee
where $\delta_p$ has been defined in \eqref{deltap}.


\subsection{Physical interpretation of the intrinsic width}
\label{sec:physint}

The non-trivial flux tube profile of the previous section corresponds to the order $\mathcal O(\gamma/N^2)$ dilaton field configuration sourced by a fundamental string in the supergravity background \eqref{NLOmetricpiuparam} at tree level (see \eqref{treedeltas}). This relation has been discussed in section \ref{sec:holodic}. Nevertheless, besides the background configuration probed by our stringy source (see \eqref{backphi0phi1}), the equation of motion for the dilaton field in \eqref{eomphi0phi1} also admits non-trivial solutions at order $\mathcal O(1)$ and $\mathcal O(\gamma)$. 
These field configurations are dual to scalar glueball states $0^{++}$ in the boundary theory.\footnote{We used the common notation $J^{PC}$ to denote a glueball state. Respectively, $J$, $P$ and $C$ refer to the spin, the parity and the charge conjugation quantum numbers.} The computation of their masses from supergravity has been a widespread topic in late 1990s literature \cite{Witten:1998zw, deMelloKoch:1998vqw, Zyskin:1998tg, Ooguri:1998hq, Csaki:1998qr}. Here, we aim to briefly review the spectrum of the dilaton modes adding some new remarks at NLO.

Let us start by finding a non-trivial solution to the $\mathcal O(1)$ equation of motion in \eqref{eomphi0}. In momentum space, it corresponds exactly to the same Heun equation discussed in section \ref{sec:LOdil}. Therefore, the $\mathcal O(1)$ on-shell dilaton field must be given by the expansions in \eqref{Heundilaton}, within the respective domain of convergence. More explicitly, with
\be\label{FourierexpLOexc}
\phi_0(v,z)=\frac{1}{2\pi} \int_{-\infty}^{+\infty} \hspace{-6pt} \rmd k \, e^{i \pt k \pt z} \, \Phi_{0}(v,k)  \, ,
\ee we have\begin{subequations}\label{LOexc}
\begin{align}
&\Phi_0^{(\infty)}(v,k) = \alpha_1(k) \, h_1 (v,k) \, , \quad 1<v<\infty \, ,\\
&\Phi_0^{(1)}(v,k) = \alpha_2(k) \, h_2 (v,k) \, , \quad 1\leq v<2 \, ,
\end{align}
\end{subequations}
$\alpha_1$ and $\alpha_2$ being two arbitrary functions of $k$. For the definition of $h_1$ and $h_2$ see \eqref{Heundilaton}.

\begin{figure}[t]
    \centering
    \captionsetup[subfigure]{labelformat=empty}
    \begin{subfigure}{0.5\textwidth} 
       \centering
        \includegraphics[width=\linewidth]{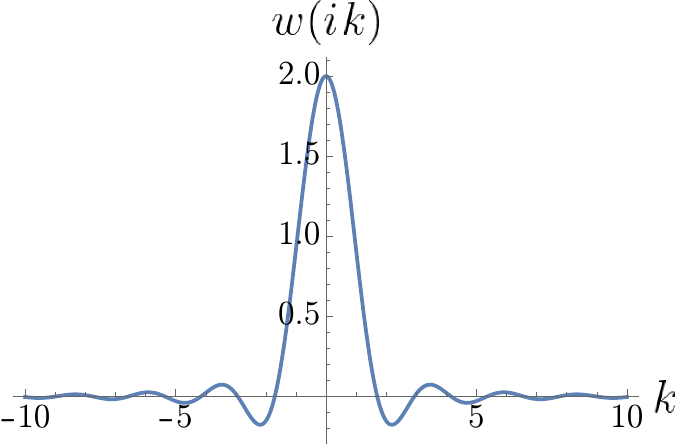} 
        \caption[]{}
    \end{subfigure}
    \hfill
    \begin{subfigure}{0.45\textwidth} 
        \centering
        \begin{tabular}{cc}
            \toprule
            $m_0$ & $4 \, m_0^2$ \\
            \midrule
            1.702 & 11.59 \\
            2.938 & 34.53 \\
            4.153 & 68.98 \\
            5.360 & 114.9 \\
            6.564 & 172.3 \\
            7.766 & 241.2 \\
            8.967 & 321.6 \\
            \bottomrule
            \vspace{1.5pt}
        \end{tabular}
        \caption[]{}
    \end{subfigure}
    \caption{\label{plotW}Plot of the function $w$ defined in \eqref{W}, evaluated along the imaginary axis. To get this result, we truncated the series in \eqref{Heundilaton} keeping the contributions for $n\leq 100$. Let us stress that the latter has been chosen to reproduce the well-known masses of the glueballs probed for $|k|\leq10$ (see the table on the right).}\captionof{table}{\label{tablemasses}The first column of the table contains the leading order $0^{++}$ glueball masses $m_0$ in units of the leading order value of $M_{\text{KK}}=2\pi T_c$ (see \eqref{TcR0}, \eqref{MKKR0} and \eqref{correctedR0}). Notice that they are computed as the zeros of $w$ in \eqref{W}, adopting the same truncation introduced to produce the plot on the left. The second column displays a particular function of $m_0$ which is ready to be compared with the well-known results listed, \eg in \cite{deMelloKoch:1998vqw, Csaki:1998qr, Zyskin:1998tg, Minahan:1998tm, Brower:2000rp}. For a comparison with the lattice results in the large $N$ limit, see Table 1 and Table 2 of~\cite{Csaki:1998qr}.}
\end{figure}
In this case, we do not have any delta-like source introducing some gluing conditions. As a consequence, $\alpha_1$ and $\alpha_2$ remain arbitrary; furthermore, the solutions \eqref{LOexc} are meant to be valid at any allowed $v$. For consistency, they must describe the very same field configuration --- \ie they must be linearly dependent --- in the overlap region $1<v<2$. In other words, we must require their Wronskian W defined in \eqref{Wronskian} to vanish for any $v\in(1,2)$. This provides a condition which cannot be solved by arbitrary values of $k$ and thus a discrete glueball spectrum. Notice that we can apply the Abel identity to get the $v$-independent combination $w$ introduced in \eqref{reducedW}. Then, the above condition reduces to the vanishing of $w$. More in details, the zeros $k_0$ of $w$ are such that
\be\label{LOprop}
\Phi_0^{(\infty)}(v,k_0) = \Phi_0^{(1)}(v,k_0) \, , \quad \forall \, v \in (1,2) \, ,
\ee
and turn out to be arranged on the imaginary axis of the complex $k$-plane (see figure \ref{plotW}). The values
\be
m_0^2 = -k_0^2>0 \, ,
\ee
listed in table \ref{tablemasses}, give the masses squared of the dilaton modes in the expansion \eqref{FourierexpLOexc}. From a boundary perspective, they correspond to the masses squared of the scalar glueball excitations in units of the leading order value of $M_{\text{KK}}^2=4\pi^2T_c^2$ (see \eqref{TcR0}, \eqref{MKKR0} and \eqref{correctedR0}).

Crucially, the reduced Wronskian $w$ is what appears in the denominator of the leading order Fourier transform in \eqref{completeprofile}. Therefore, the leading order flux tube profile --- selected by $\gamma=0$ --- shall be dominated by the lowest-lying zero of $w$ in the large $z$ limit, leading to
\be\label{LOasymprof}
\left . \frac{\left\langle \Tr F^2(x,z) \right \rangle}{2 \pt g_3^2 \pt T_c^3 \sqrt{\lambda_3/T_c}} \right |_{\gamma=0} \approx e^{-m_{0,\text{lgh}}|z|} \, , \quad |z|\to\infty \, ,
\ee 
where $m_{0,\text{lgh}}$ represents the leading order mass of the lightest glueball in the spectrum reported in table \ref{tablemasses}.\footnote{In the holographic limit, this is not the lightest state in the whole spectrum of the theory. Rather, it is the lightest mode in the harmonic expansion of the dilaton field. See, \eg \cite{Brower:2000rp} for a fully detailed discussion about the correspondence of glueball states and supergravity modes. Notice that there is only one lighter $0^{++}$ mode sourced by $\Tr F^2$ in the full spectrum. The latter has been dubbed ``exotic'' in \cite{Constable:1999gb}, since it arises as a graviton excitation along (primarily) the KK direction $\theta$. As a consequence, among the two, just the above dilaton mode should survive the KK mode decoupling limit. We conclude that it is the smallest mass reported in table \ref{tablemasses} that should be compared with the lattice data. See, \eg \cite{Bigazzi:2015bna} for a similar discussion.} Remember that $z$ parameterizes the distance from the inter-quark axis. Thus, at leading order, we can immediately identify the intrinsic width of the flux tube profile with the inverse mass of the lightest scalar glueball state in the spectrum of the confining gauge theory. This leading order behavior has been already reported in \cite{Danielsson:1998wt}.

Now, let us discuss the NLO problem. To begin with, let us express the solution to the $\mathcal O(\gamma)$ dilaton equation of motion \eqref{eomphi1} as
\be
\phi_1^{\text{(tree)}}(v,z)=\phi_{1,\text{bck}}^{\text{(tree)}}(v) + \frac{1}{2\pi} \int_{-\infty}^{+\infty} \hspace{-6pt} \rmd k \, e^{i \pt k \pt z} \, \Phi_1^{\text{(tree)}}(v,k)  \, .
\ee
Notice that $\phi_{1,\text{bck}}^{\text{(tree)}}$ is the background field configuration introduced in \eqref{phi1treebck}. Formally, once $\Phi_{\text{LO}}$ is replaced with $\Phi_0$, the \emph{linearized} equation of motion for $\Phi_1^{\text{(tree)}}$ corresponds exactly to the equation of motion for $\Phi_{\text{NLO}}$ without the delta-like source (see \eqref{NLOeom}). The very same equation already appears in \cite{Csaki:1998qr, Ooguri:1998hq}. As we saw before, the Dirac delta in the source term just affects the normalization of the solutions. Therefore, the formalism developed in section \ref{sec:NLOdil} can be also used here to study the glueball spectrum at NLO in the strong coupling expansion. 

The solution for $\Phi_1^{\text{(tree)}}$ can be parameterized as
\be
\Phi_1^{\text{(tree)}}(v,k) =
\begin{cases}
 \beta_1(k) \, h_1(v,k) + \alpha_1(k) P_1(v,k) \, , \quad  1 < v < \infty \, , \\[1ex]
  \beta_2(k) \, h_2(v,k) + \alpha_1(k) P_2(v,k) \, , \quad 1 \leq  v < 2 \, .
\end{cases}
\ee
Again, $\beta_1$ and $\beta_2$ are arbitrary functions of $k$. Formally, the particular solutions $P_1$ and $P_2$ can be expressed as in formula \eqref{ansatzP} and --- as long as $c_1$ and $\Phi_{\text{LO}}$ are mapped to $\alpha_1$ and $\Phi_0$ --- the following discussion about the free parameters still holds. Indeed, the latter are fixed by requiring normalizability and regularity at the tip, which we do not want to be spoiled here.  Notice that now $\bar v_c$ defined in \eqref{barvc} has no meaning and can be replaced with an arbitrary parameter, let us say, $a$. All in all, we can write
\begin{align}\label{P1P2exc}
&\displaystyle P_1(v,k) = - h_1(v,k) \hspace{-3pt}\int_{a}^v \hspace{-8pt} \rmd v' \frac{1}{(h_1(v',k))^2 v' \pt (v'^2-1)} \hspace{-1pt} \int_{v'}^{\infty} \hspace{-10pt} \rmd v'' \pt h_1(v'',k) \pt \mathcal J\bigl [\Phi_0^{(\infty)}(v'',k)\bigr ]/\alpha_1(k) , \nonumber \\[0.5ex]
&\displaystyle P_2(v,k) = h_2(v,k) \hspace{-3pt} \int_1^v \hspace{-8pt} \rmd v' \frac{1}{(h_2(v',k))^2 v' \pt (v'^2-1)} \hspace{-1pt}\int_1^{v'} \hspace{-10pt} \rmd v'' \pt h_2(v'',k) \pt \mathcal J\bigl [\Phi_0^{(1)}(v'',k)\bigr ]/\alpha_1(k),
\end{align}
for some $a>1$. The latter just sets the location of the zero of $P_1$. As we will see, its value will play no role in what concerns this section. Remember that the current $\mathcal J$ has been defined in \eqref{J}.

All in all, the on-shell dilaton modes describing the physical fluctuations above the fixed background configuration can be expressed as
\begin{align}
&\Phi_{{fl}}^{(\infty)}(v,k) =\bigl (\alpha_1(k) + \gamma \, \beta_1(k)\bigr ) h_1(v,k) + \gamma \,\alpha_1(k) P_1(v,k) + \mathcal O(\gamma^2) \, , \quad  1 < v < \infty \, , \nonumber\\[0.5ex]
&\Phi_{{fl}}^{(1)}(v,k) =\bigl (\alpha_2(k) + \gamma \, \beta_2(k)\bigr ) h_2(v,k) + \gamma \,\alpha_1(k) P_2(v,k) + \mathcal O(\gamma^2) \, , \quad 1 \leq  v < 2 \, .
\end{align}
Again, these two field configurations must be equivalent on the overlap region $1<v<2$. In other words, up to $\mathcal O(\gamma)$, their Wronskian must vanishes, that is
\be\label{NLOprop}
\Phi_{{fl}}^{(\infty)} \partial_v \Phi_{{fl}}^{(1)} - \Phi_{{fl}}^{(1)} \partial_v \Phi_{{fl}}^{(\infty)} = 0 \, , \quad \forall \, v \in (1,2) \, .
\ee
Let us assume that it is the case on
\be \label{k1}
k_1 = k_0 \(1+ \gamma \pt \delta_k\) \, , \quad \delta_k \in \mathcal O(1) \, .
\ee
Then, expanding \eqref{NLOprop} around $k_0$ and making use of the leading order relation in \eqref{LOprop}, we get
\be
k_0 \pt w'(k_0) \, \delta_k -  v\pt(v^2-1) \(W_1(v,k_0)-W_2(v,k_0)\) + \mathcal O(\gamma^2) = 0\, .
\ee
The quantities denoted as $W_1$ and $W_2$ have the very same formal definition as in \eqref{startingW1W2}, with $P_1$ and $P_2$ given in \eqref{P1P2exc}. Notice that the prime denotes the  derivative with respect to $k$.

We can thus follow the same steps of appendix \ref{app:deltaJ}, getting\footnote{Again, the reader just has to replace $c_1$, $\Phi_{\text{LO}}$ and $\bar v_c$ respectively with $\alpha_1$, $\Phi_0$ and $a$.}
\begin{subequations}\label{deltak}
\be
\boxed{
\delta_k = \frac{\delta_J(k_0)}{k_0 \pt w'(k_0)}
} \, ,
\ee
where
\be
\boxed{
\begin{split}
\delta_J(k_0) = - \frac{h_2(a,k_0)}{h_1(a,k_0)} \hspace{-1pt} \int_a^{\infty} \hspace{-8pt} \rmd v \, h_1(v,k_0) \pt \mathcal J\bigl [h_1(v,k_0)\bigr ] \hspace{-1pt} - \frac{h_1(a,k_0)}{h_2(a,k_0)} \hspace{-1pt} \int_1^a \hspace{-6pt}\rmd v \, h_2(a,k_0) \, {\mathcal J}\bigl [h_2(v,k_0)\bigr ]\, ,\\
a\in(1,2) \, .\\[1ex]
\end{split}
}
\ee
\end{subequations}
Crucially,
\be
\partial_a \delta_J(k_0) \propto w(k_0) = 0 \, .
\ee
Therefore, the parametric dependence on $a$ vanishes and so its value does not affect the final result. This is why we omit $a$ among the arguments of $\delta_J$ above. The reader can find an explicit expressions for the integrals appearing above in \eqref{finalintegrals}, by mapping $\bar v_c$ to $a$ in each formula. 

\begin{table}
    \centering
       \begin{tabular}{ccc}
            \toprule
            $\delta_k$ & $(\delta_k-15)/4$ \\
            \midrule
          3.88 & -2.78\\
          5.30 & -2.43\\
	5.87 & -2.28\\
	6.08 & -2.23\\
	6.17 & -2.21\\
	6.22 & -2.20\\
	6.24 & -2.19\\   
            \bottomrule
            \vspace{1.5pt}
        \end{tabular}
        \caption{The first column gives the NLO corrections to the first seven leading order glueball masses listed in table \ref{tablemasses}, according to \eqref{deltak} and \eqref{finalm2}. The second column provides a quantity that is easily comparable with the numerical results in the literature. In particular, our entries correspond exactly to the numerical factors in the NLO terms of formula (4.12) of \cite{Csaki:1998qr}. Notice that the values in the last row are new. In principle, corrections for states even more excited than the sixth would be easily computable with our formula. Let us stress that we make use of the series in \eqref{finalintegrals}, including terms whose indices sum up to an integer that is less than or equal to 100.}
        \label{tabledeltak}
\end{table}
All in all, our proposal for the masses squared of the scalar glueball excitations in units of the leading order value of $M_{\text{KK}}^2$ is
\be\label{finalm2}
m^2 = m_0^2 \, \bigl(1 + 2 \pt \gamma \pt \delta_k\bigr ) \, .
\ee
In units of the full $M_{\text{KK}}^2$ (see \eqref{MKKR0}, \eqref{correctedR0} and \eqref{deltaT}), the result reads
\be
M^2 =  m_0^2 \, \bigl[1 + 2 \pt \gamma \pt (\delta_k-15)\bigr ] \, .
\ee
It is a semi-analytical formula in the sense that, according to \eqref{deltak}, the value of the subleading correction relies on the knowledge of $m_0$, which is derived numerically (although with extreme precision). In table \ref{tabledeltak}, we report the results of our prediction for the first seven scalar glueballs in the spectrum. The agreement with the data in the literature is impressive. We believe that this correspondence represents a strong check of our formalism.

The analogies between this section and appendix \ref{app:deltaJ} are fundamental for what concerns the physical interpretation of our result and do not end here. Indeed, in \eqref{derDeltaJ}, we also proved that the parametric dependence of $\Delta_J$ on $\bar v_c$ vanishes on the zeros of $w$. It follows that
\be
\Delta_J(k_0; \bar v_c) = \delta_J(k_0) \, .
\ee
As a consequence, up to $\mathcal O (\gamma)$, the zeros of the denominator in the Fourier transform in \eqref{completeprofile} correspond to \eqref{k1}, with $\delta_k$ given in \eqref{deltak}.

The asymptotic behavior in \eqref{LOasymprof} can be thus extended up to NLO as
\be\label{NLOasymprof}
\frac{\left\langle \Tr F^2(x,z) \right \rangle}{2 \pt g_3^2 \pt T_c^3 \sqrt{\lambda_3/T_c}} \approx e^{-m_{\text{lgh}}|z|} \, , \quad |z|\to\infty \, ,
\ee 
where $m_{\text{lgh}}$ is the square root of \eqref{finalm2} for the lightest state among the dilaton modes. We conclude that the identification of the intrinsic width of the flux tube with the inverse mass of the lowest-lying glueball of the confining gauge theory holds even at NLO in the strong coupling expansion.


\subsection{Large $L$ limit and the classical broadening}
\label{sec:largeLprofile}

The profile in \eqref{completeprofile} is completely general, but also very complicated. It would be nice to derive a more insightful form to fully understand the underlying physics. In this section, we aim to study its leading order contribution --- selected by $\gamma=0$ ---  in the large inter-quark separation limit.  This allows us to discuss general aspects of the flux tube analytically, in addition to providing some useful results for the following. Let us stress that, for what concerns this section, also the critical temperature $T_c$ and the mass $m_{\text{lgh}}$ of the lighter glueball in the spectrum are evaluated at $\gamma=0$.

First, let us address the case in which the inter-quark distance $L$ is strictly equal to infinity. Then, the holographic coordinate $v_c$ of the classical string --- and so its minimum $v_m$ --- is fixed to one for all $x$ belonging to $[-L/2, +L/2]$ (cf.~\eqref{umasym}). The same applies to the function $h_2$ defined in \eqref{Heundilaton}. All in all, as $L$ goes to infinity, the boundary observer spots an infinite string of finite width described by the profile
\be\label{limitprof}
\lim_{L\to\infty}\frac{\left\langle \Tr F^2(x,z) \right \rangle}{2 \pt g_3^2 \pt T_c^3 \sqrt{\lambda_3/T_c}}  \approx \pi^2 P(z)/2 + \mathcal O(\gamma) \, ,
\ee
where
\be\label{P}
P(z) = \frac1\pi \int_{-\infty}^{+\infty} \hspace{-6pt} \rmd k \, \, e^{i \pt k \pt z} \frac{1}{w(k)} \, .
\ee
Here, $P$ has been chosen such that\footnote{To prove it, we used the integral representation of the Dirac delta $2\pi\delta(k)=\int_{-\infty}^{+\infty} \hspace{-2pt} \rmd z \,  \text{Exp}(i \pt k \pt z)$, together with the knowledge of $w(0)=2$ from \eqref{watzero}.}
\be\label{normP}
\int_{-\infty}^{+\infty} \hspace{-6pt} \rmd z \, P(z) = 1 \, .
\ee
The above result reproduces exactly formula (4.101) of \cite{Danielsson:1998wt}. Nevertheless, we derived it in a more refined way starting from the finite $L$ case, \ie with $v_c \neq 1$. In this way, we avoided many technical problems. Notice that $P$ features a fixed squared width given by
\be\label{widthP}
w^2_P  = w''(0)/\pt2 \approx 1.386 \, .
\ee
For its computation see appendix \ref{app:widthP}. The primes denote the derivative with respect to $k$. Finally, the same arguments at the beginning of the previous section can be applied here. Therefore, at large distances from the inter-quark axis, $P$ decays exponentially as in \eqref{LOasymprof}.

\begin{figure}[t]
    \centering
    \begin{subfigure}{0.475\textwidth}
       \centering
        \includegraphics[width=\linewidth]{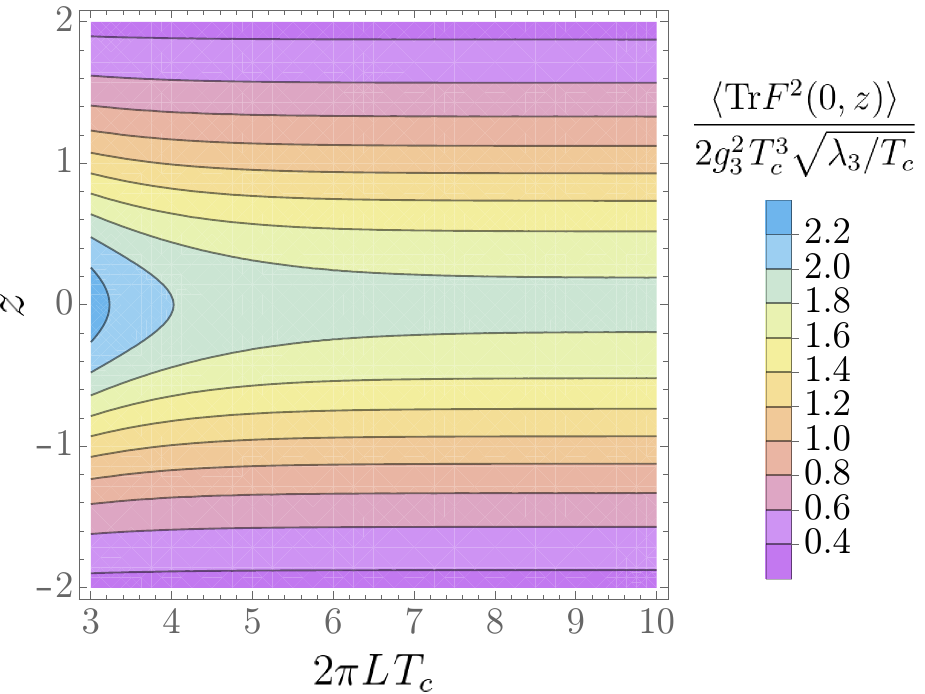} 
        \caption{}
        \label{}
    \end{subfigure}
    \begin{subfigure}{0.45\textwidth}
       \centering
        \includegraphics[width=\linewidth]{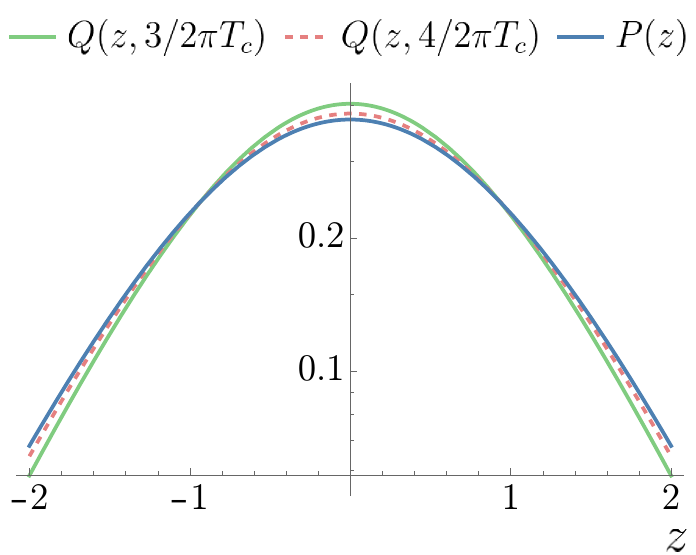} 
        \caption{}
        \label{}
    \end{subfigure}
    \caption{(a)  Classical flux tube profile at finite $L$ reported in \eqref{proflargeL}, taking into account the first subleading correction with respect to the infinite inter-quark distance case. Here, the transverse spatial direction $z$ runs from $-2$ to $+2$, in units of $2\pi T_c$. (b) Plots of $Q(L,z)$ defined in \eqref{profQ} for $2\pi L T_c = 3 \text{ (green line)}, 4 \text{ (dashed red line)}$, along with $P(z)$ reported in \eqref{P} (blue line). As it should, the latter represents the limit shape in the large $L$ limit. Let us stress that all these profiles shares the same normalization, due to the vanishing behavior of $P$ at infinity. Notice that $P$, and so $Q$, has been computed by truncating the domain of integration to $|k|\leq10$; moreover, we use the same $w$ plotted in figure \ref{plotW}.}
    \label{plotfiniteLexp}
\end{figure}
Now, let us try to introduce the first $L$-dependent correction to the flux tube in the large $L$ limit. Let us focus on the profile in the middle of the flux tube, \ie at $x=0$. From \eqref{umasym}, we know that\footnote{At leading order in the strong coupling expansion, the holographic dictionary states $2u_0/R^2=2\pi T_c$ (see \eqref{TcR0} and \eqref{LOconical}).}
\be\label{vmforfit}
v_c(0) = v_m \approx 1 + 2 \pt c \, e^{-2 \pi T_c L } + \mathcal O\bigl(e^{-4 \pi T_c L }\bigr) \, ,   \quad L\to\infty \, .
\ee
Remember that $c$ is defined in \eqref{defcd}. It thus follow that
\be
h_2(v_c(0),k) = 1 + c \, k^2 \pt e^{-2\pi T_c L} + \mathcal O \bigl ( e^{-4\pi T_c L}\bigr ) \, .
\ee
All in all, we can rephrase the profile in \eqref{completeprofile} as
\begin{subequations}\label{proflargeL}
\be
\frac{\left\langle \Tr F^2(0,z) \right \rangle}{2 \pt g_3^2 \pt T_c^3 \sqrt{\lambda_3/T_c}}  \approx \pi^2 \, v_m(L) \, Q(z,L) / \pt 2 + \mathcal O(\gamma) \, ,
\ee
where
\be\label{profQ}
Q(z,L) = \Bigl (1 - c \, e^{-2\pi T_c L} \pt\partial^2_z\Bigr) P(z) + \mathcal O \bigl ( e^{-4\pi T_c L}\bigr ) \, , \quad L \to \infty \, .
\ee
\end{subequations}
Notice that $Q$ is normalized to one as $P$.\footnote{Remember that $P$, along with its derivatives, goes exponentially to zero at large distance. See \eqref{LOasymprof}.} In figure \ref{plotfiniteLexp} we plot the above profile, along with a comparison among $Q$ at different values of $L$.

\begin{figure}[t]
    \centering
        \includegraphics[width=0.4\textwidth]{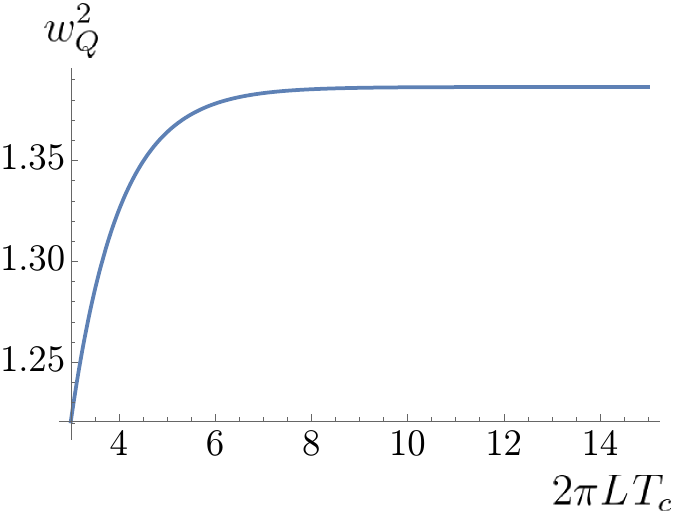} 
    \caption{Classical broadening of the flux tube at finite $L$, where the squared width $w_Q^2$ is defined in \eqref{classbroad}.}
    \label{figbroad}
\end{figure}
The above approximation provides an analytical playground that allows us to give some interesting predictions. For instance, the squared width of $Q$ turns out to be
\be\label{classbroad}
w^2_Q = w^2_P -2 \, c \,e^{-2 \pt \pi \pt T_c \pt L} + \mathcal O \bigl ( e^{-4\pi T_c L} \bigr ) \, ,
\ee
where $w^2_P$ is defined \eqref{widthP}. Its computation is deferred to appendix \ref{app:classbroad}. We conclude that the finiteness of the inter-quark distance produces a classical broadening, as shown in figure \ref{figbroad}. This is a different phenomenon compared to the logarithmic broadening coming from the quantum effects, which are not captured by the on-shell classical Nambu-Goto action discussed so far. The quantum behavior of the width of the profile has been discussed in a holographic framework in \cite{Greensite:2000cs, Giataganas:2015xna, Giataganas:2015yaa, Ridgway:2009tca, Armoni:2008sy, Loewy:2001pq}, relying on the prescription introduced in \cite{Luscher:1980iy}. In the next section, we will show how to achieve this result by following our strategy. For the time being, notice that the subleading correction in \eqref{classbroad} does not depend on how we decide to model $P$. We can use its formal definition in \eqref{P}, or we can decide to approximate it with some more easy-to-handle expression. Indeed, as it is clear from appendix \ref{app:classbroad}, any $P$ normalized to one produces the same subleading effect. To the contrary, the first contribution $w^2_P$ is strongly $P$-dependent.


\subsection{Some insights into quantum fluctuations}
\label{sec:quantfluc}

In this section, we aim to give some insights into how to include the quantum fluctuations in our predictions. In particular, we will see how the convolutional structure proposed in \cite{Cardoso:2013lla} naturally arises in gravity, getting some results that seem to mimic those obtained in \cite{Aharony:2024ctf} for $\text{QED}_3$.

Let us suppose that the world-sheet of the fundamental string sourcing the dilaton field of section \ref{sec:dilatoneom} --- and so the flux tube profile in the boundary theory --- is not classical anymore. Rather, it fluctuates around the reference configuration discussed in section \ref{sec:fundstring}. If this is the case, then we need to extend the bulk partition function appearing in section \ref{sec:holodic} as\footnote{This expression has already appeared in \cite{Vyas:2019kvy} in the same framework.}
\be
Z_{\text{bulk}} = \hspace{-4pt} \int \hspace{-4pt} \mathcal D \phi \, \mathcal D \Sigma \, e^{i \pt \hat S[\phi,\Sigma]} \, ,
\ee
where
\be\label{hatS}
\hat S[\phi,\Sigma] = S_{\text{dil}}[\phi]+S_{\text{NG}}[\phi, \Sigma] \, .
\ee
Here, we are summing over all the possible dilaton field configurations $\phi$ and over all the possible world-sheets $\Sigma$ whose boundaries are given by the world-lines of the quarks. Moreover, $S_{\text{dil}}$ and $S_{\text{NG}}$ are respectively the scalar sector \eqref{scalarsectorsugra} of the supergravity action \eqref{sugraaction} and the Nambu-Goto action in \eqref{NGaction}; both the functionals are in Einstein frame. Let us stress that now the Nambu-Goto action is not evaluated on the classical world-sheet solution discussed in section \ref{sec:fundstring}.

The entry of the holographic dictionary discussed in section \ref{sec:holodic} must be generalized as
\be
\frac{1}{2g_3^2} \left\langle \Tr F^2 \right \rangle =  \frac{i}{Z_{\text{bulk}}} \frac{\delta}{\delta \phi} Z_{\text{bulk}} \biggr |_{\left . \phi \right |_{\partial AdS_5} = 0} \, .
\ee
In the saddle point approximation, we can put the dilaton field on-shell getting
\be\label{avarageprofile}
\frac{1}{2g_3^2} \left\langle \Tr F^2 \right \rangle \approx  \frac{1}{Z_{\text{NG}}} \hspace{-1pt}\int \hspace{-1pt} \mathcal D \Sigma \, e^{i \pt S_{\text{NG}}[\phi_c, \Sigma]}\, \(- \frac{\delta}{\delta \phi} \hat S[\phi_c, \Sigma] \biggr |_{\left . \phi \right |_{\partial AdS_5} = 0}\) \, ,
\ee
where
\be
Z_{\text{NG}} = \hspace{-4pt} \int \hspace{-4pt} \mathcal D \Sigma \, e^{i \pt S_{\text{NG}}[\phi_c,\Sigma]} \, .
\ee
As usual, we denote as $\phi_c$ the solution to the dilaton equation of motion. Looking at \eqref{dictprofile}, we may interpret the full quantum profile as a functional average over all the possible flux-tubes sourced by any world-sheet connecting the quark world-lines at the boundary.

Here, we are still interested in the supergravity modes sourced by $\Tr F^2$ in the KK decoupling limit. Therefore, we can reduce over the compact spaces as in section \ref{sec:dilatoneom}. In fact, $\hat S$ above corresponds to the total action $S$ in \eqref{Stotclass}, once the Dirac deltas in the source term \eqref{sourceterm} are shifted according to the physical position of the fluctuating stringy source. The latter can be parameterized in the static gauge \eqref{staticgauge} as\footnote{Remember the change of coordinates in \eqref{dimensionlessvar}.}
\be\label{fluctws}
\hat v = v_c(x) + \chi(t,x) \,  , \quad \hat z = \zeta(t,x) \,  ,
\ee
where $\chi$ and $\zeta$ represent the (dimensionless) quantum perturbations around the classical world-sheet of section \ref{sec:fundstring}, respectively along the holographic direction and orthogonally to the inter-quark axis. Notice that, in general, the fluctuations of the world-sheet are time-dependent.

Therefore, for a given quantum world-sheet $\Sigma$ described by \eqref{fluctws}, $\phi_c$ solves the equation of motion\footnote{For the sake of simplicity, we introduced $\sqrt{\menodet{\tilde g}{10}} =  (1+\delta_{10} ) \sqrt{\menodet{g}{5}{}_{{\hspace{-1.5pt}}_{\scaleto{, 0}{5pt}}}}$. Cf.~equation \eqref{fulleom}.}
\begin{align}\label{quanphieom}
\partial_\mu \Bigl ( \sqrt{\menodet{\tilde g}{10}} \hspace{0.5pt}g^{\mu\nu} \partial_\nu \phi\Bigr ) = & \frac32 \gamma \hspace{0.5pt} R^6 \hspace{0.5pt} W \sqrt{\menodet{\tilde g}{10}} e^{-3\phi/2} \(1-e^{2\phi}\frac{\pi^2 g_s^2}{9 \zeta(3)}\) + \nonumber\\
&+ \gamma \hspace{0.5pt} g_s^2 \, \frac{512 \pi^2}{R \zeta(3)} {\textstyle{\sqrt{\menodet{g}{2}}}} \,  e^{\phi/2} \, \delta(z-\zeta(t,x)) \,\sqrt{v} \pt \delta(v-v_c(x)-\chi(t,x)) \, ,
\end{align}
with $\mu, \nu = t,x,z,v$. Remember that $R$ is the radius of the $AdS_5$ soliton and that $g_s \sim 1/N$ is the string coupling. Let us stress that the determinant of the induced metric on the world-sheet depends now on all the fluctuations of the world-sheet itself. Moreover, the results in \eqref{varS} and \eqref{Sprime} formally holds even for $\hat S$. Then, the functional derivative in \eqref{avarageprofile} can be computed as \be\label{derhatS}
\frac{\delta}{\delta \phi} \hat S[\phi_c, \Sigma] \biggr |_{\left . \phi \right |_{\partial AdS_5} = 0} \propto \lim_{v\to\infty} v^3 \, \partial_v \phi_c
\ee
for any world-sheet $\Sigma$.

In principle, we may thus adopt the same strategy of section \ref{sec:dilatoneom}. In particular, the dependence on $z$ can be traded away by switching to the momentum space. In this case, the integral representation of the above Dirac delta centered at $z=\zeta(t,x)$ makes convenient to define a shifted Fourier transform $\Phi_c$ of the on-shell dilaton field as
\be\label{flucPhi}
\phi_c(v,z;\Sigma) = \frac{1}{2\pi} \int \hspace{-4pt} \rmd k \, e^{i \pt k \pt (z-\zeta(t,x))} \Phi_c(v,k;\Sigma) \, ,
\ee 
for any $x$ and $\Sigma$. This removes all the dependence on $z$ from the equation of motion. Once solved for $\Phi_c$, it is clear how the functional derivative of $\hat S$ in \eqref{avarageprofile} turns out to be a function of $z-\zeta(t,x)$ (look at \eqref{derhatS}). 

Even if an explicit solution for $\Phi_c$ in \eqref{flucPhi} were known, the computation of the functional integral in \eqref{avarageprofile} would be still highly non-trivial in full generality. So, let us gather our thoughts to figure out what we can say about it. First, notice that the fundamental string in the bulk breaks one translation symmetry in the $2+1$-dimensional Minkowskian sector of the bulk geometry \eqref{NLOmetric}. Therefore, a massless Nambu-Goldstone boson should appear in the world-sheet theory. It is well-known that its expectation value provides the position of the string along the transverse direction to the world-sheet itself. Therefore, in the static gauge \eqref{staticgauge}, we can identify the Nambu-Goldstone boson on the world-sheet with the fluctuation $\zeta$ introduced in \eqref{fluctws}. In addition, we expect other world-sheet modes --- both bosonic and fermionic --- to acquire mass at the scale $M_{\text{KK}}$ due to the curvature of the target space. At low energies, we can imagine to integrate all them out. This computation has been performed by Aharony et al.~in \cite{Aharony:2009gg, Aharony:2010cx} for a long Type II superstring in a large class of confining backgrounds.

The low energy action for a long fundamental superstring satisfying Dirichlet boundary conditions and sitting in the IR region of the background in \eqref{LOback} --- in our notations --- reads\footnote{Let us stress that our string is localized on the five-sphere sector. Then, from a perturbative point of view, the world-sheet sigma model would feature other five massless bosons describing the position of the string in those directions. Nevertheless, in the quantum theory there is a beta function for the radius of the five-sphere that drives it towards small values, generating a non-perturbative scale and thus a mass gap at that scale. All in all, there is no massless modes on the sphere and in the deep low energy limit the only word-sheet boson that survives is $\zeta$ above. We are indebted to Ofer Aharony for a clarification about this point. Finally, notice that this result holds up to constant terms and that the fluctuation $\zeta$ is expected to be $\mathcal O \Bigl ( \sqrt{\alpha'} M_{\text{KK}}\sim\bigl ( T_c/\lambda_3\bigr )^{1/4}\Bigr )$.}
\begin{subequations}\label{effwsact}
\be
S_{\text{eff}} = - \frac{T_s}{2} \int \hspace{-4pt} \rmd t \rmd x \, \Bigl (\partial_\alpha \zeta \, \partial^\alpha \zeta + \mathcal O(T_c/\lambda_3)\Bigr) \, ,
\ee
where
\be
T_s = T_{s,cl} \, \Bigl ( 1+ \mathcal O \bigl (\sqrt{T_c/\lambda_3} \bigr ) \Bigr ) \, , \quad T_{s,cl} = \frac{1}{8\pi} \, \sqrt{\lambda_3/T_c} \, ,
\ee
\end{subequations}
denotes respectively the effective string tension in units of $M_{\text{KK}}^2$ and its classical value.\footnote{The latter corresponds to the flux tube tension studied in \cite{Callan:1998iq, Callan:1999zf} in an holographic framework, when just one quark is pulled out from a color-singlet.} Moreover, we have $\alpha=t,x$. Then, the correlator in \eqref{avarageprofile} reduces to
\be\label{afterint}
\frac{1}{2g_3^2} \left\langle \Tr F^2 (x,z) \right \rangle \approx  \frac{1}{Z_{\text{eff}}} \hspace{-1pt}\int \hspace{-2pt} \mathcal D \zeta \, \, e^{i \pt S_{\text{eff}}[\zeta]}\, \mathcal P(x,z-\zeta(t,x);\mu) \, ,
\ee
where we defined
\be
Z_{\text{eff}} = \hspace{-4pt} \int \hspace{-2pt} \mathcal D \zeta \, \, e^{i \pt S_{\text{eff}}[\zeta]} \, .
\ee
Moreover, $\mathcal P$ is some function of the boundary coordinates alone. Let us stress that the dependence on $z-\zeta$ comes from the phase factor in \eqref{flucPhi}. Furthermore, the massive fluctuations have been traded with the parametric dependence on the world-sheet mass-scale $\mu$.

We can now perform some useful manipulations. Inspired by \cite{Aharony:2024ctf}, we can massage the above correlator evaluated at the middle of the flux-tube and at $t=0$ as
\be\label{massaging}
\begin{split}
\frac{1}{2g_3^2} \left\langle \Tr F^2 (0,z) \right \rangle 
&\approx \frac{1}{Z_{\text{eff}}} \int \hspace{-4pt} \rmd z' \hspace{-4pt} \int \hspace{-2pt} \mathcal D \zeta \, \, e^{i \pt S_{\text{eff}}[\zeta]}\, \mathcal P(0,z-z';\mu) \,\delta(z' - \zeta(0,0))\\
&= \hspace{-2pt} \int \hspace{-4pt} \rmd z' \, \mathcal P(0,z-z';\mu) \,\frac{1}{Z_{\text{eff}}} \int \hspace{-2pt} \mathcal D \zeta \, \, e^{i \pt S_{\text{eff}}[\zeta]}\, \delta(z' - \zeta(0,0)) \\
& = \hspace{-2pt} \int \hspace{-4pt} \rmd z' \hspace{-2pt} \int \hspace{-4pt} \frac{\rmd p}{2\pi} \, e^{i \pt p \pt z'} \, \mathcal P(0,z-z';\mu) \,\frac{1}{Z_{\text{eff}}} \int \hspace{-2pt} \mathcal D \zeta \, \, e^{i \pt S_{\text{eff}}[\zeta] -i \pt p \pt \zeta(0,0)}\, .
\end{split}
\ee
The last line can be computed through the well-known Gaussian path integral for a $d$-dimensional scalar QFT in the presence of a source (see appendix \ref{app:scalarQFT}). Moreover, the profile $\mathcal P$ can be expanded in powers of the dimensionless parameter 
\be\label{smallparam}
\mu^2/T_{s,cl} \in O(\sqrt{T_c/\lambda_3}) \, .
\ee
For consistency, the leading order contribution must share the very same functional structure of the LO term in \eqref{completeprofile}, which captures the classical shape of the flux tube in the strong coupling limit. As we will see, this guarantees to have the right classical limit. Nevertheless, it is clear how the $\mathcal O((T_c/\lambda_3)^{3/2})$ higher derivative terms discussed in section \ref{sec:classprof} are overshadowed by the quantum corrections introduced here. Therefore, we just keep the leading term shared by both the expansions and discussed in section \ref{sec:largeLprofile} for large inter-quark separation.

In the large $L$ limit, we thus get\footnote{The order of magnitude of the corrections comes from combining the subleading terms in the scalar path integral and the overall $\sqrt{\Lambda}\sim\sqrt{T_{s,cl}}$ factor.}
\be
\begin{split}
\frac{\left\langle \Tr F^2 (0,z) \right \rangle}{2g_3^2 \pt T_c^3 \sqrt{\lambda_3/T_c}}
&\approx  \frac{\pi^2}{2} \int \hspace{-4pt} \rmd z'  \, v_m(L) \, Q(z-z',L) \hspace{-2pt} \int \hspace{-4pt} \frac{\rmd p}{2\pi} \, \, e^{i \pt p \pt z'} e^{-\frac{1}{2} p^2 / \Lambda } \[ 1 + \mathcal O \bigl((T_c/\lambda_3)^{1/2}\bigr)\]\\
& = \frac{\pi^2}{2}\sqrt{\frac{\Lambda}{2 \pi}} \, \hspace{-2pt} \int \hspace{-4pt} \rmd z' \, e^{-\frac{1}{2} \Lambda \pt z'^2} \, v_m(L) \, Q(z-z',L)   + \mathcal O \bigl((T_c/\lambda_3)^{1/4}\bigr) \, ,
\end{split}
\ee
where $v_m$ and $Q$ can be found in \eqref{vmforfit} and \eqref{profQ} respectively. Let us stress that, at this level, we gave up on the finite coupling corrections coming from higher derivative terms in the supergravity action. To the contrary, we kept the effect produced by the finiteness of the inter-quark distance $L$. Moreover, we introduced
\be
\Lambda^{-1} = i \pt G(0,0) / T_{s,cl} \, ,
\ee
where $G$ is the scalar propagator defined in \eqref{scalarprop}. The computation of $G(0,0)$ has been performed in full details in \cite{Aharony:2024ctf}, paying close attention to the cancellation of divergences for an infinite/compact confining flux tube in massive $\text{QED}_3$. It turns out that the finite part grows like\footnote{Here, we are assuming that divergences cancel out as in \cite{Aharony:2024ctf}}
\be
i \pt G(0,0) \approx \log \(L \pt M_{\text{KK}}\)\hspace{-2pt}/2\pi \, , \quad L \to \infty \, ,
\ee
where we chose $M_{\text{KK}}=2\pi T_c$ as UV cut-off scale. In the same limit, $Q$ reduces to \eqref{P}.

\begin{figure}[t]
    \centering
    \begin{subfigure}{0.495\textwidth}
       \centering
        \includegraphics[width=\linewidth]{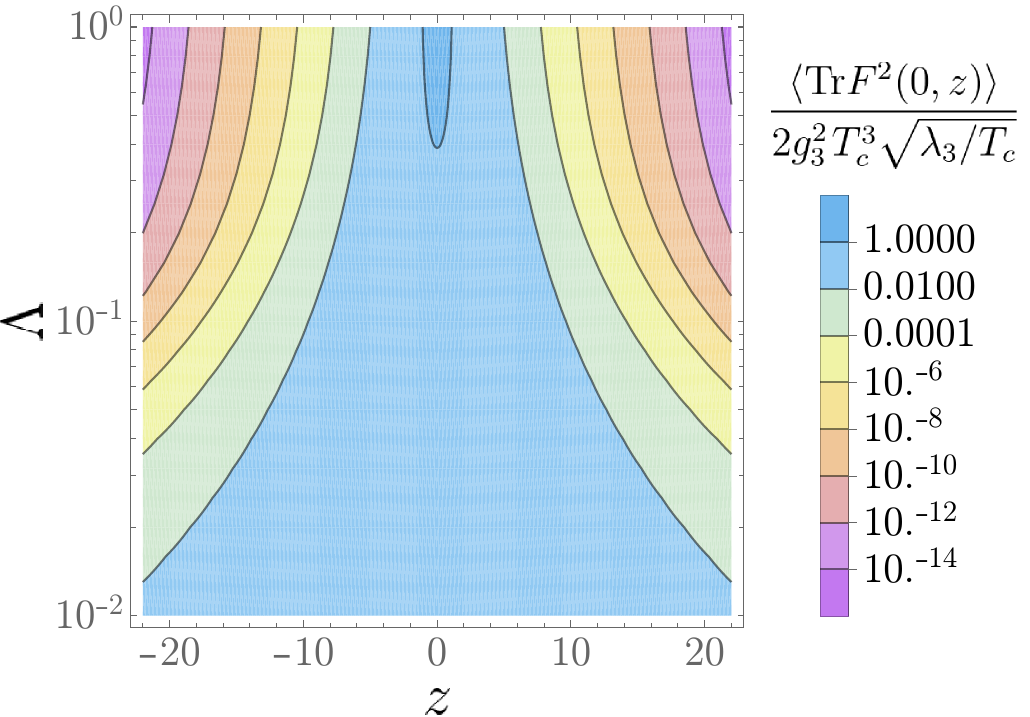} 
        \caption{}
        \label{}
    \end{subfigure}
    \begin{subfigure}{0.495\textwidth}
       \centering
        \includegraphics[width=\linewidth]{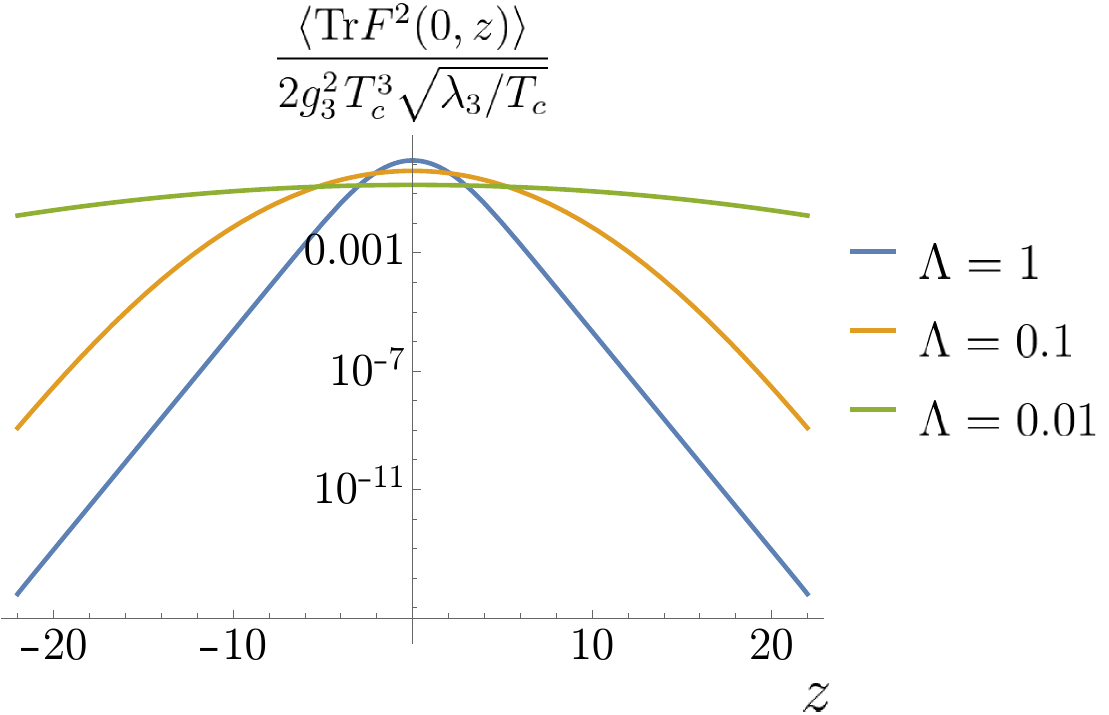} 
        \caption{}
        \label{}
    \end{subfigure}
    \caption{(a) Plot of the quantum flux tube profile proposed in \eqref{finalquantprof}, as the distance from the inter-quark axis $z$ and the parameter $\Lambda$ vary. (b) Same subject of the plot on the left. Here, we report different profiles at fixed $\Lambda$ as a function of $z$.}
    \label{plotqnt}
\end{figure}
All in all, at leading order in the strong coupling expansion in our confining model, we propose that the profile of the quantum flux tube for large inter-quark separation $L$ reads
\begin{subequations}\label{finalquantprof}
\be
\frac{\left\langle \Tr F^2 (0,z) \right \rangle}{2g_3^2 T_c^3 \sqrt{\lambda_3/T_c}} \approx \frac{\pi^2}{2} \sqrt{\frac{\Lambda}{2\pi}} \, \hspace{-2pt} \int \hspace{-4pt} \rmd z' \, e^{-\frac{1}{2} \Lambda \, z'^2} \, P(z-z')   + \mathcal O \bigl((T_c/\lambda_3)^{1/4}\bigr) \, , \quad L \to \infty \, ,
\ee
with
\be\label{largeLLambda}
\Lambda^{-1} \approx \frac{\log \(L M_{\text{KK}}\)}{2\pi T_{s,cl}} \, , \quad L\to\infty.
\ee
\end{subequations}
It is the convolution of the classical fixed-width profile $P$ presented in \eqref{P} with an $L$-dependent Gaussian wave-function. In figure \ref{plotqnt}, the reader can find the plot of the above profile as a function of both $z$ and $\Lambda$. The structure of this result resembles exactly what has been computed for confining strings in massive $\text{QED}_3$ (see \cite{Aharony:2024ctf}), with $P$ mimicking the classical solution for the electric field. Let us stress that, if $L$ is comparable to the inverse UV mass scale $M_{\text{KK}}$ and so to the intrinsic width of the flux tube,\footnote{In section \ref{sec:physint}, we identified the intrinsic width of the flux tube with the inverse mass of the lowest lying glueball in the spectrum, and so with something of order $M_{\text{KK}}^{-1}$ (see table \ref{tablemasses}).} then the Gaussian wave function localizes the integral around $z'=0$. A saddle point integration thus reduces the above profile to its classical component $P$, which decays exponentially as the distance from the inter-quark axis increases and displays a $L$-independent squared width (see the previous section).

\begin{figure}[t]
    \centering
        \includegraphics[width=0.6\textwidth]{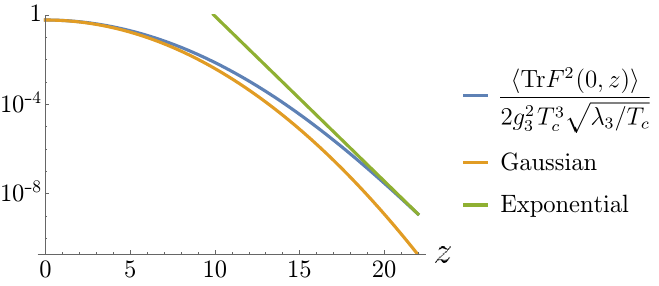} 
    \caption{Plot of the profile in \eqref{finalquantprof} as the distance from the inter-quark distance $z$ varies, at fixed $\Lambda=0.1$ (blue line). It clearly interpolates between a Gaussian shape at small $z$ (yellow line) and an exponential tail at large $z$ (green line). The above is completely analogous to figure 1 in \cite{Aharony:2024ctf}.}
    \label{figqntcomparison}
\end{figure}
We conclude that, for inter-quark separations similar to the intrinsic width of the flux tube, the above profile features the Abrikosov-like behavior 
discussed in the introduction. On the other hand, if the inter-quark separation is much greater than the intrinsic width, 
the flux tube behaves like a real quantum string having an increasing variance with $L$, but still exponentially suppressed at large distances from the inter-quark axis.\footnote{This behavior is inherited from $P$ and follows from a saddle-point analysis of the above integral far away from the inter-quark axis. See \cite{Aharony:2024ctf} for an analogous discussion.} The explicit computation of its squared width (see appendix \ref{app:quantborad}) leads to
\be\label{quantbroad}
w^2 \approx \Lambda^{-1} \, , \quad L\to\infty \, ,
\ee
reproducing the well-known result \eqref{quantumbroad} from EST. Notice that, using a classical profile at large but finite $L$, the width is expected to gain some $\mathcal O \bigl ( e^{-2\pt\pi \pt T_c \pt L}\bigr )$ corrections (cf.~\eqref{classbroad}). Our proposal thus provides an analytical solution to the problem of bridging the Abrikosov and the EST descriptions of the flux-tube in the confining model at hand. 
In figure \ref{figqntcomparison}, the transition from a Gaussian-like shape --- related to the quantum fluctuations --- and the exponential behavior far away from the inter-quark axis is manifest.


\section{Conclusions}

In this paper, we computed the classical flux tube profile in a strongly-coupled large $N$ three-dimensional $SU(N)$ gauge theory at finite inter-quark distance from Holography, including also the first subleading correction in the strong coupling expansion. In a few words, a non-trivial profile at the boundary theory is sourced by a fundamental string diving into the dual bulk geometry. The final result is presented in full generality in section \ref{sec:classprof}, while section \ref{sec:setup} provides all the technical details about its computation. Let us stress that our prediction 
is non-analytical in the ‘t Hooft coupling, as it is clear from \eqref{completeprofile}. Therefore, it would be hardly achievable from a pure boundary perspective. Furthermore, the formal derivation of all its features can be found in section \ref{sec:physint} and section \ref{sec:largeLprofile}. Here, we avoid to report complicated formulae. Rather, we wrap up some conclusions relying on their graphical realization. In this way, we aim to discuss the main findings of this paper without getting lost in technical aspects.

The pages following these conclusions are devoted to several plots of our holographic proposal, as a function of the boundary spatial coordinates. We collect them in table \ref{plotfullprofile} and table \ref{logplotfullprofile}. Each plot refers to a specific choice of inter-quark distance and 't Hooft coupling. Graphically, it is clear that the expectation value of the YM Lagrangian density is localized within a tube-like subregion of the space. Our result includes both the Coulomb-like behavior at the location of the color sources and the longitudinal fairly-constant contribution discussed in \cite{Baker:2018mhw, Baker:2019gsi, Baker:2022cwb, Baker:2023dnn, Baker:2024peg, Baker:2024rjq}.\footnote{The former has been analyzed in appendix \ref{app:coulomb}.} There, through Monte Carlo simulations, the authors found out the spatial distribution of all the color field components sourced by a static pair in a $3+1$-dimensional $SU(3)$ gauge theory (with or without dynamical fermions). Here, we derived a very similar behavior in the $2+1$-dimensional case. Moreover, from the log plots in table \ref{logplotfullprofile}, the exponential behavior along the transverse direction to the inter-quark axis predicted in \eqref{NLOasymprof} is now manifest. In section \ref{sec:physint}, we proposed an analytical method at NLO in the strong coupling expansion for identifying the decay length scale --- namely, the intrinsic width of the flux tube --- with the (inverse) mass of the lowest-lying glueball in the spectrum. This is an analytical prediction that may be corroborated through large $N$ lattice simulations. We defer the comparison to an upcoming related paper \cite{toappear}.


En passant, in this paper we also present other interesting findings. To begin with, in section \ref{sec:physint}, we derive a semi-analytical formula for the first correction to the scalar glueball masses in the strong coupling expansion. We report it in formula \eqref{deltak}, to be equipped with the results in \eqref{finalintegrals}. By ``semi-analytical'' we mean that our prediction relies on the knowledge of the leading-order values of the masses, which are derived numerically (see table \ref{tablemasses}). We believe that the agreement with the literature shown in table \ref{tabledeltak} represents a strong check of our formalism. Notice that, with our formula, we can easily compute the corrections for the higher excited states that are not present elsewhere.

Then, in section \ref{sec:largeLprofile}, we discuss what happens in the large inter-quark separation limit, at leading order in the finite coupling corrections. We provide some analytical results that include the first correction at large $L$. The latter turns out to be exponentially suppressed, but still significant if $L$ is not too large with respect to a multiple of the critical temperature of the model.

Finally, in section \ref{sec:quantfluc}, we try to promote our formulae to the quantum world. In \cite{Cardoso:2013lla}, the authors took into account the quantum fluctuations by  convolving a typical classical flux tube profile with a Gaussian distribution. By ``typical'' we mean that the classical component decays exponentially as the distance from the inter-quark axis increases and features a fixed width. Here, we derive the very same structure from the gravitational path integral. See \eqref{finalquantprof} for the final version of our proposal. Notably, a similar result has been derived in \cite{Aharony:2024ctf}, for long confining strings in massive $\text{QED}_3$. Intriguingly, the width of the full quantum profile reproduces exactly the logarithmic widening coming from the EST approach to confinement \cite{Luscher:1980iy}. Indeed, it is easy to verify that the combination of formulae \eqref{largeLLambda} and \eqref{quantbroad} gives equation \eqref{quantumbroad} for the three-dimensional case. We conclude that our analytical proposal bridges the gap between the Abrikosov and the EST descriptions of the flux-tube discussed in the introduction.

\newpage

\begin{table}[H]
    \centering
       \begin{tabular}{>{\centering\arraybackslash}m{0.0575\textwidth}| >{\centering\arraybackslash}m{0.425\textwidth}>{\centering\arraybackslash}m{0.425\textwidth}}
            \toprule
            \diagbox{$\ell$}{$\lambda$} & $10$ & $20$ \\
            \midrule
            5   & \includegraphics[width=0.4\textwidth]{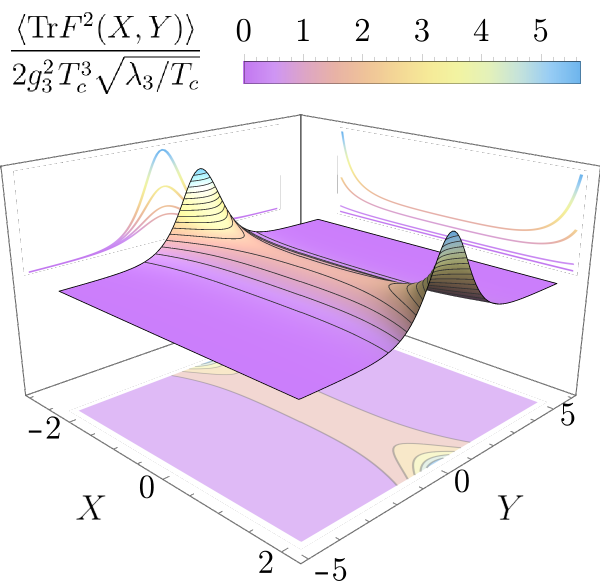} & \includegraphics[width=0.4\textwidth]{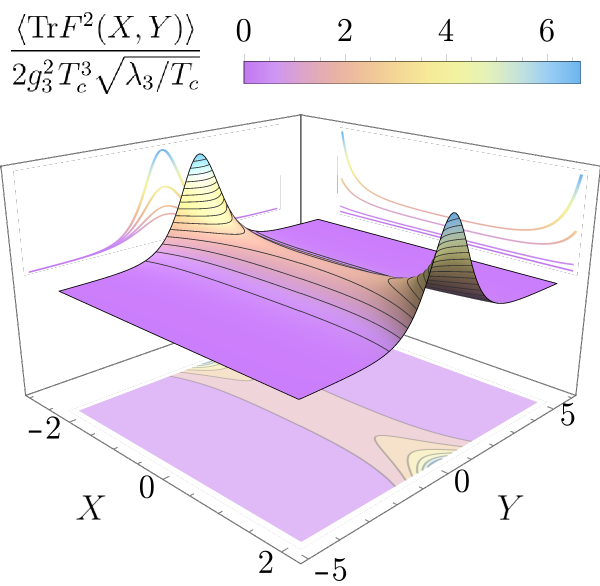} \\
            10 & \includegraphics[width=0.4\textwidth]{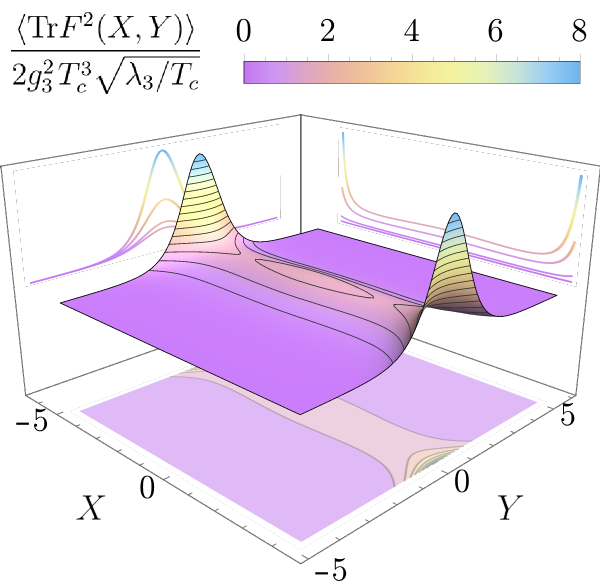} & \includegraphics[width=0.4\textwidth]{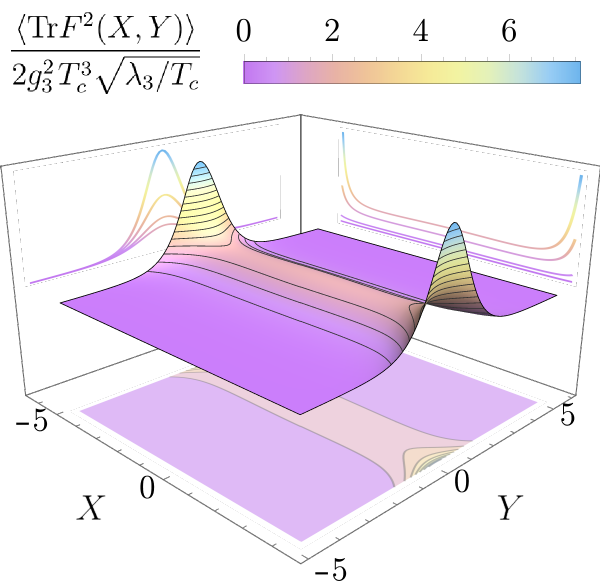} \\ 
            15 & \includegraphics[width=0.4\textwidth]{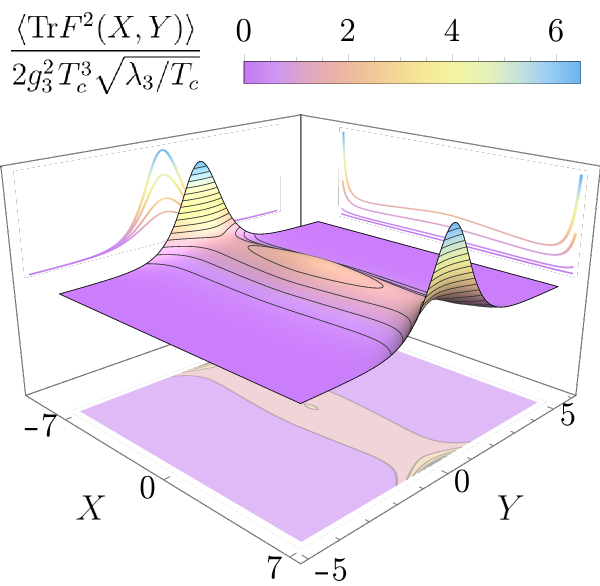} & \includegraphics[width=0.4\textwidth]{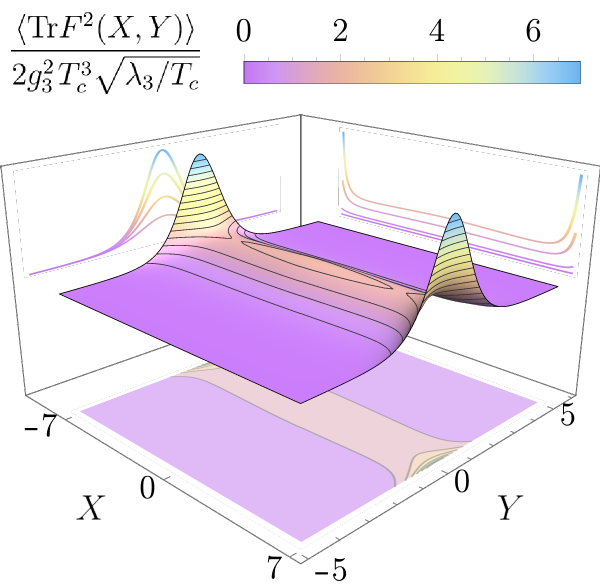} \\ 
            \bottomrule
        \end{tabular}
        \caption{Plots of the NLO profile reported in \eqref{completeprofile} as a function of the spatial boundary coordinates. We denote as $X=2\pi x \pt T_c (1-15 \gamma)$ ($Y=2\pi y \pt T_c (1-15 \gamma)$) the dimensionless coordinate along (transverse to) the inter-quark axis. $Y$ corresponds to the variable $z$ in the main body. We decide to switch notation for the sake of aesthetics. 
Each entry of the table corresponds to a different value of the dimensionless inter-quark distance $\ell=2\pi L \pt T_c  (1-15 \gamma)$ and of the dimensionless `t Hooft coupling $\lambda=\lambda_3/T_c$.}
        \label{plotfullprofile}
\end{table}

\newpage

\begin{table}[H]
    \centering
       \begin{tabular}{>{\centering\arraybackslash}m{0.0575\textwidth}| >{\centering\arraybackslash}m{0.425\textwidth}>{\centering\arraybackslash}m{0.425\textwidth}}
            \toprule
            \diagbox{$\ell$}{$\lambda$} & $10$ & $20$ \\
            \midrule
            5   & \includegraphics[width=0.4\textwidth]{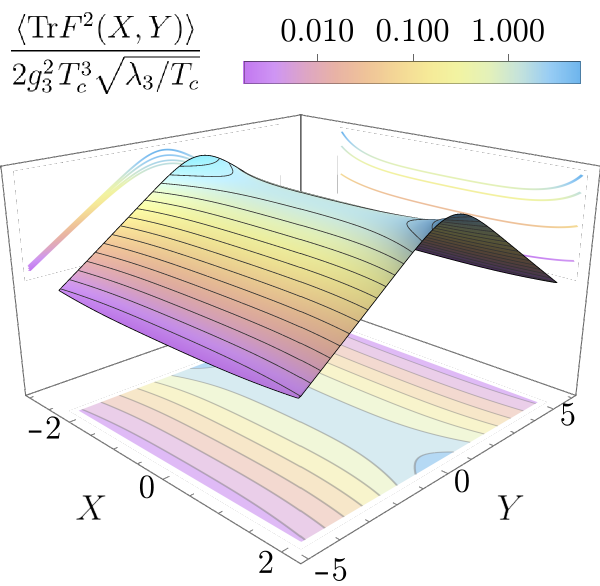} & \includegraphics[width=0.4\textwidth]{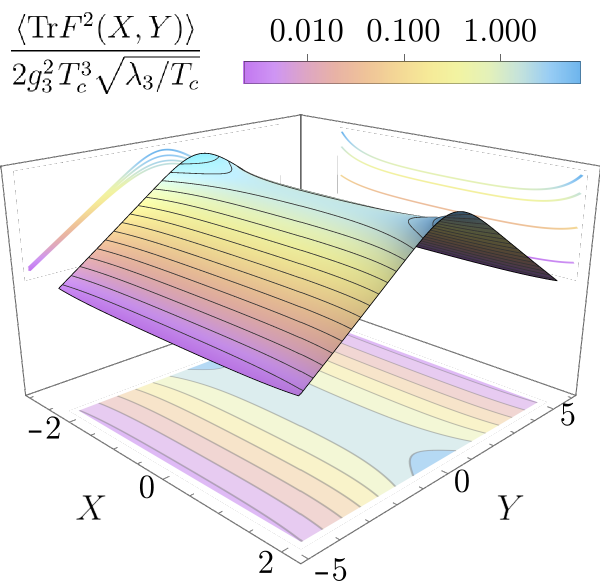} \\
            10 & \includegraphics[width=0.4\textwidth]{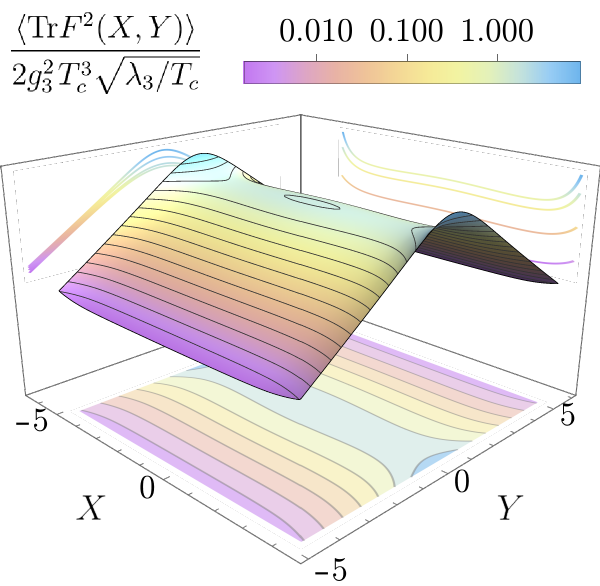} & \includegraphics[width=0.4\textwidth]{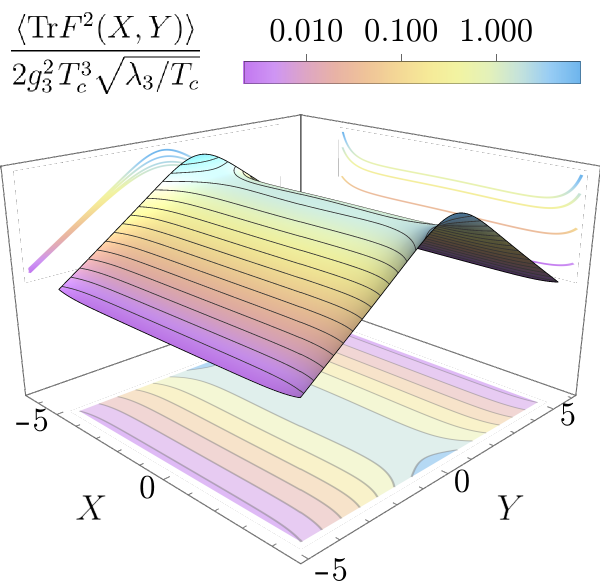} \\ 
            15 & \includegraphics[width=0.4\textwidth]{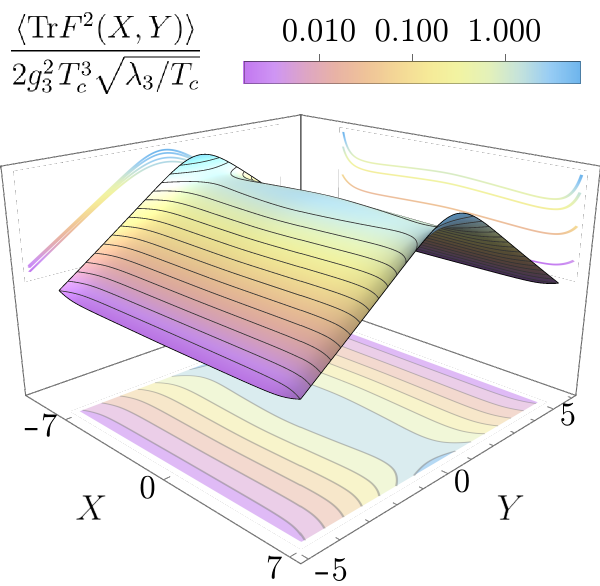} & \includegraphics[width=0.4\textwidth]{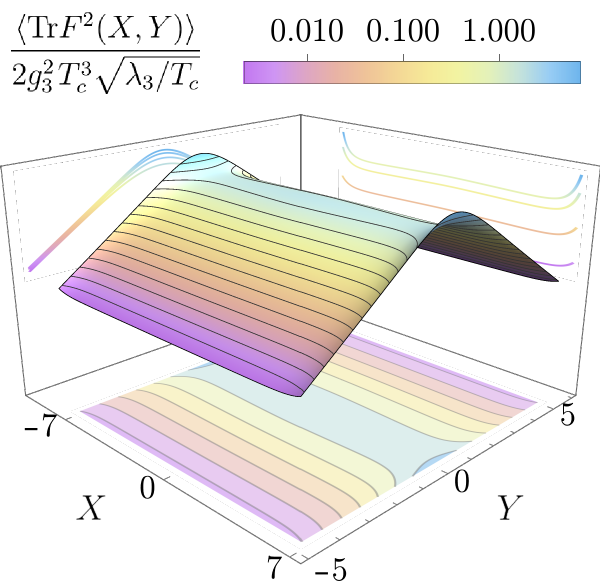} \\ 
            \bottomrule
        \end{tabular}
        \caption{Same subject as table \ref{plotfullprofile}. Here, we report the NLO profile in logarithmic scale.}
        \label{logplotfullprofile}
\end{table}

\newpage

\vskip 15pt \centerline{\bf Acknowledgments} \vskip 10pt 

\noindent 
We are grateful to Ofer Aharony, Francesco Bigazzi, Andrea Bulgarelli, Chiara Cabras, Michele Caselle, Elia Cellini, Matteo Ciardi, Aldo L. Cotrone, Alessandro Mariani, José Matos, Alessandro Nada, Marco Panero, Dario Panfalone and Lorenzo Verzichelli for comments and very helpful discussions. This work was supported by the Simons Foundation grant  994300 (``Simons Collaboration on Confinement and QCD Strings”). 


\appendix
\section{The classical string profile in the large $L$ limit}
\label{app:largeL}

Here, we aim to discuss the behavior of the classical string profile at large inter-quark separation. To this end, let us study equations \eqref{equmdeltam} through the machinery developed in \cite{Kinar:1999xu, Greensite:1999jw, Kinar:1998vq, Greensite:1998bp}.

Notice that the integral in equation \eqref{LOLum} diverges in the small $\varepsilon$ limit. Consistently, from a numerical point of view, it is clear that the minimum value of the profile collapses onto $u_0$ as $L$ increases. E.~g., see figure \ref{fundshape}. Therefore, to begin with, let us delve into the analysis of equation \eqref{LOLum} in the small $\varepsilon$ regime. This step is not just a review of \cite{Kinar:1999xu, Greensite:1999jw, Kinar:1998vq, Greensite:1998bp}, but we will also take into account some missing terms in the final result. Then, we will discuss also equation \eqref{NLOLum} in the same regime, on the tree-level solution provided in \eqref{treedeltas}. The latter analysis is completely new.

Up to $\mathcal O (\varepsilon^2)$, we can write
\be\label{Lumasym}
L = \frac{2 \pt R^2}{u_{m,0}} \int_1^\infty \hspace{-6pt} \rmd \eta \, \frac{\Psi(\eta)}{\displaystyle\sqrt{\bigl ( \eta - 1 + \varepsilon/4 \bigr ) \bigl ( \eta -1 \bigr )}} + \mathcal O(\varepsilon^2)\, , \quad \varepsilon \to 0 \, ,
\ee
where we defined
\be
\Psi(\eta) = \frac{1}{\displaystyle \sqrt{\bigl ( \eta +1 \bigr ) \bigl ( \eta -i \bigr ) \bigl ( \eta +i \bigr ) \bigl ( \eta +1 - \varepsilon/4 \bigr ) \bigl ( \eta -i + i \pt \varepsilon/4 \bigr ) \bigl ( \eta + i - i \pt \varepsilon/4 \bigr )}} \, .
\ee
Making use of
\begin{subequations}
\begin{align}
&\hspace{-12pt}\partial_\eta \[ 2 \, \log \(\sqrt{\eta-1} + \sqrt{\eta - 1 + \varepsilon / 4} \)\] = \frac{1}{\displaystyle\sqrt{\bigl ( \eta - 1 + \varepsilon/4 \bigr ) \bigl ( \eta -1 \bigr )}} \, , \\
&\hspace{-12pt}\frac{\partial_\eta\[ - \sqrt{\eta-1}\sqrt{\eta - 1 + \varepsilon/4}  + \(\eta-1+\varepsilon/8\) 2 \, \log \(\sqrt{\eta-1} + \sqrt{\eta - 1 + \varepsilon/4}\)\]}{2 \, \log \(\sqrt{\eta-1} + \sqrt{\eta - 1 + \varepsilon/4} \)} =  1\, ,
\end{align}
\end{subequations}
we can integrate by parts \eqref{Lumasym} twice getting \eqref{asympLum1}.

Now, we can proceed to the analysis of equation \eqref{NLOLum} at tree level. As before, the numerator can be rephrased as
\be
\int_1^\infty \hspace{-8pt} \rmd \eta \, \frac{\bigl ( \eta^4-1 \bigr ) \delta_{u,1}^{\text{(tree)}}(\eta) - \eta^4 \delta_{h,1}^{\text{(tree)}}(\eta)}{\bigl (\eta^4-1 \bigr )^{3/2} \sqrt{\eta^4-1+\varepsilon} } = \int_1^\infty \hspace{-6pt} \rmd \eta \, \frac{\tilde\Psi(\eta)}{\displaystyle\sqrt{\bigl ( \eta - 1 + \varepsilon/4 \bigr ) \bigl ( \eta -1 \bigr )}} + \mathcal O(\varepsilon^2)\, , \quad \varepsilon \to 0 \, , 
\ee
given
\be
\tilde \Psi(\eta) = \frac{\delta_{u,1}^{\text{(tree)}}(\eta)-\eta^4 \delta_{h,1}^{\text{(tree)}}(\eta)/\bigl (\eta^4 - 1 \bigr )}{\displaystyle \sqrt{\bigl ( \eta +1 \bigr ) \bigl ( \eta -i \bigr ) \bigl ( \eta +i \bigr ) \bigl ( \eta +1 - \varepsilon/4 \bigr ) \bigl ( \eta -i + i \pt \varepsilon/4 \bigr ) \bigl ( \eta + i - i \pt \varepsilon/4 \bigr )}}\, .
\ee
Notice that
\begin{subequations}
\begin{align}
&\delta_{u,1}^{\text{(tree)}}(\eta) = 15 \[ \frac{5}{\eta^4} (1-\varepsilon) + \frac{5}{\eta^8} (1-2\pt\varepsilon) - \frac{19}{\eta^{12}} (1-3\pt\epsilon)\] + \mathcal O (\varepsilon^2) \, ,\\
&\frac{\eta^4 \delta_{h,1}^{\text{(tree)}}}{\eta^4 - 1} = \frac58 \[ 9 \pt (1 - \varepsilon) + \frac{19}{2} \pt (1-2\pt\varepsilon) \pt \frac{\eta^4+1}{\eta^4} + 8 \pt (1-3\pt\varepsilon)\pt \frac{\eta^8+\eta^4+1}{\eta^8} \] + \mathcal O (\varepsilon^2) \, .
\end{align}
\end{subequations}
The same strategy adopted for the leading order equation leads to
\be
\int_1^\infty \hspace{-8pt} \rmd \eta \, \scaleto{\frac{\bigl ( \eta^4-1 \bigr ) \delta_{u,1}^{\text{(tree)}}(\eta) - \eta^4 \delta_{h,1}^{\text{(tree)}}(\eta)}{\bigl (\eta^4-1 \bigr )^{3/2} \sqrt{\eta^4-1+\varepsilon} }}{35pt} = {\textstyle{\frac{335}{8}}} \log {\textstyle{\frac\varepsilon4}} +{\textstyle{\frac{5}{16}}}  \({\textstyle{\frac{26349}{77}}} + 67 \pi - 402 \log 2\) + \mathcal O(\varepsilon \log \varepsilon) \, ,
\ee
as $\varepsilon \to 0$.

The last step is the discussion of the denominator in \eqref{NLOLum}, which can be recast as
\be
\int_1^\infty \hspace{-8pt} \rmd \eta \, \frac{\eta^4 + 1 - \varepsilon}{\bigl (\eta^4 - 1 + \varepsilon \bigr )^{3/2} \sqrt{\eta^4-1}} = \int_1^\infty \hspace{-6pt} \rmd \eta \, \frac{\bar\Psi(\eta)}{\displaystyle \bigl ( \eta - 1 + \varepsilon/4 \bigr )^{3/2}\sqrt{\eta -1}} + \mathcal O(\varepsilon^2)\, , \quad \varepsilon \to 0 \, , 
\ee
where we introduced
\be
\bar \Psi(\eta) = \frac{\eta^4 + 1 -\varepsilon}{\displaystyle \bigl [\bigl ( \eta +1 - \varepsilon/4 \bigr ) \bigl ( \eta -i + i \pt \varepsilon/4 \bigr ) \bigl ( \eta + i - i \pt \varepsilon/4 \bigr )\bigr ]^{3/2} \sqrt{\bigl ( \eta +1 \bigr ) \bigl ( \eta -i \bigr ) \bigl ( \eta +i \bigr )}}\, .
\ee
Since
\begin{subequations}
\begin{align}
&\hspace{-5pt}\scaleto{\partial_\eta \hspace{-2pt}\[ \frac8\varepsilon \sqrt{\frac{\eta-1}{\eta-1+\varepsilon/4}}\] = \frac{1}{\displaystyle\bigl ( \eta - 1 + \varepsilon/4 \bigr )^{3/2}\sqrt{\eta -1}}}{32.5pt} \, ,\\
&\hspace{-5pt}\scaleto{\partial_\eta \hspace{-2pt}\[ \frac8\varepsilon \sqrt{\eta - 1 + \varepsilon/4} \sqrt{\eta -1} - 2 \, \log \bigl (\sqrt{\eta-1} + \sqrt{\eta - 1 + \varepsilon / 4} \bigr ) \] = \frac8\varepsilon \sqrt{\frac{\eta-1}{\eta-1+\varepsilon/4}}}{32.5pt} ,\\
&\hspace{-5pt}\scaleto{\frac{\partial_\eta \hspace{-2pt}\[ \frac{8 \eta - 8 + 3\varepsilon}{2\varepsilon} \sqrt{\eta - 1 + \varepsilon/4} \sqrt{\eta -1} - \(\eta-1+3\varepsilon/16\) 2 \, \log \(\sqrt{\eta-1} + \sqrt{\eta - 1 + \varepsilon / 4} \)\]}{\frac8\varepsilon \sqrt{\bigl ( \eta - 1 + \varepsilon/4 \bigr ) \bigl ( \eta -1 \bigr )} - 2 \, \log \(\sqrt{\eta-1} + \sqrt{\eta - 1 + \varepsilon / 4} \)} = 1}{42.5pt} ,
\end{align}
\end{subequations}
three integrations by parts give
\be
\int_1^\infty \hspace{-8pt} \rmd \eta \, \frac{\eta^4 + 1 - \varepsilon}{\bigl (\eta^4 - 1 + \varepsilon \bigr )^{3/2} \sqrt{\eta^4-1}} = \frac1\varepsilon \biggl(1 + \frac{\varepsilon}{8} \log\frac{\varepsilon}{4} + \mathcal O (\varepsilon)\biggr) \, , \quad \varepsilon \to 0 \, .
\ee
All in all, \eqref{asympLum2} follows as well.

\section{About the homogeneous dilaton equation of motion}
\label{app:homoeom}

Let us consider the homogeneous equation associated to \eqref{LOeom}, that is
\be \label{homoHeun}
\partial_v^2 \Phi + \(\frac1v + \frac{1}{v-1} + \frac{1}{v+1}\) \partial_v \Phi - \frac{k^2}{v(v^2-1)} \Phi = 0 \, .
\ee
In this appendix, we aim to find an analytic and normalizable expression for the solution to the above equation, expanded about $v=+\infty$ and $v=+1$.

First, at very large $v$, the above equation reduces to
\be
\partial_v^2 \Phi + \frac3v \, \partial_v \Phi - \frac{k^2}{v^3} \, \Phi = 0 \, ,
\ee
which is solved by
\be
\Phi(v,k) = \frac{a_1(k)}{v} I_2\(2,\frac{2 k}{\sqrt{v}}\) + \frac{a_2(k)}{v} K_2\(2,\frac{2 k}{\sqrt{v}}\) \, ,
\ee
for some functions $a_1,a_2$ of $k$. Here, $I_2$ and $K_2$ are modified Bessel functions. The normalizability condition enforces
\be
a_2(k)=0
\ee
and so the solution to equation \eqref{homoHeun} can be expanded about $v=+\infty$ as
\be
\Phi(v,k) = \sum_{n=-1}^\infty \frac{a(n,k)}{v^{n+2}} \, , \quad a(-1, k)=0 \, , \quad a(0, k)=+1 \, .
\ee
Plugging back this expression into \eqref{homoHeun}, we find the recurrence relation reported in \eqref{reca}.

Similarly, we can evaluate equation \eqref{homoHeun} around $v=1$, getting
\be
\partial_v^2 \Phi + \frac{1}{v-1} \partial_v \Phi - \frac{k^2}{2(v-1)} \Phi = 0 \, .
\ee
Its solution can be written as
\be
\Phi(v,k) = \tilde a_1(k) \, I_0\(\sqrt{2\, k^2(v-1)}\) + \tilde a_2(k) \, K_0\(\sqrt{2 \, k^2(v-1)}\) \, ,
\ee
for some functions $\tilde a_1$, $\tilde a_2$ of $k$. Again, $I_0$ and $K_0$ are modified Bessel functions. Crucially, regularity at $v=1$ selects
\be
\tilde a _2(k)=0 \, .
\ee
As a consequence, the solution to equation \eqref{homoHeun} takes a Taylor series form like
\be
\Phi(v,k) = \sum_{n=-1}^\infty b(n,k) (v-1)^{n} \, , \quad b(-1, k)=0 \, , \quad b(0, k)=+1 \, .
\ee
Again, inserting this expansion in equation \eqref{homoHeun}, we get to the recurrence relation in \eqref{recb}.

Notice that, for $k=0$, the coefficients $a(n,0)$ and $b(n,0)$ can be expressed in a closed form. Indeed, the recurrence relations in \eqref{recrel} reduces to
\begin{subequations}\label{recrel0}
\begin{align}
& a(n+1,0) = \frac{n+1}{n+3} a(n-1,0) \, , \\
& b(n+1,0) = -\frac{1}{2(n+1)^2} \Bigr [ b(n-1,0) (n^2-1)+ 3n(n+1) b(n,0) \Bigl ] \, .
\end{align}
\end{subequations}
with
\begin{subequations}
\begin{align}
& a(-1, 0) = 0 \, , \quad \hspace{-1pt}a(0,0) = 1 \, , \\
& b(-1, 0) = 0 \, , \quad b(0,0) = 1 \,
\end{align}
\end{subequations}
The latter are solved by
\begin{subequations}
\begin{align}
& a(2m-1, 0) = 0 \, , \quad a(2m,0) = \frac{1}{m+1} \, , \quad \forall m \ge 0 \, , \\
&  b(-1, 0) = 0 \, , \quad b(0,0) = 1 \, , \quad b(n, 0) = 0 \, , \quad \forall n \ge 1 \, .
\end{align}
\end{subequations}
We conclude that
\be
\sum_{n=0}^\infty a(n,0) / v^{n+2} = 2 \log v - \log (v^2 - 1) \, , \quad \sum_{n=0}^\infty b(n,0) (v-1)^{n} = 1 \, .
\ee
It thus follows that the Wronskian defined in \eqref{Wronskian} reduces to
\be
W(v,0) = \frac{2}{v(v^2-1)} \, ,
\ee
from which
\be\label{watzero}
w(0)=2 \, .
\ee
This value turns out to be useful both in the main body and in appendix \ref{app:broad}.

The same equations \eqref{recrel0} holds for $\partial_k a(n+1,0)$ and $\partial_k b(n+1,0)$. What changes are the initial values, namely
\begin{subequations}
\begin{align}
&\partial_k a(-1,0) = 0 \, , \quad \hspace{-1pt}\partial_k a(0,0) = 0 \, , \\
&\partial_k b(-1,0) = 0 \, , \quad \partial_k b(0,0) = 0 \, .
\end{align}
\end{subequations}
Therefore we conclude that
\be
\partial_k a(n,0) = 0 \, , \quad \partial_k b(n,0) \, , \quad \forall n \, ,
\ee
from which
\be\label{wprimeatzero}
w'(0)=0 \, .
\ee
Again, this is something helpful for appendix \ref{app:broad}.


\section{Details on the NLO corrections}
\label{app:deltaJ}

In this appendix we collect some technical aspects related to the correction $\Delta_J$ defined in \eqref{deltaJ}. As already noted in the main body of the paper, the latter does not depend on the holographic coordinate. Nevertheless, it features a parametric dependence on the radial position of the stringy source due to the gluing boundary conditions.

To begin with, let us start by defining
\begin{subequations}\label{startingW1W2}
\begin{align}
&W_1(v,k; \bar v_c) = h_2(v,k) \, \partial_v P_1(v,k; \bar v_c) - P_1(v,k; \bar v_c) \, \partial_v h_2(v,k) \, , \\
&W_2(v,k; \bar v_c) = h_2(v,k) \, \partial_v P_2(v,k; \bar v_c) - P_2(v,k; \bar v_c) \, \partial_v h_2(v,k) \, .
\end{align}
\end{subequations}
Remember that $h_1$ and $h_2$ are defined in \eqref{Heundilaton} and solve the homogeneous equation associated to the leading order problem \eqref{LOeom}. On the other hand, $P_1$ and $P_2$ are defined in \eqref{ansatzP} as solutions to equation \eqref{eomP}. It thus follows that
\be
v(v^2-1) \partial_v W_n(v,k; \bar v_c) = - (3v^2-1) W_n(v,k; \bar v_c) + h_2(v,k) \, \mathcal K\bigl [\Phi_{\text{LO}} (v,k; \bar v_c)\bigr ] \, ,
\ee
for $n=1,2$, which is solved by
\be\label{w1w2def}
v(v^2-1) W_n (v,k; \bar v_c) = w_n(b,k; \bar v_c) + \int_b^v \hspace{-8pt}\rmd v' \, h_2(v',k) \, \mathcal K\bigl [\Phi_{\text{LO}} (v',k; \bar v_c)\bigr ]\, , \quad n=1,2 \, .
\ee
Here, $w_1$ and $w_2$ are defined as $v$-independent constant of integration. In the above parameterization, we choose them such that $w_n(b,k; \bar v_c) = b(b^2-1) W_n (b,k; \bar v_c)$ for some $b\in(1,2)$.

Then, $\Delta_J$ can be rephrased as
\be
\Delta_J(k; \bar v_c) = v(v^2-1) \left . \Bigl (W_1(v,k; \bar v_c) - W_2(v,k; \bar v_c)\Bigr ) \right |_{v=v^*} \, ,
\ee
that is
\be\label{deltaJw1w2}
\Delta_J(k; \bar v_c) = w_1(b,k; \bar v_c)-w_2(b,k; \bar v_c)\, .
\ee
The dependence on $v$ is easily canceled out, as it should. For consistency, this result must be also $b$-independent. Let us show it explicitly.

Collecting all the results of section \ref{sec:NLOdil}, we can express $P_1$ and $P_2$ as
\begin{subequations}\label{finalP1P2}
\begin{empheq}[left=\empheqlbrace]{align}
&\scaleto{\displaystyle P_1(v,k; \bar v_c) = - h_1(v,k) \hspace{-3pt}\int_{\bar v_c}^v \hspace{-7pt} \frac{\rmd v'}{(h_1(v',k))^2 v' \pt (v'^2-1)} \hspace{-1pt} \int_{v'}^{\infty} \hspace{-10pt} \rmd v'' \pt h_1(v'',k) \pt \mathcal K\bigl [\Phi_{\text{LO}}(v'',k;\bar v_c)\bigr ] , }{27.5pt}\\[0.5ex]
&\scaleto{\displaystyle P_2(v,k; \bar v_c) = h_2(v,k) \hspace{-3pt} \int_1^v \hspace{-8pt} \frac{\rmd v'}{(h_2(v',k))^2 v' \pt (v'^2-1)} \hspace{-1pt}\int_1^{v'} \hspace{-10pt} \rmd v'' \pt h_2(v'',k) \pt \mathcal K\bigl [\Phi_{\text{LO}}(v'',k; \bar v_c)\bigr ] .}{27.5pt}
\end{empheq}
\end{subequations}
On these solutions, it easily follows that
\begin{subequations}
\begin{empheq}[left=\empheqlbrace]{align}
&\scaleto{\displaystyle v(v^2-1) W_1(v,k; \bar v_c) = - \frac{h_2(v,k)}{h_1(v,k)} \int_v^{\infty} \hspace{-10pt} \rmd v' \pt h_1(v',k) \pt \mathcal K\bigl [\Phi_{\text{LO}}(v',k; \bar v_c)\bigr ] - \frac{w(k)P_1(v,k; \bar v_c)}{h_1(v,k)},}{26pt} \\[0.5ex]
&\scaleto{\displaystyle v(v^2-1) W_2(v,k; \bar v_c) = \int_1^v \hspace{-8pt} \rmd v' \pt h_2(v',k) \pt \mathcal K\bigl [\Phi_{\text{LO}}(v',k; \bar v_c)\bigr ] .}{26pt}
\end{empheq}
\end{subequations}
Then, from a comparison with \eqref{w1w2def}, we get
\begin{subequations}\label{finalw1w2}
\begin{empheq}[left=\empheqlbrace]{align}
&\displaystyle w_1(b,k;\bar v_c) = - \frac{h_2(v,k)}{h_1(v,k)} \int_v^{\infty} \hspace{-10pt} \rmd v' \pt h_1(v',k) \pt \mathcal K\bigl [\Phi_{\text{LO}}(v',k;\bar v_c)\bigr ] - \frac{w(k)P_1(v,k;\bar v_c)}{h_1(v,k)} +\\
&\qquad\qquad\qquad\qquad\qquad\qquad\qquad\qquad \qquad- \int_b^v \hspace{-4pt}\rmd v' \, h_2(v',k) \, \mathcal K\bigl [\Phi_{\text{LO}}(v'',k;\bar v_c)\bigr ], \nonumber\\
&\displaystyle w_2(b,k;\bar v_c) = \int_1^b \hspace{-8pt} \rmd v \pt h_2(v,k) \pt \mathcal K\bigl [\Phi_{\text{LO}}(v,k;\bar v_c)\bigr ] .
\end{empheq}
\end{subequations}
Remember that $w$ has been defined in \eqref{W}. The reader can check that the expression for $w_1$ above is actually $v$-independent, by taking the derivative with respect to $v$. Therefore, without loss of generality, we can evaluate it on $\bar v_c$. This choice is convenient, since, by definition, $P_1(\bar v_c,k; \bar v_c)=0$. All in all, the above results, along with the definition of $\Phi_{\text{LO}}$ in \eqref{soldilLO}, lead to formula \eqref{finaldeltaJ}. As it should, the dependence on $b$ cancels out.

Finally, the integrals in \eqref{finaldeltaJ} can be computed explicitly making use of the series expansions of $h_1$ and $h_2$ in \eqref{Heundilaton}. The computation is a bit tedious, but straightforward. The final result is
\begin{subequations}\label{finalintegrals}
\begin{align}
&\int_{\bar v_c}^{\infty} \hspace{-8pt} \rmd v \, h_1(v,k) \pt \mathcal J\bigl [h_1(v,k)\bigr ] = 15 \hspace{-2pt}\sum_{m,\pt n=0}^\infty \hspace{-3pt} J_1(m,n,k,\bar v_c) \,\frac{1}{\bar v_c^{m+n+4}} \, , \\
&\int_1^{\bar v_c} \hspace{-8pt} \rmd v \, h_2(v,k) \pt \mathcal J\bigl [h_2(v,k)\bigr ] = \hspace{-2pt}\sum_{m,\pt n, \pt \ell=0}^\infty \hspace{-3pt} J_2(m,n,\ell,k,\bar v_c) \, (\bar v_c - 1)^{m+n+\ell+1} \, ,
\end{align}
where
\begin{align}
&\frac{J_1(m,n,k,\bar v_c)}{a(m,k)a(n,k)} = A_1(m+n,\bar v_c)(n+2)+B_1(m+n,\bar v_c)\frac{k^2}{\bar v_c} + C_1(m+n,\bar v_c) \, , \\[1ex]
&\frac{J_2(m,n,\ell,k,\bar v_c)}{b(m,k)b(n,k)} = \frac{k^2 A_2(\ell) - B_2(\ell) - 2 \pt m \pt C_2(\ell) }{m+n+\ell+1} - \frac{m\pt C_2(\ell)(\bar v_c - 1)}{m+n+\ell+2}  \, ,
\end{align}
and
\begin{align}
&A_1(\ell,\bar v_c) = \frac{10}{\ell+4}+\frac{10}{(\ell+6)\pt\bar v_c^2}-\frac{86}{(\ell+8)\pt\bar v_c^4} + \frac{66}{(\ell+10)\pt\bar v_c^6} \, , \\
&B_1(\ell,\bar v_c) = \frac{5}{\ell+5}+\frac{5}{(\ell+7)\pt\bar v_c^2}-\frac{19}{(\ell+9)\pt\bar v_c^4} \, , \\
&C_1(\ell,\bar v_c) =  - \frac{27}{4\pt(\ell+10)\pt\bar v_c^6} \, ,\\
&A_2(\ell,\bar v_c) = \frac18 (-1)^{\ell+1} (\ell+1)(19\pt\ell^4+266\pt\ell^3+1249\pt\ell^2+2426\pt\ell+1080) \, , \\
&B_2(\ell,\bar v_c) = \frac{9}{64} (-1)^\ell (\ell+1)(\ell+2)(\ell+3)(\ell+4)(\ell+5)(\ell+6) \, , \\
&C_2(\ell,\bar v_c) = \frac14 (-1)^{\ell+1} (\ell+1) (33\pt\ell^4+462\pt\ell^3+2143\pt\ell^2+4082\pt\ell+2160) \, .
\end{align}
\end{subequations}

\section{The width of the flux tube}
\label{app:broad}

Given a profile $\mathcal P$, we define its squared width as
\be\label{generalwidth}
w^2 = \frac{\int^{+\infty}_{-\infty} \hspace{-4pt} \rmd z \, z^2 \,  \mathcal P(z)}{\int^{+\infty}_{-\infty} \hspace{-4pt} \rmd z  \, \mathcal P(z)} \, .
\ee
Let us apply this formula to the profiles appearing in the main body

\subsection{Infinite inter-quark separation}
\label{app:widthP}

The classical profile of the flux tube established between infinitely distant charges is given in formula \eqref{limitprof}, at leading order in the strong coupling expansion. Then, its squared width is reduced to
\be
w^2_P = \int^{+\infty}_{-\infty} \hspace{-4pt} \rmd z \, z^2 \, P(z) \, ,
\ee
where $P$ is defined in \eqref{P}. Therefore, we have
\be
w^2_P = - 2 \int^{+\infty}_{-\infty} \hspace{-8pt} \rmd k \, \frac{1}{w(k)} \, \partial^2_k \pt \delta(k) = - \left . \partial^2_k \(\frac{2}{w(k)}\) \right |_{k=0} \, .
\ee
Using the results in \eqref{watzero} and \eqref{wprimeatzero}, we finally reproduce equation \eqref{widthP}.

\subsection{Classical broadening}
\label{app:classbroad}
Here, we focus on the classical profile \eqref{proflargeL} at leading order in the strong coupling expansion, which includes the first $L$-dependent correction in the large limit. Since $Q$ defined in \eqref{profQ} is normalized to one, its mean squared width reads
\be
w^2_Q = \int^{+\infty}_{-\infty} \hspace{-8pt} \rmd z \, z^2 \, P(z) - c \, e^{-2\pt\pi T_c \pt L} \hspace{-2pt} \int^{+\infty}_{-\infty} \hspace{-8pt} \rmd z \, z^2 \, \partial^2_z P(z) \, .
\ee
Notice that $P$ goes exponentially to zero as $z$ goes to infinity. And so do its derivatives. Therefore, using the result of the previous section and integrating by parts twice, we get the final result reported in \eqref{classbroad}. 

\subsection{Quantum broadening}
\label{app:quantborad}

In this appendix we want to show how the logarithmic broadening introduced in \eqref{quantumbroad} arises from a direct computation, starting from our proposal for the quantum profile of the flux tube in three dimensions.

So, let us define
\be
w^2 = \frac{\int^{+\infty}_{-\infty} \hspace{-4pt} \rmd z \, z^2 \, \left\langle \text{Tr} F^2(0,z) \right\rangle}{\int^{+\infty}_{-\infty} \hspace{-4pt} \rmd z  \, \left\langle \text{Tr} F^2(0,z) \right\rangle} \, ,
\ee
where, in the large inter-quark distance limit, the above correlation function is given in \eqref{finalquantprof}. Using the integral representation of the Dirac delta and the definition of $P$ in \eqref{P}, it easily follows that
\be
\mathcal N \int^{+\infty}_{-\infty} \hspace{-4pt} \rmd z  \, \left\langle \text{Tr} F^2(0,z) \right\rangle = \frac{2\pi}{w(0)}\sqrt{\frac{2\pi}{\Lambda}} \, ,
\ee
where $\mathcal N = \sqrt{2 \, T_c}/(g^2_3 \, T^3_c \sqrt{\pi \lambda_3 \Lambda})$. Furthermore, well-known results about Gaussian integration lead to
\be
\begin{split}
\mathcal N \int^{+\infty}_{-\infty} \hspace{-8pt} \rmd z  \, z^2 \left\langle \text{Tr} F^2(0,z) \right\rangle
&=  - 2 \pi \, \int^{+\infty}_{-\infty} \hspace{-8pt} \rmd z \, e^{-\Lambda z^2/2} \left . \partial^2_k \(\frac{1}{w(k)} e^{-i \pt k \pt z}\) \right |_{k=0}\\
&= \frac{2\pi}{w(0)} \sqrt{\frac{2\pi}{\Lambda}} \[ \frac{1}{\Lambda} + \frac{w''(0)}{w(0)}  - 2 \(\frac{w'(0)}{w(0)}\)^{\hspace{-1pt}2} \] \, .
\end{split}
\ee
Given the values reported in \eqref{watzero} and \eqref{wprimeatzero}, we conclude that
\be
w^2 = \Lambda^{-1} + \frac12 \pt w''(0) \, .
\ee
In the large $L$ limit, this result reproduces \eqref{quantbroad}. The expected quantum broadening is thus derived (see \eqref{largeLLambda}). Notice that the constant term represents the classical contribution coming from $P$ (cf.~\eqref{widthP}).

\section{Gaussian path integral in scalar QFT}
\label{app:scalarQFT}

The partition function of a (free) scalar QFT in $d$ dimensions coupled to a current $j$ reads
\be
\mathcal Z[\pt j \pt ] 
= \hspace{-2pt} \int \hspace{-3pt} \mathcal D \varphi \, \text{Exp}\[-\frac{i}{2} \int \hspace{-3pt}\rmd^d \sigma \, \varphi(\sigma) \(-\partial^2 + m^2\) \varphi(\sigma) + i \int \hspace{-3pt} \rmd^d \sigma \, j(\sigma)\pt\varphi(\sigma)\] \, .
\ee
It is well-known that the above functional integration results in
\be
\mathcal Z[\pt j \pt ] = \mathcal Z [0] \, \text{Exp}\[-\frac{i}{2} \int \hspace{-3pt} \rmd^d \sigma \int \hspace{-3pt} \rmd^d \sigma'  \, j(\sigma) \pt G(\sigma, \sigma') \pt j(\sigma')\] \, ,
\ee
where
\be\label{scalarprop}
G(\sigma, \sigma') = - i \int \frac{\rmd^d q}{(2\pi)^d} \frac{e^{i \pt q \pt (\sigma-\sigma')}}{q^2 + m^2}
\ee
is the scalar propagator. These formulae can be applied to \eqref{massaging} for $d=2$, $m=0$, $\varphi=\sqrt{T_{s,cl}} \, \zeta$ and $j(t,x) = - p \, \delta(t)\delta(x)/\sqrt{T_{s,cl}}$.

\section{Coulomb-like behavior near the color charges}
\label{app:coulomb}

Near the color charges, the resulting chromoelectric field is expected to feature a Coulomb-like behavior. See, \eg \cite{Cardoso:2013lla, Baker:2018mhw,Baker:2019gsi} for a discussion at numerical level. With our formalism, we cannot arbitrarily approach the quarks at $x=\pm L/2$. This follows from the constraint on the holographic coordinate of the classical string-like source introduced in \eqref{barvc} and reported below formula \eqref{completeprofile}. Therefore, along the inter-quark axis at $z=0$ and as close as possible to the static charges, we may try to reproduce our data with a function like
\be\label{testfunction}
\mathcal C_\pm (x) = \mathfrak a \[ \frac{1+\mathfrak b \, \mathfrak t \, (x \pm L/2) + \mathfrak c \, \mathfrak t^2 \, (x \pm L/2)^2}{\mathfrak t^2 (x \pm L/2)^2} \]^{\hspace{-1pt} 2} \, ,
\ee
where we defined
\be
\mathfrak t =  2\pi \pt T_c  (1-15 \gamma) \, ,
\ee
while $\mathfrak a$, $\mathfrak b$, $\mathfrak c$ are numerical coefficients. Expanding around $x=\pm L/2$, the leading term is nothing more than the $\mathcal O (1/(x \pm L/2)^4)$ Coulomb branch that we expect. Remember that our measure for the profile of the flux tube is the vacuum expectation value of the YM Lagrangian density.

\begin{figure}[t]
    \centering
    \begin{subfigure}{0.46\textwidth}
       \centering
        \includegraphics[width=\linewidth]{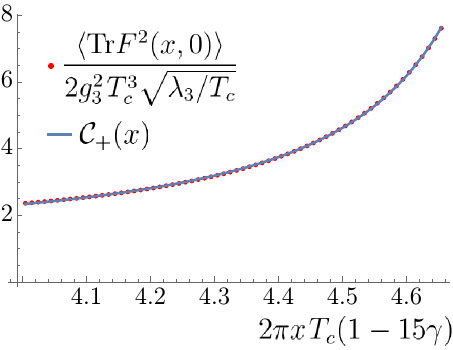} 
        \caption{}
        \label{}
    \end{subfigure}
    \hfill
    \begin{subfigure}{0.49\textwidth}
       \centering
        \includegraphics[width=\linewidth]{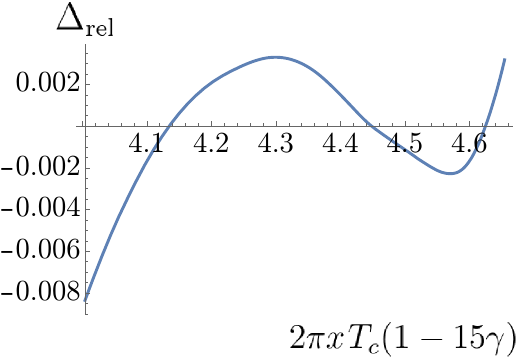} 
        \caption{}
        \label{}
    \end{subfigure}
    \caption{(a) The red dots refer to the classical flux tube profile along the inter-quark axis, up to NLO order in the strong coupling expansion. They come from the sampling of our general proposal in \eqref{completeprofile}, at $z=0$ and given $2\pi L \pt T_c (1-15\gamma)=10$, $\lambda_3/T_c=20$. On the other hand, the blue line represents the function $\mathcal C_+$ defined in \eqref{testfunction}, where the numerical coefficients has been fixed as in \eqref{fittedcoeff}. (b) The relative error defined as $\Delta_{\text{rel}} = 1 - \left\langle \Tr F^2(x,0) \right \rangle \hspace{-2pt}/ \bigl (2 \pt g_3^2 \mathcal \pt C_+ (x) \pt T_c^3 \sqrt{\lambda_3/T_c}\bigr )$, as the coordinate $x$ along the inter-quark axis varies.}
    \label{plotCoulomb}
\end{figure}
To fix ideas, let us set $2\pi L \pt T_c  (1-15 \gamma) = 10$ and $\lambda_3/T_c = 20$. We are thus referring to the fourth entry of table \ref{plotfullprofile}. Moreover, let us focus on the quark located at $x=+L/2$. The best fit to the numerical data --- coming from the sampling of our prediction in \eqref{completeprofile} for $z=0$ --- is achieved for
\be\label{fittedcoeff}
\mathfrak a \approx 0.00108 \, , \quad \mathfrak b \approx -16.0 \, , \quad \mathfrak c \approx 29.4 \, .
\ee
In figure \ref{plotCoulomb}, we report the plot of $\mathcal C_+$ with $\mathfrak a$, $\mathfrak b$, $\mathfrak c$ fixed as above. Moreover, we also show a measure of the relative error between the data from the full profile \eqref{completeprofile} at $z=0$ and the fitted curve.  All in all, we conclude that the numerical analysis of our results confirms the Coulomb-like behavior of the flux tube profile near the quarks within a very small margin of error.

\bibliographystyle{utphys}

\begin{thebibliography}{100}

\bibitem{DiGiacomo:1989yp}
A.~Di~Giacomo, M.~Maggiore, and S.~Olejnik, ``{Evidence for Flux Tubes From
  Cooled {QCD} Configurations},''
  \href{http://dx.doi.org/10.1016/0370-2693(90)90828-T}{{\em Phys. Lett. B}
  {\bfseries 236} (1990) 199--202}.

\bibitem{DiGiacomo:1990hc}
A.~Di~Giacomo, M.~Maggiore, and S.~Olejnik, ``{Confinement and Chromoelectric
  Flux Tubes in Lattice {QCD}},''
  \href{http://dx.doi.org/10.1016/0550-3213(90)90567-W}{{\em Nucl. Phys. B}
  {\bfseries 347} (1990) 441--460}.

\bibitem{Singh:1993jj}
V.~Singh, D.~A. Browne, and R.~W. Haymaker, ``{Structure of Abrikosov vortices
  in SU(2) lattice gauge theory},''
  \href{http://dx.doi.org/10.1016/0370-2693(93)91146-E}{{\em Phys. Lett. B}
  {\bfseries 306} (1993) 115--119},
  \href{http://arxiv.org/abs/hep-lat/9301004}{{\ttfamily
  arXiv:hep-lat/9301004}}.

\bibitem{Bali:1994de}
G.~S. Bali, K.~Schilling, and C.~Schlichter, ``{Observing long color flux tubes
  in SU(2) lattice gauge theory},''
  \href{http://dx.doi.org/10.1103/PhysRevD.51.5165}{{\em Phys. Rev. D}
  {\bfseries 51} (1995) 5165--5198},
  \href{http://arxiv.org/abs/hep-lat/9409005}{{\ttfamily
  arXiv:hep-lat/9409005}}.

\bibitem{Bali:2000gf}
G.~S. Bali, ``{QCD forces and heavy quark bound states},''
  \href{http://dx.doi.org/10.1016/S0370-1573(00)00079-X}{{\em Phys. Rept.}
  {\bfseries 343} (2001) 1--136},
  \href{http://arxiv.org/abs/hep-ph/0001312}{{\ttfamily arXiv:hep-ph/0001312}}.

\bibitem{Takahashi:2000te}
T.~T. Takahashi, H.~Matsufuru, Y.~Nemoto, and H.~Suganuma, ``{The Three quark
  potential in the SU(3) lattice QCD},''
  \href{http://dx.doi.org/10.1103/PhysRevLett.86.18}{{\em Phys. Rev. Lett.}
  {\bfseries 86} (2001) 18--21},
  \href{http://arxiv.org/abs/hep-lat/0006005}{{\ttfamily
  arXiv:hep-lat/0006005}}.

\bibitem{Bissey:2006bz}
F.~Bissey, F.-G. Cao, A.~R. Kitson, A.~I. Signal, D.~B. Leinweber, B.~G.
  Lasscock, and A.~G. Williams, ``{Gluon flux-tube distribution and linear
  confinement in baryons},''
  \href{http://dx.doi.org/10.1103/PhysRevD.76.114512}{{\em Phys. Rev. D}
  {\bfseries 76} (2007) 114512},
  \href{http://arxiv.org/abs/hep-lat/0606016}{{\ttfamily
  arXiv:hep-lat/0606016}}.

\bibitem{Andersson:1980vk}
B.~Andersson, G.~Gustafson, and T.~Sjostrand, ``{How to Find the Gluon Jets in
  e+ e- Annihilation},''
  \href{http://dx.doi.org/10.1016/0370-2693(80)90861-8}{{\em Phys. Lett. B}
  {\bfseries 94} (1980) 211--215}.

\bibitem{JADE:1981ofk}
{\bfseries JADE} Collaboration, W.~Bartel {\em et~al.}, ``{Experimental Study
  of Jets in electron - Positron Annihilation},''
  \href{http://dx.doi.org/10.1016/0370-2693(81)90505-0}{{\em Phys. Lett. B}
  {\bfseries 101} (1981) 129--134}.

\bibitem{JADE:1983ihf}
{\bfseries JADE} Collaboration, W.~Bartel {\em et~al.}, ``{Particle
  Distribution in Three Jet Events Produced by e+ e- Annihilation},''
  \href{http://dx.doi.org/10.1007/BF01648774}{{\em Z. Phys. C} {\bfseries 21}
  (1983) 37}.

\bibitem{JADE:1983mvo}
{\bfseries JADE} Collaboration, W.~Bartel {\em et~al.}, ``{Test of
  Fragmentation Models by Comparison with Three Jet Events Produced in e+ e-
  ---{\ensuremath{>}} Hadrons},''
  \href{http://dx.doi.org/10.1016/0370-2693(84)90687-7}{{\em Phys. Lett. B}
  {\bfseries 134} (1984) 275}.

\bibitem{CELLO:1983pbu}
{\bfseries CELLO} Collaboration, H.~J. Behrend {\em et~al.}, ``{On the Model
  Dependence of the Determination of the Strong Coupling Constant in Second
  Order {QCD} From $e^+ e^-$ Annihilation Into Hadrons},''
  \href{http://dx.doi.org/10.1016/0370-2693(84)91667-8}{{\em Phys. Lett. B}
  {\bfseries 138} (1984) 311--316}.

\bibitem{TASSO:1985cia}
{\bfseries TASSO} Collaboration, M.~Althoff {\em et~al.}, ``{A Study of Three
  Jet Events in e+ e- Annihilation Into Hadrons at 34.6-GeV Center-Of-Mass
  Energy},'' \href{http://dx.doi.org/10.1007/BF01571375}{{\em Z. Phys. C}
  {\bfseries 29} (1985) 29}.

\bibitem{TPCTwoGamma:1984tkq}
{\bfseries TPC/Two Gamma} Collaboration, H.~Aihara {\em et~al.}, ``{Tests of
  models for parton fragmentation using three jet events in $e^+e^-$
  annihiliation at $\sqrt{s} = 29$ GeV},''
  \href{http://dx.doi.org/10.1103/PhysRevLett.54.270}{{\em Phys. Rev. Lett.}
  {\bfseries 54} (1985) 270}. [Erratum: Phys.Rev.Lett. 54, 1209 (1985)].

\bibitem{Azimov:1985zta}
Y.~I. Azimov, Y.~L. Dokshitzer, V.~A. Khoze, and S.~I. Troian, ``{The String
  Effect and QCD Coherence},''
  \href{http://dx.doi.org/10.1016/0370-2693(85)90709-9}{{\em Phys. Lett. B}
  {\bfseries 165} (1985) 147--150}.

\bibitem{Anisovich:2000ut}
A.~V. Anisovich, C.~A. Baker, C.~J. Batty, D.~V. Bugg, C.~Hodd, H.~C. Lu, V.~A.
  Nikonov, A.~V. Sarantsev, V.~V. Sarantsev, and B.~S. Zou, ``{I = 0 C = +1
  mesons from 1920 to 2410 MeV},''
  \href{http://dx.doi.org/10.1016/S0370-2693(00)01018-2}{{\em Phys. Lett. B}
  {\bfseries 491} (2000) 47--58},
  \href{http://arxiv.org/abs/1109.0883}{{\ttfamily arXiv:1109.0883 [hep-ex]}}.

\bibitem{Anisovich:2001pn}
A.~V. Anisovich, C.~A. Baker, C.~J. Batty, D.~V. Bugg, V.~A. Nikonov, A.~V.
  Sarantsev, V.~V. Sarantsev, and B.~S. Zou, ``{Partial wave analysis of anti-p
  p annihilation channels in flight with I = 1, C = +1},''
  \href{http://dx.doi.org/10.1016/S0370-2693(01)01017-6}{{\em Phys. Lett. B}
  {\bfseries 517} (2001) 261--272},
  \href{http://arxiv.org/abs/1110.0278}{{\ttfamily arXiv:1110.0278 [hep-ex]}}.

\bibitem{Anisovich:2001pp}
A.~V. Anisovich, C.~A. Baker, C.~J. Batty, D.~V. Bugg, V.~A. Nikonov, A.~V.
  Sarantsev, V.~V. Sarantsev, and B.~S. Zou, ``{A partial wave analysis of
  anti-p p --{\ensuremath{>}} eta eta pi0},''
  \href{http://dx.doi.org/10.1016/S0370-2693(01)01018-8}{{\em Phys. Lett. B}
  {\bfseries 517} (2001) 273--281},
  \href{http://arxiv.org/abs/1109.6817}{{\ttfamily arXiv:1109.6817 [hep-ex]}}.

\bibitem{Anisovich:2002su}
A.~V. Anisovich, C.~A. Baker, C.~J. Batty, D.~V. Bugg, L.~Montanet, V.~A.
  Nikonov, A.~V. Sarantsev, V.~V. Sarantsev, and B.~S. Zou, ``{Combined
  analysis of meson channels with I = 1, C = -1 from 1940 to 2410 MeV},''
  \href{http://dx.doi.org/10.1016/S0370-2693(02)02302-X}{{\em Phys. Lett. B}
  {\bfseries 542} (2002) 8--18},
  \href{http://arxiv.org/abs/1109.5247}{{\ttfamily arXiv:1109.5247 [hep-ex]}}.

\bibitem{Anisovich:2002xoo}
A.~V. Anisovich, C.~A. Baker, C.~J. Batty, D.~V. Bugg, L.~Montanet, V.~A.
  Nikonov, A.~V. Sarantsev, V.~V. Sarantsev, and B.~S. Zou, ``{I = 0, C = -1
  mesons from 1940 to 2410 MeV},''
  \href{http://dx.doi.org/10.1016/S0370-2693(02)02303-1}{{\em Phys. Lett. B}
  {\bfseries 542} (2002) 19--28},
  \href{http://arxiv.org/abs/1109.5817}{{\ttfamily arXiv:1109.5817 [hep-ex]}}.

\bibitem{Bugg:2004xu}
D.~V. Bugg, ``{Four sorts of meson},''
  \href{http://dx.doi.org/10.1016/j.physrep.2004.03.008}{{\em Phys. Rept.}
  {\bfseries 397} (2004) 257--358},
  \href{http://arxiv.org/abs/hep-ex/0412045}{{\ttfamily arXiv:hep-ex/0412045}}.

\bibitem{Luscher:1980ac}
M.~Luscher, ``{Symmetry Breaking Aspects of the Roughening Transition in Gauge
  Theories},'' \href{http://dx.doi.org/10.1016/0550-3213(81)90423-5}{{\em Nucl.
  Phys. B} {\bfseries 180} (1981) 317--329}.

\bibitem{Luscher:1980fr}
M.~Luscher, K.~Symanzik, and P.~Weisz, ``{Anomalies of the Free Loop Wave
  Equation in the WKB Approximation},''
  \href{http://dx.doi.org/10.1016/0550-3213(80)90009-7}{{\em Nucl. Phys. B}
  {\bfseries 173} (1980) 365}.

\bibitem{Luscher:1980iy}
M.~Luscher, G.~Munster, and P.~Weisz, ``{How Thick Are Chromoelectric Flux
  Tubes?},'' \href{http://dx.doi.org/10.1016/0550-3213(81)90151-6}{{\em Nucl.
  Phys. B} {\bfseries 180} (1981) 1--12}.

\bibitem{Caselle:1995fh}
M.~Caselle, F.~Gliozzi, U.~Magnea, and S.~Vinti, ``{Width of long color flux
  tubes in lattice gauge systems},''
  \href{http://dx.doi.org/10.1016/0550-3213(95)00639-7}{{\em Nucl. Phys. B}
  {\bfseries 460} (1996) 397--412},
  \href{http://arxiv.org/abs/hep-lat/9510019}{{\ttfamily
  arXiv:hep-lat/9510019}}.

\bibitem{Zach:1997yz}
M.~Zach, M.~Faber, and P.~Skala, ``{Investigating confinement in dually
  transformed U(1) lattice gauge theory},''
  \href{http://dx.doi.org/10.1103/PhysRevD.57.123}{{\em Phys. Rev. D}
  {\bfseries 57} (1998) 123--131},
  \href{http://arxiv.org/abs/hep-lat/9705019}{{\ttfamily
  arXiv:hep-lat/9705019}}.

\bibitem{Koma:2003gi}
Y.~Koma, M.~Koma, and P.~Majumdar, ``{Static potential, force, and flux tube
  profile in 4-D compact U(1) lattice gauge theory with the multilevel
  algorithm},'' \href{http://dx.doi.org/10.1016/j.nuclphysb.2004.05.024}{{\em
  Nucl. Phys. B} {\bfseries 692} (2004) 209--231},
  \href{http://arxiv.org/abs/hep-lat/0311016}{{\ttfamily
  arXiv:hep-lat/0311016}}.

\bibitem{Panero:2005iu}
M.~Panero, ``{A Numerical study of confinement in compact QED},''
  \href{http://dx.doi.org/10.1088/1126-6708/2005/05/066}{{\em JHEP} {\bfseries
  05} (2005) 066}, \href{http://arxiv.org/abs/hep-lat/0503024}{{\ttfamily
  arXiv:hep-lat/0503024}}.

\bibitem{Giudice:2006hw}
P.~Giudice, F.~Gliozzi, and S.~Lottini, ``{Quantum broadening of k-strings in
  gauge theories},''
  \href{http://dx.doi.org/10.1088/1126-6708/2007/01/084}{{\em JHEP} {\bfseries
  01} (2007) 084}, \href{http://arxiv.org/abs/hep-th/0612131}{{\ttfamily
  arXiv:hep-th/0612131}}.

\bibitem{Rajantie:2012zn}
A.~Rajantie, K.~Rummukainen, and D.~J. Weir, ``{Form factor and width of a
  quantum string},'' \href{http://dx.doi.org/10.1103/PhysRevD.86.125040}{{\em
  Phys. Rev. D} {\bfseries 86} (2012) 125040},
  \href{http://arxiv.org/abs/1210.1106}{{\ttfamily arXiv:1210.1106 [hep-lat]}}.

\bibitem{Amado:2013rja}
A.~Amado, N.~Cardoso, and P.~Bicudo, ``{Flux tube widening in compact U (1)
  lattice gauge theory computed at $T < T_c$ with the multilevel method and
  GPUs},'' \href{http://arxiv.org/abs/1309.3859}{{\ttfamily arXiv:1309.3859
  [hep-lat]}}.

\bibitem{Pennanen:1997qm}
P.~Pennanen, A.~M. Green, and C.~Michael, ``{Flux tube structure and beta
  functions in SU(2)},'' \href{http://dx.doi.org/10.1103/PhysRevD.56.3903}{{\em
  Phys. Rev. D} {\bfseries 56} (1997) 3903--3916},
  \href{http://arxiv.org/abs/hep-lat/9705033}{{\ttfamily
  arXiv:hep-lat/9705033}}.

\bibitem{Chernodub:2007wi}
M.~N. Chernodub and F.~V. Gubarev, ``{Confining string and its widening in
  HP**1 embedding approach},''
  \href{http://dx.doi.org/10.1103/PhysRevD.76.016003}{{\em Phys. Rev. D}
  {\bfseries 76} (2007) 016003},
  \href{http://arxiv.org/abs/hep-lat/0703007}{{\ttfamily
  arXiv:hep-lat/0703007}}.

\bibitem{Bakry:2010zt}
A.~S. Bakry, D.~B. Leinweber, P.~J. Moran, A.~Sternbeck, and A.~G. Williams,
  ``{String effects and the distribution of the glue in mesons at finite
  temperature},'' \href{http://dx.doi.org/10.1103/PhysRevD.82.094503}{{\em
  Phys. Rev. D} {\bfseries 82} (2010) 094503},
  \href{http://arxiv.org/abs/1004.0782}{{\ttfamily arXiv:1004.0782 [hep-lat]}}.

\bibitem{Cardoso:2013lla}
N.~Cardoso, M.~Cardoso, and P.~Bicudo, ``{Inside the SU(3) quark-antiquark QCD
  flux tube: screening versus quantum widening},''
  \href{http://dx.doi.org/10.1103/PhysRevD.88.054504}{{\em Phys. Rev. D}
  {\bfseries 88} (2013) 054504},
  \href{http://arxiv.org/abs/1302.3633}{{\ttfamily arXiv:1302.3633 [hep-lat]}}.

\bibitem{Gliozzi:2010zv}
F.~Gliozzi, M.~Pepe, and U.~J. Wiese, ``{The Width of the Confining String in
  Yang-Mills Theory},''
  \href{http://dx.doi.org/10.1103/PhysRevLett.104.232001}{{\em Phys. Rev.
  Lett.} {\bfseries 104} (2010) 232001},
  \href{http://arxiv.org/abs/1002.4888}{{\ttfamily arXiv:1002.4888 [hep-lat]}}.

\bibitem{Gliozzi:2010zt}
F.~Gliozzi, M.~Pepe, and U.~J. Wiese, ``{The Width of the Color Flux Tube at
  2-Loop Order},'' \href{http://dx.doi.org/10.1007/JHEP11(2010)053}{{\em JHEP}
  {\bfseries 11} (2010) 053}, \href{http://arxiv.org/abs/1006.2252}{{\ttfamily
  arXiv:1006.2252 [hep-lat]}}.

\bibitem{nambu1970duality}
Y.~Nambu, ``Duality and hydrodynamics,'' 1970.
\newblock Lectures at the Copenhagen symposium.

\bibitem{Goto:1971ce}
T.~Goto, ``{Relativistic quantum mechanics of one-dimensional mechanical
  continuum and subsidiary condition of dual resonance model},''
  \href{http://dx.doi.org/10.1143/PTP.46.1560}{{\em Prog. Theor. Phys.}
  {\bfseries 46} (1971) 1560--1569}.

\bibitem{Hara:1971ur}
O.~Hara, ``{On origin and physical meaning of ward-like identity in
  dual-resonance model},'' \href{http://dx.doi.org/10.1143/PTP.46.1549}{{\em
  Prog. Theor. Phys.} {\bfseries 46} (1971) 1549--1559}.

\bibitem{Caselle:2024ent}
M.~Caselle, E.~Cellini, and A.~Nada, ``{Numerical determination of the width
  and shape of the effective string using Stochastic Normalizing Flows},''
  \href{http://dx.doi.org/10.1007/JHEP02(2025)090}{{\em JHEP} {\bfseries 02}
  (2025) 090}, \href{http://arxiv.org/abs/2409.15937}{{\ttfamily
  arXiv:2409.15937 [hep-lat]}}.

\bibitem{Cea:2012qw}
P.~Cea, L.~Cosmai, and A.~Papa, ``{Chromoelectric flux tubes and coherence
  length in QCD},'' \href{http://dx.doi.org/10.1103/PhysRevD.86.054501}{{\em
  Phys. Rev. D} {\bfseries 86} (2012) 054501},
  \href{http://arxiv.org/abs/1208.1362}{{\ttfamily arXiv:1208.1362 [hep-lat]}}.

\bibitem{Verzichelli:2025cqc}
L.~Verzichelli, M.~Caselle, E.~Cellini, A.~Nada, and D.~Panfalone, ``{Intrinsic
  width of the flux tube in 2+1 dimensional Yang-Mills theories},''
  \href{http://dx.doi.org/10.22323/1.466.0403}{{\em PoS} {\bfseries
  LATTICE2024} (2025) 403}, \href{http://arxiv.org/abs/2501.01740}{{\ttfamily
  arXiv:2501.01740 [hep-lat]}}.

\bibitem{Caselle:2026coc}
M.~Caselle, E.~Cellini, A.~Nada, D.~Panfalone, and L.~Verzichelli, ``{Intrinsic
  Width of the Flux Tube as a tool to explore confining mechanisms in Lattice
  Gauge Theories},'' \href{http://dx.doi.org/10.1007/JHEP07(2026)027}{{\em JHEP} {\bfseries 07}
  (2026) 027}, \href{http://arxiv.org/abs/2601.19520}{{\ttfamily
  arXiv:2601.19520 [hep-lat]}}.


\bibitem{Amorosso:2026mdo}
R.~Amorosso, S.~Syritsyn, and R.~Venugopalan, ``{Entanglement Enabled
  Tomography of Flux Tubes in (2+1)D Yang-Mills Theory},'' \href{http://dx.doi.org/10.1007/JHEP08(2026)167}{{\em JHEP} {\bfseries 08}
  (2026) 167}, \href{http://arxiv.org/abs/2601.17199}{{\ttfamily arXiv:2601.17199
  [hep-th]}}.

\bibitem{Nambu:1974zg}
Y.~Nambu, ``{Strings, Monopoles and Gauge Fields},''
  \href{http://dx.doi.org/10.1103/PhysRevD.10.4262}{{\em Phys. Rev. D}
  {\bfseries 10} (1974) 4262}.

\bibitem{Mandelstam:1974pi}
S.~Mandelstam, ``{Vortices and Quark Confinement in Nonabelian Gauge
  Theories},'' \href{http://dx.doi.org/10.1016/0370-1573(76)90043-0}{{\em Phys.
  Rept.} {\bfseries 23} (1976) 245--249}.

\bibitem{tHooft:1979rtg}
G.~'t~Hooft, ``{A Property of Electric and Magnetic Flux in Nonabelian Gauge
  Theories},'' \href{http://dx.doi.org/10.1016/0550-3213(79)90595-9}{{\em Nucl.
  Phys. B} {\bfseries 153} (1979) 141--160}.

\bibitem{Baker:1989qp}
M.~Baker, J.~S. Ball, and F.~Zachariasen, ``{{QCD} Flux Tubes for SU(3)},''
  \href{http://dx.doi.org/10.1103/PhysRevD.41.2612}{{\em Phys. Rev. D}
  {\bfseries 41} (1990) 2612}.

\bibitem{Baker:1991bc}
M.~Baker, J.~S. Ball, and F.~Zachariasen, ``{Dual QCD: A Review},''
  \href{http://dx.doi.org/10.1016/0370-1573(91)90123-4}{{\em Phys. Rept.}
  {\bfseries 209} (1991) 73--127}.

\bibitem{Cea:1992sd}
P.~Cea and L.~Cosmai, ``{Lattice investigation of dual superconductor mechanism
  of confinement},'' \href{http://dx.doi.org/10.1016/0920-5632(93)90276-C}{{\em
  Nucl. Phys. B Proc. Suppl.} {\bfseries 30} (1993) 572--575}.

\bibitem{Cea:1992vx}
P.~Cea and L.~Cosmai, ``{Dual superconductor mechanism of confinement on the
  lattice},'' \href{http://dx.doi.org/10.1007/BF02768788}{{\em Nuovo Cim. A}
  {\bfseries 107} (1994) 541--548},
  \href{http://arxiv.org/abs/hep-lat/9210030}{{\ttfamily
  arXiv:hep-lat/9210030}}.

\bibitem{Cea:1993pi}
P.~Cea and L.~Cosmai, ``{On The Meissner effect in SU(2) lattice gauge
  theory},'' \href{http://dx.doi.org/10.1016/0920-5632(94)90350-6}{{\em Nucl.
  Phys. B Proc. Suppl.} {\bfseries 34} (1994) 219--221},
  \href{http://arxiv.org/abs/hep-lat/9311023}{{\ttfamily
  arXiv:hep-lat/9311023}}.

\bibitem{Cea:1994ed}
P.~Cea and L.~Cosmai, ``{Dual Meissner effect and string tension in SU(2)
  lattice gauge theory},''
  \href{http://dx.doi.org/10.1016/0370-2693(95)00299-Z}{{\em Phys. Lett. B}
  {\bfseries 349} (1995) 343--347},
  \href{http://arxiv.org/abs/hep-lat/9404017}{{\ttfamily
  arXiv:hep-lat/9404017}}.

\bibitem{Cea:1994aj}
P.~Cea and L.~Cosmai, ``{London penetration length and string tension in SU(2)
  lattice gauge theory},''
  \href{http://dx.doi.org/10.1016/0920-5632(95)00208-Q}{{\em Nucl. Phys. B
  Proc. Suppl.} {\bfseries 42} (1995) 225--227},
  \href{http://arxiv.org/abs/hep-lat/9411048}{{\ttfamily
  arXiv:hep-lat/9411048}}.

\bibitem{Cea:1995zt}
P.~Cea and L.~Cosmai, ``{Dual superconductivity in the SU(2) pure gauge vacuum:
  A Lattice study},'' \href{http://dx.doi.org/10.1103/PhysRevD.52.5152}{{\em
  Phys. Rev. D} {\bfseries 52} (1995) 5152--5164},
  \href{http://arxiv.org/abs/hep-lat/9504008}{{\ttfamily
  arXiv:hep-lat/9504008}}.

\bibitem{Cea:1995ga}
P.~Cea and L.~Cosmai, ``{The SU(2) confining vacuum as a dual
  superconductor},'' \href{http://dx.doi.org/10.1016/0920-5632(96)00065-5}{{\em
  Nucl. Phys. B Proc. Suppl.} {\bfseries 47} (1996) 318--321},
  \href{http://arxiv.org/abs/hep-lat/9509007}{{\ttfamily
  arXiv:hep-lat/9509007}}.

\bibitem{Cardaci:2010tb}
M.~S. Cardaci, P.~Cea, L.~Cosmai, R.~Falcone, and A.~Papa, ``{Chromoelectric
  flux tubes in QCD},''
  \href{http://dx.doi.org/10.1103/PhysRevD.83.014502}{{\em Phys. Rev. D}
  {\bfseries 83} (2011) 014502},
  \href{http://arxiv.org/abs/1011.5803}{{\ttfamily arXiv:1011.5803 [hep-lat]}}.

\bibitem{Cea:2014uja}
P.~Cea, L.~Cosmai, F.~Cuteri, and A.~Papa, ``{Flux tubes in the SU(3) vacuum:
  London penetration depth and coherence length},''
  \href{http://dx.doi.org/10.1103/PhysRevD.89.094505}{{\em Phys. Rev. D}
  {\bfseries 89} no.~9, (2014) 094505},
  \href{http://arxiv.org/abs/1404.1172}{{\ttfamily arXiv:1404.1172 [hep-lat]}}.

\bibitem{Cea:2015wjd}
P.~Cea, L.~Cosmai, F.~Cuteri, and A.~Papa, ``{Flux tubes at finite
  temperature},'' \href{http://dx.doi.org/10.1007/JHEP06(2016)033}{{\em JHEP}
  {\bfseries 06} (2016) 033}, \href{http://arxiv.org/abs/1511.01783}{{\ttfamily
  arXiv:1511.01783 [hep-lat]}}.

\bibitem{Cea:2017ocq}
P.~Cea, L.~Cosmai, F.~Cuteri, and A.~Papa, ``{Flux tubes in the QCD vacuum},''
  \href{http://dx.doi.org/10.1103/PhysRevD.95.114511}{{\em Phys. Rev. D}
  {\bfseries 95} no.~11, (2017) 114511},
  \href{http://arxiv.org/abs/1702.06437}{{\ttfamily arXiv:1702.06437
  [hep-lat]}}.

\bibitem{Aharony:2024ctf}
O.~Aharony, N.~Barel, and T.~Sheaffer, ``{Effective strings in QED$_{3}$},''
  \href{http://dx.doi.org/10.1007/JHEP03(2025)143}{{\em JHEP} {\bfseries 03}
  (2025) 143}, \href{http://arxiv.org/abs/2412.01313}{{\ttfamily
  arXiv:2412.01313 [hep-th]}}.

\bibitem{Caselle:2025vhx}
M.~Caselle and A.~Mariani, ``{The finite temperature ground state energy of the
  confining string in three-dimensional U(1) gauge theory},''
  \href{http://dx.doi.org/10.1007/JHEP06(2025)010}{{\em JHEP} {\bfseries 06}
  (2025) 010}, \href{http://arxiv.org/abs/2503.05354}{{\ttfamily
  arXiv:2503.05354 [hep-lat]}}.

\bibitem{tHooft:1993dmi}
G.~'t~Hooft, ``{Dimensional reduction in quantum gravity},'' {\em Conf. Proc.
  C} {\bfseries 930308} (1993) 284--296,
  \href{http://arxiv.org/abs/gr-qc/9310026}{{\ttfamily arXiv:gr-qc/9310026}}.

\bibitem{Susskind:1994vu}
L.~Susskind, ``{The World as a hologram},''
  \href{http://dx.doi.org/10.1063/1.531249}{{\em J. Math. Phys.} {\bfseries 36}
  (1995) 6377--6396}, \href{http://arxiv.org/abs/hep-th/9409089}{{\ttfamily
  arXiv:hep-th/9409089}}.

\bibitem{Maldacena:1997re}
J.~M. Maldacena, ``{The Large N limit of superconformal field theories and
  supergravity},'' \href{http://dx.doi.org/10.4310/ATMP.1998.v2.n2.a1}{{\em
  Adv. Theor. Math. Phys.} {\bfseries 2} (1998) 231--252},
  \href{http://arxiv.org/abs/hep-th/9711200}{{\ttfamily arXiv:hep-th/9711200}}.

\bibitem{Gubser:1998bc}
S.~S. Gubser, I.~R. Klebanov, and A.~M. Polyakov, ``{Gauge theory correlators
  from noncritical string theory},''
  \href{http://dx.doi.org/10.1016/S0370-2693(98)00377-3}{{\em Phys. Lett. B}
  {\bfseries 428} (1998) 105--114},
  \href{http://arxiv.org/abs/hep-th/9802109}{{\ttfamily arXiv:hep-th/9802109}}.

\bibitem{Witten:1998qj}
E.~Witten, ``{Anti-de Sitter space and holography},''
  \href{http://dx.doi.org/10.4310/ATMP.1998.v2.n2.a2}{{\em Adv. Theor. Math.
  Phys.} {\bfseries 2} (1998) 253--291},
  \href{http://arxiv.org/abs/hep-th/9802150}{{\ttfamily arXiv:hep-th/9802150}}.

\bibitem{Witten:1998zw}
E.~Witten, ``{Anti-de Sitter space, thermal phase transition, and confinement
  in gauge theories},''
  \href{http://dx.doi.org/10.4310/ATMP.1998.v2.n3.a3}{{\em Adv. Theor. Math.
  Phys.} {\bfseries 2} (1998) 505--532},
  \href{http://arxiv.org/abs/hep-th/9803131}{{\ttfamily arXiv:hep-th/9803131}}.

\bibitem{Polchinski:1991ax}
J.~Polchinski and A.~Strominger, ``{Effective string theory},''
  \href{http://dx.doi.org/10.1103/PhysRevLett.67.1681}{{\em Phys. Rev. Lett.}
  {\bfseries 67} (1991) 1681--1684}.

\bibitem{Aharony:2009gg}
O.~Aharony and E.~Karzbrun, ``{On the effective action of confining strings},''
  \href{http://dx.doi.org/10.1088/1126-6708/2009/06/012}{{\em JHEP} {\bfseries
  06} (2009) 012}, \href{http://arxiv.org/abs/0903.1927}{{\ttfamily
  arXiv:0903.1927 [hep-th]}}.

\bibitem{Aharony:2010cx}
O.~Aharony and M.~Field, ``{On the effective theory of long open strings},''
  \href{http://dx.doi.org/10.1007/JHEP01(2011)065}{{\em JHEP} {\bfseries 01}
  (2011) 065}, \href{http://arxiv.org/abs/1008.2636}{{\ttfamily arXiv:1008.2636
  [hep-th]}}.

\bibitem{tHooft:1973alw}
G.~'t~Hooft, ``{A Planar Diagram Theory for Strong Interactions},''
  \href{http://dx.doi.org/10.1016/0550-3213(74)90154-0}{{\em Nucl. Phys. B}
  {\bfseries 72} (1974) 461}.

\bibitem{Gross:1998gk}
D.~J. Gross and H.~Ooguri, ``{Aspects of large N gauge theory dynamics as seen
  by string theory},'' \href{http://dx.doi.org/10.1103/PhysRevD.58.106002}{{\em
  Phys. Rev. D} {\bfseries 58} (1998) 106002},
  \href{http://arxiv.org/abs/hep-th/9805129}{{\ttfamily arXiv:hep-th/9805129}}.

\bibitem{Rey:1998ik}
S.-J. Rey and J.-T. Yee, ``{Macroscopic strings as heavy quarks in large N
  gauge theory and anti-de Sitter supergravity},''
  \href{http://dx.doi.org/10.1007/s100520100799}{{\em Eur. Phys. J. C}
  {\bfseries 22} (2001) 379--394},
  \href{http://arxiv.org/abs/hep-th/9803001}{{\ttfamily arXiv:hep-th/9803001}}.

\bibitem{Maldacena:1998im}
J.~M. Maldacena, ``{Wilson loops in large N field theories},''
  \href{http://dx.doi.org/10.1103/PhysRevLett.80.4859}{{\em Phys. Rev. Lett.}
  {\bfseries 80} (1998) 4859--4862},
  \href{http://arxiv.org/abs/hep-th/9803002}{{\ttfamily arXiv:hep-th/9803002}}.

\bibitem{Kinar:1999xu}
Y.~Kinar, E.~Schreiber, J.~Sonnenschein, and N.~Weiss, ``{Quantum fluctuations
  of Wilson loops from string models},''
  \href{http://dx.doi.org/10.1016/S0550-3213(00)00238-8}{{\em Nucl. Phys. B}
  {\bfseries 583} (2000) 76--104},
  \href{http://arxiv.org/abs/hep-th/9911123}{{\ttfamily arXiv:hep-th/9911123}}.

\bibitem{Greensite:1999jw}
J.~Greensite and P.~Olesen, ``{World sheet fluctuations and the heavy quark
  potential in the AdS / CFT approach},''
  \href{http://dx.doi.org/10.1088/1126-6708/1999/04/001}{{\em JHEP} {\bfseries
  04} (1999) 001}, \href{http://arxiv.org/abs/hep-th/9901057}{{\ttfamily
  arXiv:hep-th/9901057}}.

\bibitem{Kinar:1998vq}
Y.~Kinar, E.~Schreiber, and J.~Sonnenschein, ``{Q anti-Q potential from strings
  in curved space-time: Classical results},''
  \href{http://dx.doi.org/10.1016/S0550-3213(99)00652-5}{{\em Nucl. Phys. B}
  {\bfseries 566} (2000) 103--125},
  \href{http://arxiv.org/abs/hep-th/9811192}{{\ttfamily arXiv:hep-th/9811192}}.

\bibitem{Greensite:1998bp}
J.~Greensite and P.~Olesen, ``{Remarks on the heavy quark potential in the
  supergravity approach},''
  \href{http://dx.doi.org/10.1088/1126-6708/1998/08/009}{{\em JHEP} {\bfseries
  08} (1998) 009}, \href{http://arxiv.org/abs/hep-th/9806235}{{\ttfamily
  arXiv:hep-th/9806235}}.

\bibitem{Klebanov:1997kc}
I.~R. Klebanov, ``{World volume approach to absorption by nondilatonic
  branes},'' \href{http://dx.doi.org/10.1016/S0550-3213(97)00235-6}{{\em Nucl.
  Phys. B} {\bfseries 496} (1997) 231--242},
  \href{http://arxiv.org/abs/hep-th/9702076}{{\ttfamily arXiv:hep-th/9702076}}.

\bibitem{Klebanov:1999xv}
I.~R. Klebanov, W.~Taylor, and M.~Van~Raamsdonk, ``{Absorption of dilaton
  partial waves by D3-branes},''
  \href{http://dx.doi.org/10.1016/S0550-3213(99)00448-4}{{\em Nucl. Phys. B}
  {\bfseries 560} (1999) 207--229},
  \href{http://arxiv.org/abs/hep-th/9905174}{{\ttfamily arXiv:hep-th/9905174}}.

\bibitem{Armoni:2008sy}
A.~Armoni and J.~M. Ridgway, ``{Quantum Broadening of k-Strings from the
  AdS/CFT Correspondence},''
  \href{http://dx.doi.org/10.1016/j.nuclphysb.2008.04.022}{{\em Nucl. Phys. B}
  {\bfseries 801} (2008) 118--127},
  \href{http://arxiv.org/abs/0803.2409}{{\ttfamily arXiv:0803.2409 [hep-th]}}.

\bibitem{Giataganas:2015yaa}
D.~Giataganas and N.~Irges, ``{On the holographic width of flux tubes},''
  \href{http://dx.doi.org/10.1007/JHEP05(2015)105}{{\em JHEP} {\bfseries 05}
  (2015) 105}, \href{http://arxiv.org/abs/1502.05083}{{\ttfamily
  arXiv:1502.05083 [hep-th]}}.

\bibitem{Giataganas:2015xna}
D.~Giataganas, ``{Properties of Confinement in Holography},''
  \href{http://dx.doi.org/10.22323/1.231.0150}{{\em PoS} {\bfseries CORFU2014}
  (2015) 150}, \href{http://arxiv.org/abs/1505.07065}{{\ttfamily
  arXiv:1505.07065 [hep-th]}}.

\bibitem{Greensite:2000cs}
J.~Greensite and P.~Olesen, ``{Broadening of the QCD(3) flux tube from the AdS
  / CFT correspondence},''
  \href{http://dx.doi.org/10.1088/1126-6708/2000/11/030}{{\em JHEP} {\bfseries
  11} (2000) 030}, \href{http://arxiv.org/abs/hep-th/0008080}{{\ttfamily
  arXiv:hep-th/0008080}}.

\bibitem{Loewy:2001pq}
A.~Loewy and J.~Sonnenschein, ``{On the holographic duals of N=1 gauge
  dynamics},'' \href{http://dx.doi.org/10.1088/1126-6708/2001/08/007}{{\em
  JHEP} {\bfseries 08} (2001) 007},
  \href{http://arxiv.org/abs/hep-th/0103163}{{\ttfamily arXiv:hep-th/0103163}}.

\bibitem{Ridgway:2009tca}
J.~M. Ridgway, {\em {Aspects of $k$-strings $\& k$-domain walls in the AdS/CFT
  correspondence}}.
\newblock PhD thesis, Swansea U., 2009.

\bibitem{Cardoso:2006mf}
M.~Cardoso, P.~Bicudo, and P.~D. Sacramento, ``{Confinement of monopole field
  lines in a superconductor at T does not equal 0},''
  \href{http://dx.doi.org/10.1016/j.aop.2007.02.007}{{\em Annals Phys.}
  {\bfseries 323} (2008) 337--355},
  \href{http://arxiv.org/abs/hep-ph/0607218}{{\ttfamily arXiv:hep-ph/0607218}}.

\bibitem{Cardoso:2010kw}
N.~Cardoso, M.~Cardoso, and P.~Bicudo, ``{SU(3) gauge invariant lattice QCD
  exploration of the dual superconductor picture in flux tube fusion, in the
  dual gluon mass, and in the dual Ginzburg-Landau parameters},''
  \href{http://arxiv.org/abs/1004.0166}{{\ttfamily arXiv:1004.0166 [hep-lat]}}.

\bibitem{Danielsson:1998wt}
U.~H. Danielsson, E.~Keski-Vakkuri, and M.~Kruczenski, ``{Vacua, propagators,
  and holographic probes in AdS / CFT},''
  \href{http://dx.doi.org/10.1088/1126-6708/1999/01/002}{{\em JHEP} {\bfseries
  01} (1999) 002}, \href{http://arxiv.org/abs/hep-th/9812007}{{\ttfamily
  arXiv:hep-th/9812007}}.

\bibitem{Gubser:1998nz}
S.~S. Gubser, I.~R. Klebanov, and A.~A. Tseytlin, ``{Coupling constant
  dependence in the thermodynamics of N=4 supersymmetric Yang-Mills theory},''
  \href{http://dx.doi.org/10.1016/S0550-3213(98)00514-8}{{\em Nucl. Phys. B}
  {\bfseries 534} (1998) 202--222},
  \href{http://arxiv.org/abs/hep-th/9805156}{{\ttfamily arXiv:hep-th/9805156}}.

\bibitem{Pawelczyk:1998pb}
J.~Pawelczyk and S.~Theisen, ``{AdS(5) x S**5 black hole metric at
  O(alpha-prime**3)},''
  \href{http://dx.doi.org/10.1088/1126-6708/1998/09/010}{{\em JHEP} {\bfseries
  09} (1998) 010}, \href{http://arxiv.org/abs/hep-th/9808126}{{\ttfamily
  arXiv:hep-th/9808126}}.

\bibitem{Vyas:2019kvy}
V.~Vyas, ``{Flux-tubes in confining gauge theories with gravitational dual},''
  \href{http://arxiv.org/abs/1904.06777}{{\ttfamily arXiv:1904.06777
  [hep-th]}}.

\bibitem{Ooguri:1998hq}
H.~Ooguri, H.~Robins, and J.~Tannenhauser, ``{Glueballs and their Kaluza-Klein
  cousins},'' \href{http://dx.doi.org/10.1016/S0370-2693(98)00877-6}{{\em Phys.
  Lett. B} {\bfseries 437} (1998) 77--81},
  \href{http://arxiv.org/abs/hep-th/9806171}{{\ttfamily arXiv:hep-th/9806171}}.

\bibitem{Csaki:1998qr}
C.~Csaki, H.~Ooguri, Y.~Oz, and J.~Terning, ``{Glueball mass spectrum from
  supergravity},'' \href{http://dx.doi.org/10.1088/1126-6708/1999/01/017}{{\em
  JHEP} {\bfseries 01} (1999) 017},
  \href{http://arxiv.org/abs/hep-th/9806021}{{\ttfamily arXiv:hep-th/9806021}}.

\bibitem{deMelloKoch:1998vqw}
R.~de~Mello~Koch, A.~Jevicki, M.~Mihailescu, and J.~P. Nunes, ``{Evaluation of
  glueball masses from supergravity},''
  \href{http://dx.doi.org/10.1103/PhysRevD.58.105009}{{\em Phys. Rev. D}
  {\bfseries 58} (1998) 105009},
  \href{http://arxiv.org/abs/hep-th/9806125}{{\ttfamily arXiv:hep-th/9806125}}.

\bibitem{toappear}
L.~Verzichelli, T.~Canneti, M.~Caselle, E.~Cellini, A.~Nada, and D.~Panfalone.
  To appear.

\bibitem{Grisaru:1986vi}
M.~T. Grisaru and D.~Zanon, ``{$\sigma$ Model Superstring Corrections to the
  Einstein-hilbert Action},''
  \href{http://dx.doi.org/10.1016/0370-2693(86)90765-3}{{\em Phys. Lett. B}
  {\bfseries 177} (1986) 347--351}.

\bibitem{Freeman:1986zh}
M.~D. Freeman, C.~N. Pope, M.~F. Sohnius, and K.~S. Stelle, ``{Higher Order
  $\sigma$ Model Counterterms and the Effective Action for Superstrings},''
  \href{http://dx.doi.org/10.1016/0370-2693(86)91495-4}{{\em Phys. Lett. B}
  {\bfseries 178} (1986) 199--204}.

\bibitem{Park:1987jp}
Q.-H. Park and D.~Zanon, ``{More on $\sigma$ Model Beta Functions and
  Low-energy Effective Actions},''
  \href{http://dx.doi.org/10.1103/PhysRevD.35.4038}{{\em Phys. Rev. D}
  {\bfseries 35} (1987) 4038}.

\bibitem{Gross:1986iv}
D.~J. Gross and E.~Witten, ``{Superstring Modifications of Einstein's
  Equations},'' \href{http://dx.doi.org/10.1016/0550-3213(86)90429-3}{{\em
  Nucl. Phys. B} {\bfseries 277} (1986) 1}.

\bibitem{Tseytlin:1986zz}
A.~A. Tseytlin, ``{Ambiguity in the Effective Action in String Theories},''
  \href{http://dx.doi.org/10.1016/0370-2693(86)90930-5}{{\em Phys. Lett. B}
  {\bfseries 176} (1986) 92--98}.

\bibitem{Green:1997tv}
M.~B. Green and M.~Gutperle, ``{Effects of D instantons},''
  \href{http://dx.doi.org/10.1016/S0550-3213(97)00269-1}{{\em Nucl. Phys. B}
  {\bfseries 498} (1997) 195--227},
  \href{http://arxiv.org/abs/hep-th/9701093}{{\ttfamily arXiv:hep-th/9701093}}.

\bibitem{Green:1997di}
M.~B. Green and P.~Vanhove, ``{D instantons, strings and M theory},''
  \href{http://dx.doi.org/10.1016/S0370-2693(97)00785-5}{{\em Phys. Lett. B}
  {\bfseries 408} (1997) 122--134},
  \href{http://arxiv.org/abs/hep-th/9704145}{{\ttfamily arXiv:hep-th/9704145}}.

\bibitem{Kiritsis:1997em}
E.~Kiritsis and B.~Pioline, ``{On R**4 threshold corrections in IIb string
  theory and (p, q) string instantons},''
  \href{http://dx.doi.org/10.1016/S0550-3213(97)00645-7}{{\em Nucl. Phys. B}
  {\bfseries 508} (1997) 509--534},
  \href{http://arxiv.org/abs/hep-th/9707018}{{\ttfamily arXiv:hep-th/9707018}}.

\bibitem{Green:1998by}
M.~B. Green and S.~Sethi, ``{Supersymmetry constraints on type IIB
  supergravity},'' \href{http://dx.doi.org/10.1103/PhysRevD.59.046006}{{\em
  Phys. Rev. D} {\bfseries 59} (1999) 046006},
  \href{http://arxiv.org/abs/hep-th/9808061}{{\ttfamily arXiv:hep-th/9808061}}.

\bibitem{Liu:2025uqu}
J.~Liu, R.~Minasian, R.~Savelli, and A.~Schachner, ``{Type IIB at eight
  derivatives: Five-Point Axio-Dilaton Couplings},''
  \href{http://arxiv.org/abs/2507.07934}{{\ttfamily arXiv:2507.07934
  [hep-th]}}.

\bibitem{Paulos:2008tn}
M.~F. Paulos, ``{Higher derivative terms including the Ramond-Ramond
  five-form},'' \href{http://dx.doi.org/10.1088/1126-6708/2008/10/047}{{\em
  JHEP} {\bfseries 10} (2008) 047},
  \href{http://arxiv.org/abs/0804.0763}{{\ttfamily arXiv:0804.0763 [hep-th]}}.

\bibitem{Green:2003an}
M.~B. Green and C.~Stahn, ``{D3-branes on the Coulomb branch and instantons},''
  \href{http://dx.doi.org/10.1088/1126-6708/2003/09/052}{{\em JHEP} {\bfseries
  09} (2003) 052}, \href{http://arxiv.org/abs/hep-th/0308061}{{\ttfamily
  arXiv:hep-th/0308061}}.

\bibitem{Rajaraman:2005up}
A.~Rajaraman, ``{On a supersymmetric completion of the R**4 term in type IIB
  supergravity},'' \href{http://dx.doi.org/10.1103/PhysRevD.74.085018}{{\em
  Phys. Rev. D} {\bfseries 72} (2005) 125008},
  \href{http://arxiv.org/abs/hep-th/0505155}{{\ttfamily arXiv:hep-th/0505155}}.

\bibitem{Garousi:2013tca}
M.~R. Garousi, ``{S-duality invariant dilaton couplings at order
  $\alpha'^3$},'' \href{http://dx.doi.org/10.1007/JHEP10(2013)076}{{\em JHEP}
  {\bfseries 10} (2013) 076}, \href{http://arxiv.org/abs/1306.6851}{{\ttfamily
  arXiv:1306.6851 [hep-th]}}.

\bibitem{Blau2025notes}
M.~Blau, ``Lecture notes on general relativity.''
  \url{http://www.blau.itp.unibe.ch/GRLecturenotes.html}.
\newblock Version: July 21, 2025.

\bibitem{Callan:1999ki}
C.~G. Callan, Jr. and A.~Guijosa, ``{Undulating strings and gauge theory
  waves},'' \href{http://dx.doi.org/10.1016/S0550-3213(99)00630-6}{{\em Nucl.
  Phys. B} {\bfseries 565} (2000) 157--175},
  \href{http://arxiv.org/abs/hep-th/9906153}{{\ttfamily arXiv:hep-th/9906153}}.

\bibitem{NIST:DLMF}
``{\it NIST Digital Library of Mathematical Functions}.''
  \url{https://dlmf.nist.gov/}, release 1.2.3 of 2024-12-15.
\newblock \url{https://dlmf.nist.gov/}. F.~W.~J. Olver, A.~B. {Olde Daalhuis},
  D.~W. Lozier, B.~I. Schneider, R.~F. Boisvert, C.~W. Clark, B.~R. Miller,
  B.~V. Saunders, H.~S. Cohl, and M.~A. McClain, eds.

\bibitem{Zyskin:1998tg}
M.~Zyskin, ``{A Note on the glueball mass spectrum},''
  \href{http://dx.doi.org/10.1016/S0370-2693(98)01067-3}{{\em Phys. Lett. B}
  {\bfseries 439} (1998) 373--381},
  \href{http://arxiv.org/abs/hep-th/9806128}{{\ttfamily arXiv:hep-th/9806128}}.

\bibitem{Minahan:1998tm}
J.~A. Minahan, ``{Glueball mass spectra and other issues for supergravity duals
  of QCD models},'' \href{http://dx.doi.org/10.1088/1126-6708/1999/01/020}{{\em
  JHEP} {\bfseries 01} (1999) 020},
  \href{http://arxiv.org/abs/hep-th/9811156}{{\ttfamily arXiv:hep-th/9811156}}.

\bibitem{Brower:2000rp}
R.~C. Brower, S.~D. Mathur, and C.-I. Tan, ``{Glueball spectrum for QCD from
  AdS supergravity duality},''
  \href{http://dx.doi.org/10.1016/S0550-3213(00)00435-1}{{\em Nucl. Phys. B}
  {\bfseries 587} (2000) 249--276},
  \href{http://arxiv.org/abs/hep-th/0003115}{{\ttfamily arXiv:hep-th/0003115}}.

\bibitem{GradshteynRyzhik}
I.~S. Gradshteyn and I.~M. Ryzhik, {\em {Table of Integrals, Series, and
  Products}}.
\newblock Academic Press, seventh~ed., 2007.

\bibitem{Constable:1999gb}
N.~R. Constable and R.~C. Myers, ``{Spin two glueballs, positive energy
  theorems and the AdS / CFT correspondence},''
  \href{http://dx.doi.org/10.1088/1126-6708/1999/10/037}{{\em JHEP} {\bfseries
  10} (1999) 037}, \href{http://arxiv.org/abs/hep-th/9908175}{{\ttfamily
  arXiv:hep-th/9908175}}.

\bibitem{Bigazzi:2015bna}
F.~Bigazzi, A.~L. Cotrone, and R.~Sisca, ``{Notes on Theta Dependence in
  Holographic Yang-Mills},''
  \href{http://dx.doi.org/10.1007/JHEP08(2015)090}{{\em JHEP} {\bfseries 08}
  (2015) 090}, \href{http://arxiv.org/abs/1506.03826}{{\ttfamily
  arXiv:1506.03826 [hep-th]}}.

\bibitem{Callan:1998iq}
C.~G. Callan, Jr., A.~Guijosa, and K.~G. Savvidy, ``{Baryons and string
  creation from the five-brane world volume action},''
  \href{http://dx.doi.org/10.1016/S0550-3213(99)00057-7}{{\em Nucl. Phys. B}
  {\bfseries 547} (1999) 127--142},
  \href{http://arxiv.org/abs/hep-th/9810092}{{\ttfamily arXiv:hep-th/9810092}}.

\bibitem{Callan:1999zf}
C.~G. Callan, Jr., A.~Guijosa, K.~G. Savvidy, and O.~Tafjord, ``{Baryons and
  flux tubes in confining gauge theories from brane actions},''
  \href{http://dx.doi.org/10.1016/S0550-3213(99)00312-0}{{\em Nucl. Phys. B}
  {\bfseries 555} (1999) 183--200},
  \href{http://arxiv.org/abs/hep-th/9902197}{{\ttfamily arXiv:hep-th/9902197}}.

\bibitem{Baker:2018mhw}
M.~Baker, P.~Cea, V.~Chelnokov, L.~Cosmai, F.~Cuteri, and A.~Papa, ``{Isolating
  the confining color field in the SU(3) flux tube},''
  \href{http://dx.doi.org/10.1140/epjc/s10052-019-6978-y}{{\em Eur. Phys. J. C}
  {\bfseries 79} no.~6, (2019) 478},
  \href{http://arxiv.org/abs/1810.07133}{{\ttfamily arXiv:1810.07133
  [hep-lat]}}.

\bibitem{Baker:2019gsi}
M.~Baker, P.~Cea, V.~Chelnokov, L.~Cosmai, F.~Cuteri, and A.~Papa, ``{The
  confining color field in SU(3) gauge theory},''
  \href{http://dx.doi.org/10.1140/epjc/s10052-020-8077-5}{{\em Eur. Phys. J. C}
  {\bfseries 80} no.~6, (2020) 514},
  \href{http://arxiv.org/abs/1912.04739}{{\ttfamily arXiv:1912.04739
  [hep-lat]}}.

\bibitem{Baker:2022cwb}
M.~Baker, V.~Chelnokov, L.~Cosmai, F.~Cuteri, and A.~Papa, ``{Unveiling
  confinement in pure gauge SU(3): flux tubes, fields, and magnetic
  currents},'' \href{http://dx.doi.org/10.1140/epjc/s10052-022-10848-2}{{\em
  Eur. Phys. J. C} {\bfseries 82} no.~10, (2022) 937},
  \href{http://arxiv.org/abs/2207.08797}{{\ttfamily arXiv:2207.08797
  [hep-lat]}}. [Erratum: Eur.Phys.J.C 83, 1063 (2023)].

\bibitem{Baker:2023dnn}
M.~Baker, V.~Chelnokov, L.~Cosmai, F.~Cuteri, and A.~Papa, ``{Unveiling SU(3)
  flux tubes at nonzero temperature: electric fields and magnetic currents},''
  \href{http://dx.doi.org/10.1140/epjc/s10052-024-12472-8}{{\em Eur. Phys. J.
  C} {\bfseries 84} no.~2, (2024) 150},
  \href{http://arxiv.org/abs/2310.04298}{{\ttfamily arXiv:2310.04298
  [hep-lat]}}.

\bibitem{Baker:2024peg}
M.~Baker, P.~Cea, V.~Chelnokov, L.~Cosmai, and A.~Papa, ``{Unveiling the flux
  tube structure in full QCD},''
  \href{http://dx.doi.org/10.1140/epjc/s10052-024-13725-2}{{\em Eur. Phys. J.
  C} {\bfseries 85} no.~1, (2025) 29},
  \href{http://arxiv.org/abs/2409.20168}{{\ttfamily arXiv:2409.20168
  [hep-lat]}}.

\bibitem{Baker:2024rjq}
M.~Baker, P.~Cea, V.~Chelnokov, L.~Cosmai, and A.~Papa, ``{Investigating the
  Flux Tube Structure within Full QCD},''
  \href{http://dx.doi.org/10.22323/1.466.0390}{{\em PoS} {\bfseries
  LATTICE2024} (2025) 390}, \href{http://arxiv.org/abs/2411.01886}{{\ttfamily
  arXiv:2411.01886 [hep-lat]}}.

\end{thebibliography}

\providecommand{\href}[2]{#2}\begingroup\raggedright\endgroup

\end{document}